\documentclass[fleqn,usenatbib]{mnras}

\usepackage{newtxtext,newtxmath}
\usepackage{enumitem} % For customizing enumerate

\usepackage[T1]{fontenc}

\DeclareRobustCommand{\VAN}[3]{#2}
\let\VANthebibliography\thebibliography
\def\thebibliography{\DeclareRobustCommand{\VAN}[3]{##3}\VANthebibliography}

\usepackage{graphicx}	% Including figure files
\usepackage{amsmath}	% Advanced maths commands

\usepackage{xcolor}

\newcommand{\bvresult}{b_v = 0.339\pm 0.034}

\newcommand{\wignerJ}[6]{\begin{pmatrix}#1 & #2 & #3\\
#4 & #5 & #6
\end{pmatrix}}

\title[ACT DR6+DESI LRGs kSZ velocity reconstruction]{The Atacama Cosmology Telescope: Cross-correlation of kSZ and continuity equation velocity reconstruction with photometric DESI LRGs }

\author[F. McCarthy et al.]{Fiona McCarthy$^{1,2,3}$\thanks{E-mail: fmm43@cam.ac.uk}, Boryana Hadzhiyska$^{4,2,5,6}$,  J. Richard Bond$^{7}$,  William R.~Coulton$^{1,2}$, \newauthor Jo Dunkley$^{8,9}$, 
Carmen Embil Villagra$^{1,2}$,   Matthew C. Johnson$^{10,11}$, Kavilan Moodley$^{12,13}$, \newauthor Toshiya Namikawa$^{1,14,2}$,  Bernardita Ried Guachalla$^{15,16,17}$,     Blake D.~Sherwin$^{1,2}$, Crist\'obal Sif\'on$^{18}$,   \newauthor Alexander van Engelen$^{19}$,  Eve M. Vavagiakis$^{20,21}$, 
Edward~J.~Wollack$^{22}$\\
$^{1}$DAMTP, Centre for Mathematical Sciences, Wilberforce Road, Cambridge, CB3 0WA, UK\\
$^{2}$Kavli Institute for Cosmology Cambridge, Madingley Road, Cambridge, CB3 0HA, UK\\
$^{3}$Center for Computational Astrophysics, Flatiron 
Institute, 162 5th Avenue, New York, NY 10010 USA\\
$^{4}$Institute of Astronomy, Madingley Road, Cambridge, CB3 0HA, United Kingdom\\
$^{5}$Physics Division, Lawrence Berkeley National Laboratory, Berkeley, CA 94720, USA\\
$^{6}$Berkeley Center for Cosmological Physics, Department of Physics, University of California, Berkeley, CA 94720, USA\\
$^{7}$Canadian Institute for Theoretical Astrophysics, 60 St. George Street, University of Toronto, Toronto, ON, M5S 3H8, Canada\\
$^{8}$ {Joseph Henry Laboratories of Physics, Jadwin Hall, Princeton University, Princeton, NJ, USA 08544} \\
 $^{9}${Department of Astrophysical Sciences, Peyton Hall, Princeton University, Princeton, NJ USA 08544}\\
$^{10}$Perimeter Institute for Theoretical Physics, 31 Caroline St N, Waterloo, ON N2L 2Y5, Canada\\
$^{11}$Department of Physics and Astronomy, York University, Toronto, ON M3J 1P3, Canada\\
$^{12}$A. Astrophysics Research Centre, University of KwaZulu-Natal, Westville Campus, Durban 4041, South Africa\\
$^{13}$B. School of Mathematics, Statistics \& Computer Science, University of KwaZulu-Natal, Westville Campus, Durban 4041, South Africa\\
$^{14}$Center for Data-Driven Discovery, Kavli IPMU (WPI), UTIAS, The University of Tokyo, Kashiwa, 277-8583, Japan\\
$^{15}${Department of Physics, Stanford University, Stanford, CA, USA 94305-4085}\\
$^{16}${Kavli Institute for Particle Astrophysics and Cosmology, 382 Via Pueblo Mall Stanford, CA 94305-4060, USA}\\
$^{17}${SLAC National Accelerator Laboratory 2575 Sand Hill Road Menlo Park, California 94025, USA}\\
$^{18}$Instituto de F\'isica, Pontificia Universidad Cat\'olica de Valpara\'iso, Casilla 4059, Valpara\'iso, Chile\\
$^{19}$School of Earth and Space Exploration, Arizona State University, Tempe, AZ 85287, USA\\
$^{20}$Department of Physics, Duke University, Durham, NC, 27704, USA\\
$^{21}$Department of Physics, Cornell University, Ithaca, NY, 14853, USA\\
$^{22}$NASA Goddard Space Flight Center, 8800 Greenbelt Road, Greenbelt, MD 20771, USA
}

\date{Accepted XXX. Received YYY; in original form ZZZ}

\pubyear{\the\year{}}

\begin{document}
\label{firstpage}
\pagerange{\pageref{firstpage}--\pageref{lastpage}}
\maketitle

\begin{abstract}

Over the last year, kinematic Sunyaev--Zel'dovich (kSZ) velocity reconstruction---the measurement of the large-scale velocity field using the anisotropic statistics of the small-scale kSZ-galaxy overdensity correlation--- has emerged as a statistically significant probe of the large-scale Universe.  In this work, we perform a 2-dimensional tomographic reconstruction using ACT DR6 CMB data and DESI legacy luminous red galaxies (LRGs). We measure the cross-correlation of the kSZ-reconstructed velocity $v^{\mathrm{kSZ}}$ with the velocity inferred from the continuity equation applied to the DESI LRGs $v^{\mathrm{cont}}$ 
at the $\sim 10 \sigma$ level, detecting the signal with an amplitude with respect to our theory of $\bvresult$.  We fit a scale-dependent galaxy bias model to our measurement in order to constrain local primordial non-Gaussianity $f_{\mathrm{NL}}^{\mathrm{loc}}$, finding {$f_{\mathrm{NL}}^{\mathrm{loc}}=-180^{+61}_{-86}$} at 67\% confidence, with $f_{\mathrm{NL}}^{\mathrm{loc}}$ consistent with zero at 95\% confidence. {We also measure an auto spectrum at $2.1\sigma$ significance which provides a constraint on $b_v$ of $b_v=0.26_{-0.05}^{+0.11}$, which is consistent with the measurement from the cross spectrum. Our combined measurement is $b_v=0.33\pm0.03$, an $11\sigma$ measurement.  We find a good fit of our model to the data in all cases. Finally, we use different ACT frequency combinations to explore foreground contamination, finding no evidence for foreground contamination in our velocity cross correlation. We compare to a similar measurement where $v^{\mathrm{kSZ}}$ is directly cross correlated with the large-scale galaxy field, and find signs of foreground contamination which is contained in the equal-redshift spectra. }
\end{abstract}


%%%%%%%%%%%%%%%%%%%%%%%%%%%%%%%%%%%%%%%%%%%%%%%%%%

%%%%%%%%%%%%%%%%% BODY OF PAPER %%%%%%%%%%%%%%%%%%

\section{Introduction}

The kinematic Sunyaev-Zel'dovich (kSZ) effect~\citep{1969Ap&SS...4..301Z,1970Ap&SS...7....3S,1980MNRAS.190..413S} is a foreground signal to the cosmic microwave background (CMB) sourced when a CMB photon Thomson scatters off ionized electron gas in the late Universe moving with respect to the CMB rest frame. The signal contains information about the bulk velocity as well as the density of electrons along the line of sight. 

On small scales (angular multipoles $\ell>\sim4000$), the kSZ signal is expected to dominate over the primary CMB anisotropies, and it is emerging as a new probe of cosmic structure and astrophysics as we make higher sensitivity and resolution CMB measurements. We have moved beyond the era of relatively low and medium significance detections with \textit{Planck} and early ACT and SPT datasets  to a phase in which kSZ is a statistically significant probe of cosmology and astrophysics with the Atacama Cosmology Telescope (ACT)~(e.g~\citealt{2012PhRvL.109d1101H,2013A&A...557A..52P,2016PhRvD..93h2002S,2016MNRAS.461.3172S,2016PhRvL.117e1301H,2021PhRvD.103f3513S,2021PhRvD.104d3518K,2021A&A...645A.112T,2021PhRvD.104d3502C,2024arXiv240707152H,2025JCAP...05..057M,Lague:2024czc,2025PhRvD.112j3512G,2025arXiv250621684L,2025arXiv250621657H,2025arXiv251012553R}). The Simons Observatory~\citep{2019JCAP...02..056A,2025JCAP...08..034A} Large Aperture Telescope (LAT) survey has recently begun, and the kSZ will soon become a precision cosmology probe as well as a highly significant probe of astrophysics.

\begin{figure*}
\includegraphics[width=\columnwidth]{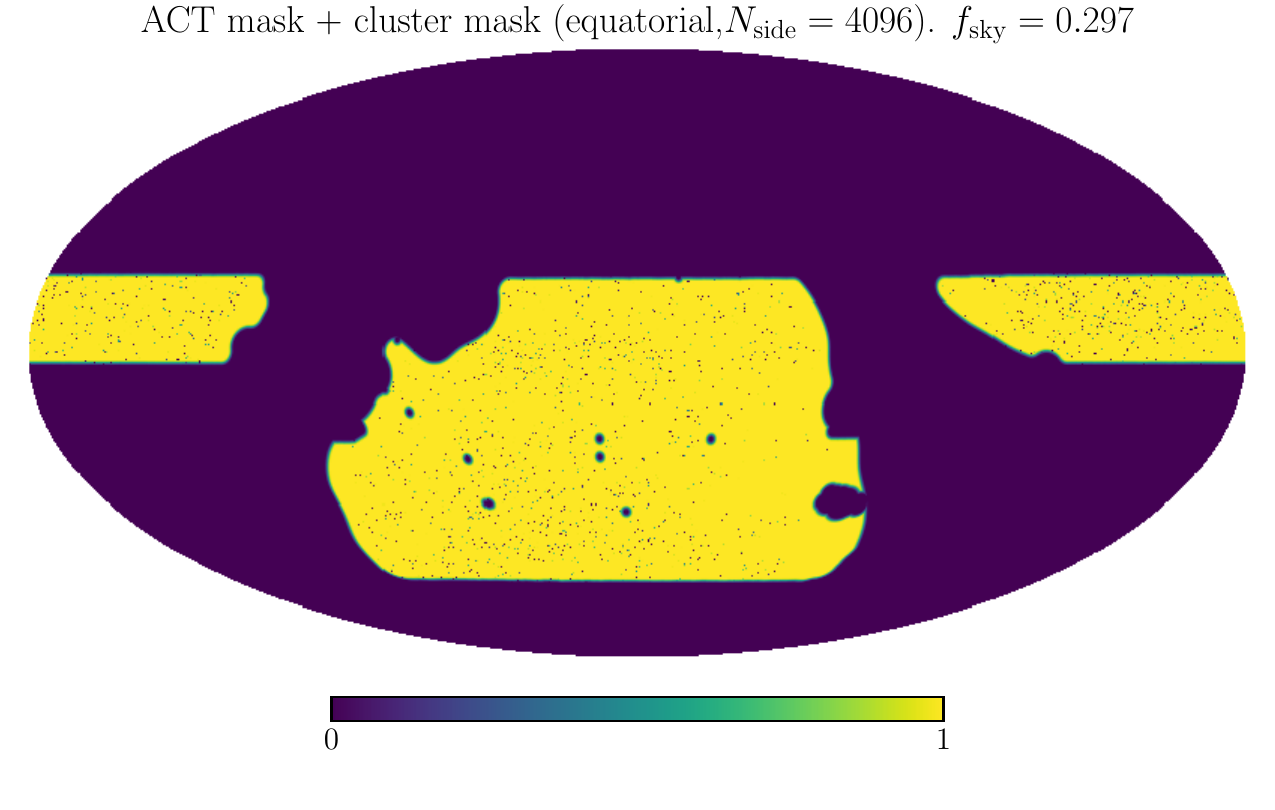}
\includegraphics[width=\columnwidth]{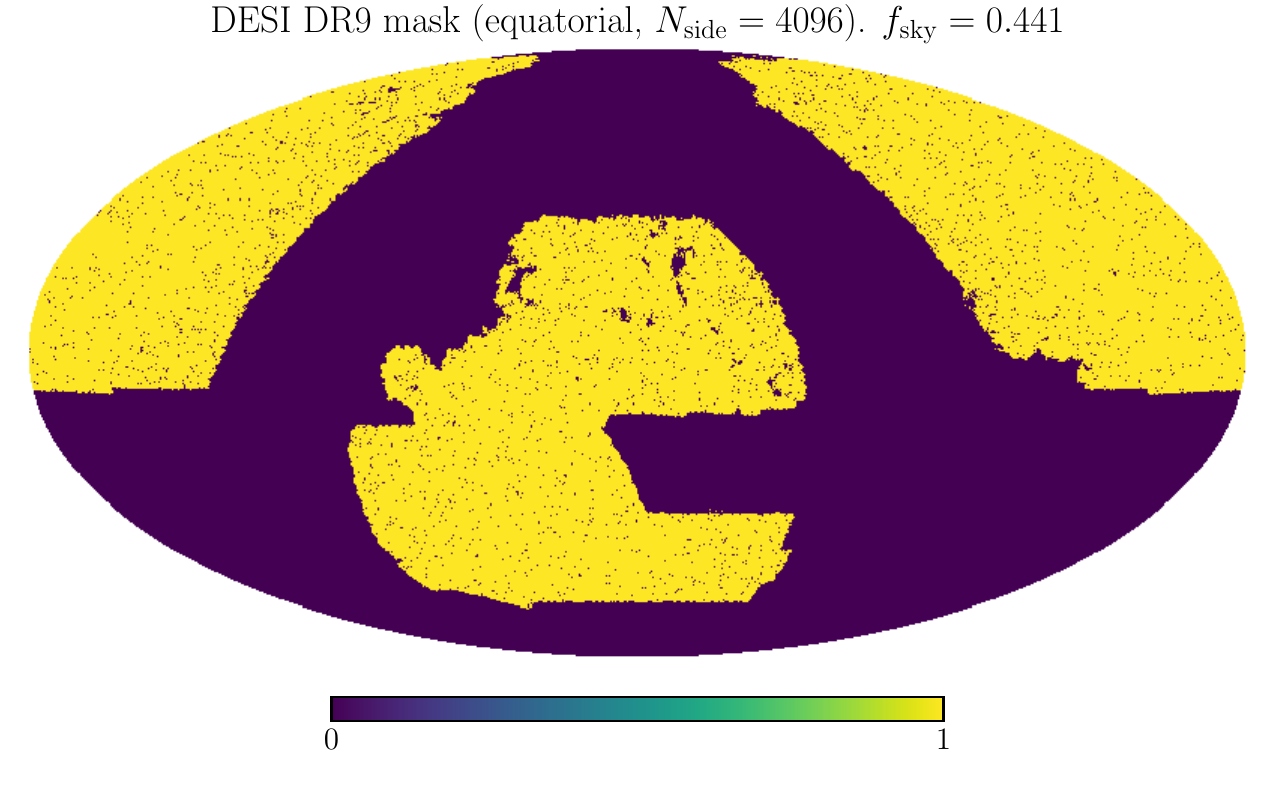}
\includegraphics[width=\columnwidth]{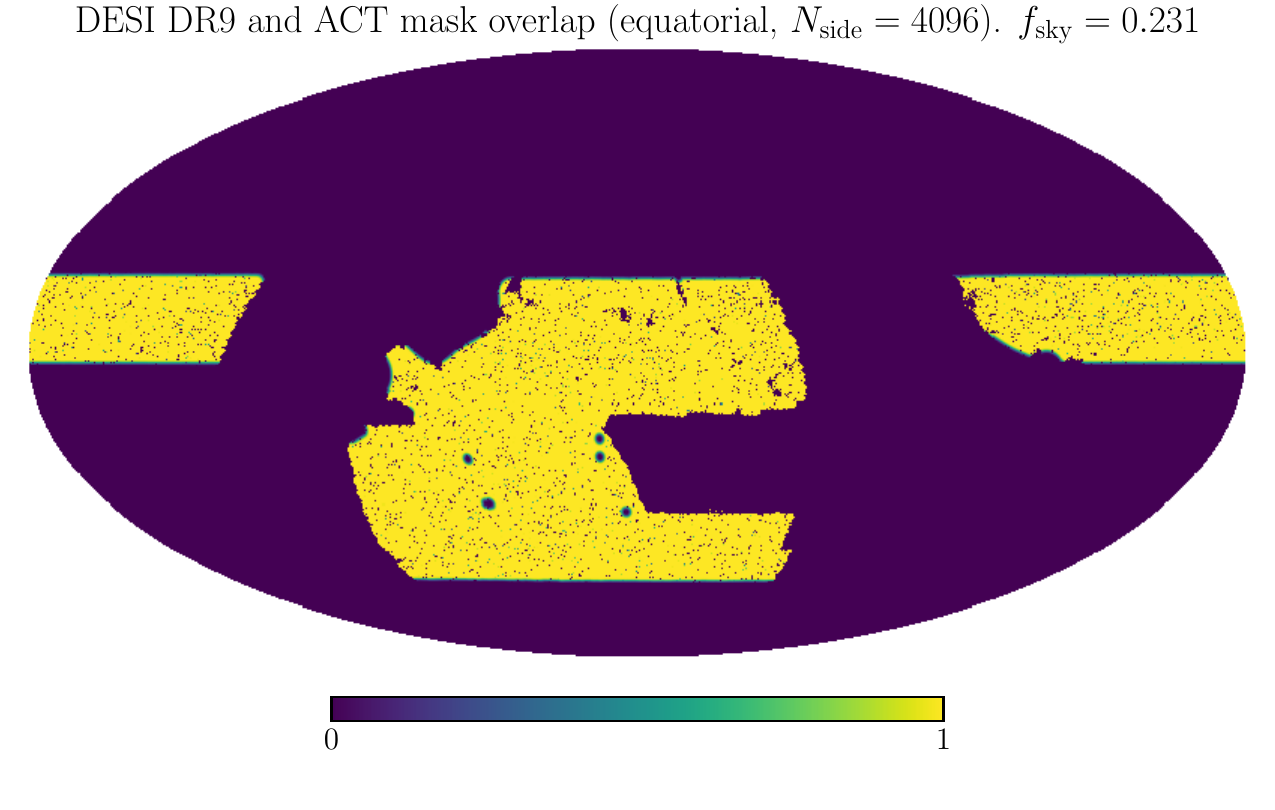}
\includegraphics[width=\columnwidth]{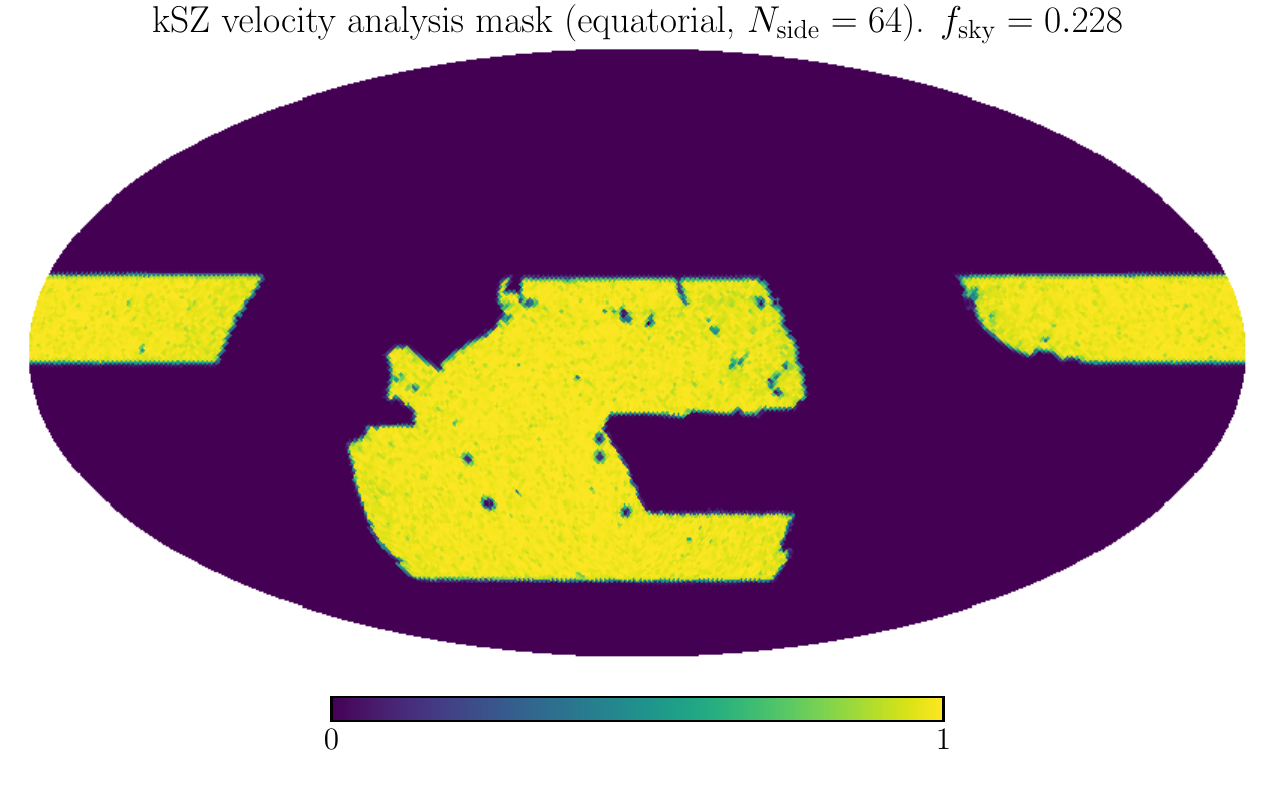}
\includegraphics[width=\columnwidth]{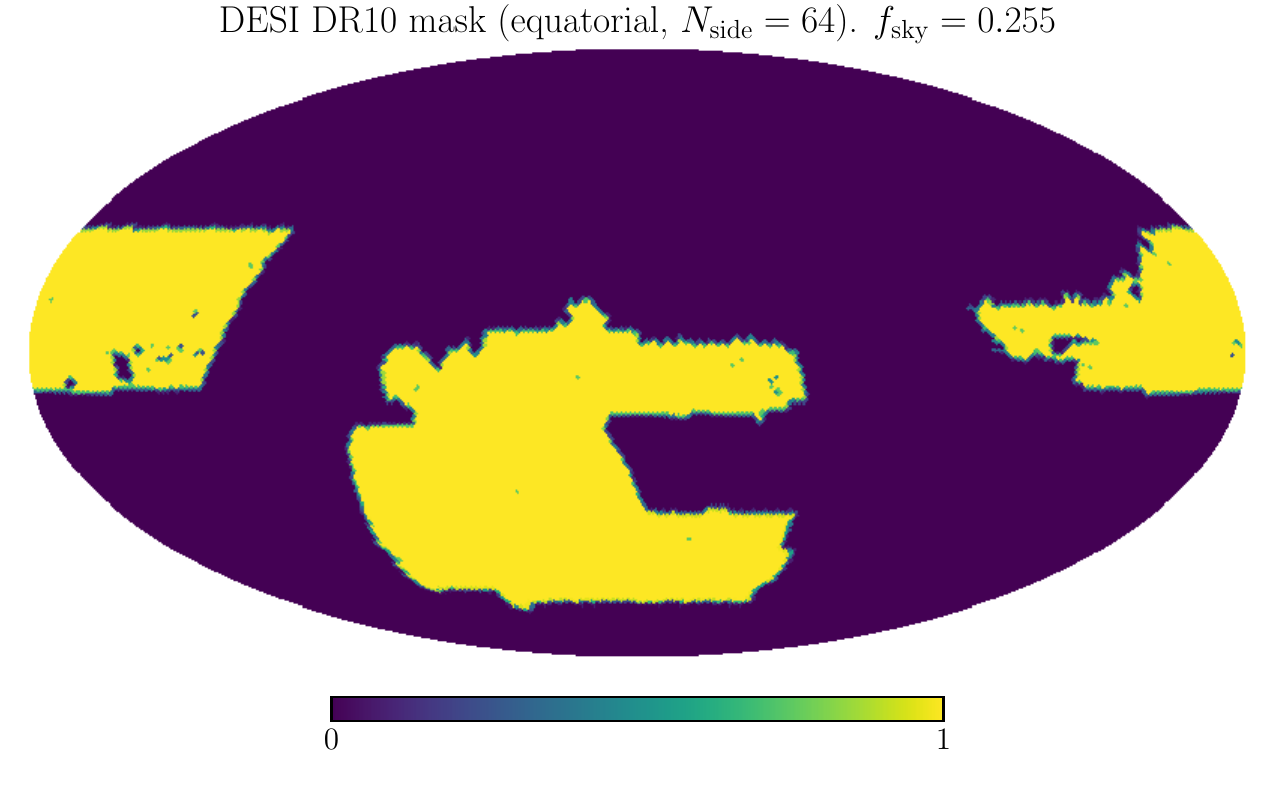}
\includegraphics[width=\columnwidth]{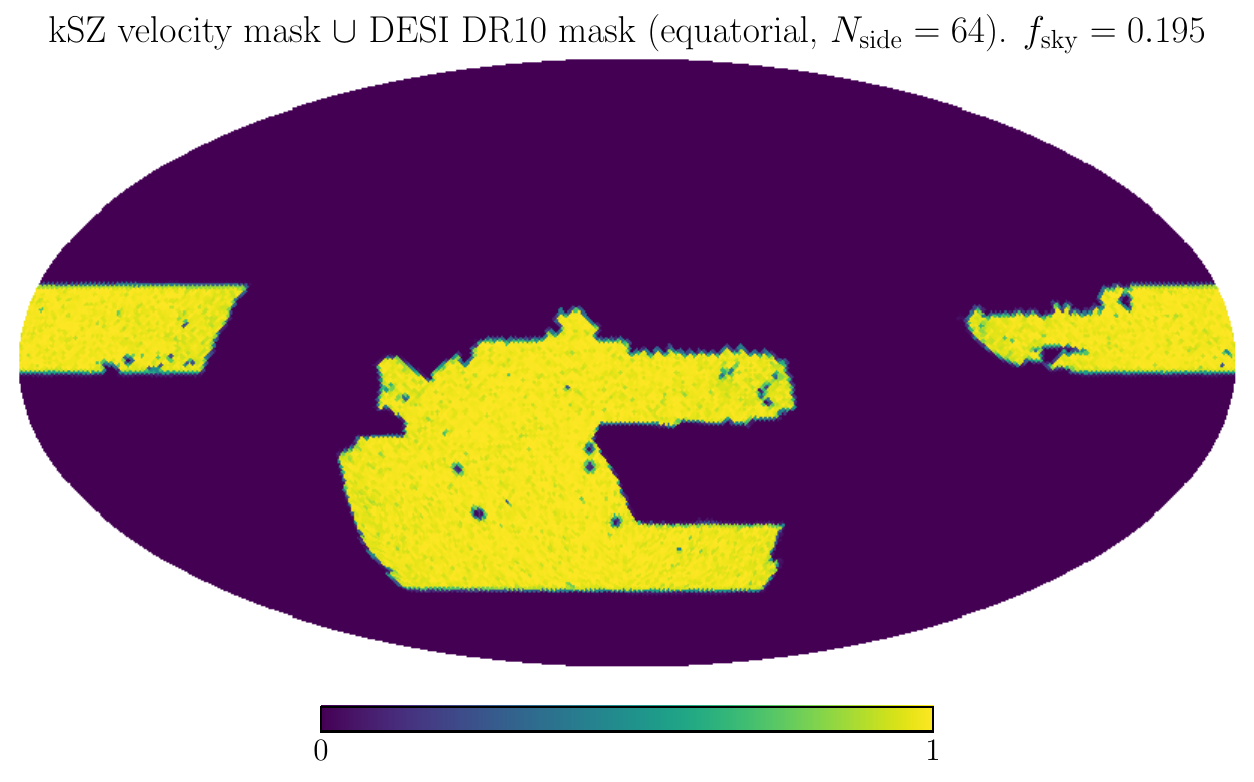}

\caption{Masks used in this analysis.  The analysis mask used for the CMB dataset is shown on the top left; this is the ACT mask combined with the cluster mask created from the ACT DR5 cluster catalogue. The DESI DR9 mask that we use for the small-scale galaxies is shown on the top right. The product of this with the ACT mask + cluster mask, which we use to perform kSZ velocity reconstruction at $N_{\mathrm{side}}=4096$, is shown on middle left. We downgrade this to an $N_{\mathrm{side}}=64$ mask and apodize with a 2 degree apodization scale to perform analysis on the reconstructed kSZ velocity; this is shown on the middle right. The mask used for the large-scale DR10 DESI galaxies is shown on the bottom left. The overlap of the two large-scale maps is shown on the bottom right to indicate the overlap in sky area between both large-scale measurements, although note that we never use this mask in practice.}\label{fig:masks}
\end{figure*}

kSZ velocity reconstruction~\citep{Zhang_2010,2017JCAP...02..040T,2018PhRvD..98l3501D,2018arXiv181013423S} uses the kSZ effect to measure the large-scale velocity field of the Universe using the mode-coupling in the temperature-galaxy correlation $\left<Tg\right>$   induced by the kSZ effect.\footnote{In the ensemble average, $\left<Tg\right>$  is zero; we refer here to the average over the one realization that we see in our sky, \textit{i.e.} the average at fixed large-scale modes.} The first implementations of kSZ velocity reconstruction were performed using \textit{Planck}~\citep{2020A&A...641A...1P} data~\citep{2024arXiv240500809B} and ACT data~\citep{2025JCAP...05..057M,2025PhRvL.134o1003L,2025arXiv250621657H,2025arXiv250621684L}, with the highest signal-to-noise detections~\citep{2025arXiv250621657H,2025arXiv250621684L} coming from a combination with small- and large-scale DESI luminous red galaxies (LRGs)~\citep{2023JCAP...11..097Z}. 

In kSZ velocity reconstruction, an estimate of the large-scale velocity $\hat v$ is created from small-scale anisotropic power in $\left<Tg\right>$. There is a distinct scale separation, with large scales here corresponding to cosmological scales (multipole $L\sim \mathcal{O}(1-100)$, or comoving wavenumber $k\sim\mathcal{O}(0.01) \,\mathrm{Mpc}^{-1}$)  and small scales corresponding to intra-halo scales ($\ell\sim\mathcal{O}(1000)$ or $k\sim\mathcal{O}(1) \,\mathrm{Mpc}^{-1}$). Going forward, a superscript ``$L$'' on a field will be used to refer to a large-scale mode of it, and a superscript ``$S$'' to a small-scale mode.  

 In this work, we report a measurement of the kSZ velocity signal at $\sim10\sigma$ in cross correlation with a separate measurement of the large-scale velocity field, and $11\sigma$ when the kSZ velocity auto is included. Similarly to~\cite{2025JCAP...05..057M} and~\cite{2025arXiv250621684L}, we work in the 2-dimensional formalism of~\cite{2018PhRvD..98l3501D}. We use a $\left<T_{\mathrm{ACT-DR6}}^{{S}} g_{\mathrm{DESI-LRG}}^{{S}}g_{\mathrm{DESI-LRG}}^{{L}}\right>$ data combination similar to both~\cite{2025arXiv250621684L} and~\cite{2025arXiv250621657H}, which recently both found $11.7\sigma$ detections of the signal (although~\cite{2025arXiv250621657H} used ACT DR5 data as opposed to DR6).

 In this paper, we follow our recent work~\citep{2025JCAP...05..057M}, which used a $\left<T_{\mathrm{ACT-DR6}}^{{S}} g_{\mathrm{DESI-LRG}}^{{S}}g_{\mathrm{SDSS}}^{{L}}\right>$ combination where the large-scale Sloan Digital Sky Survey (SDSS) galaxies were filtered to produce a velocity estimate and measure the angular velocity cross-power spectrum $C_L^{vv}$. Now,  we measure $C_L^{vv}$ using a redshift-binned 2-dimensional velocity template which is constructed from the continuity equation applied to the DR10 photometric DESI LRGs~\citep{2024arXiv240707152H,2024PhRvD.109j3534H,2024PhRvD.109j3533R}. We include a demonstration of  our pipeline on $N$-body simulations.

This large-scale velocity field can then be correlated with itself or with other fields.  The  velocity-g cross-correlation  is sensitive to the scale-dependence of the large-scale galaxy bias~\citep{2019PhRvD.100h3508M} and is thus a probe of local primordial non-Gaussianity~\citep{2008PhRvD..77l3514D}, parametrized by $f_{\mathrm{NL}}^{\mathrm{loc}}$ (hereafter $f_{\mathrm{NL}}$). $f_{\mathrm{NL}}$ constraints have recently been obtained from the signal~\citep{2024arXiv240805264K,2025PhRvL.134o1003L,2025arXiv250621657H}, with  ~\cite{2025arXiv250621657H}
 finding $f_{\mathrm{NL}}=-30^{+40}_{-33}$.  In this work, we also derive $f_{\mathrm{NL}}$ constraints, finding $-264<f_{\mathrm{NL}}<-109$ at 67\% confidence, with $f_{\mathrm{NL}}$ consistent with zero at 95\% confidence. Our results are not as tight as previous analyses with a similar dataset and signal-to-noise; we take steps to compare the analyses and attribute this to our use of a higher minimum scale $k_{\mathrm{min}}$.

We include a measurement of the kSZ velocity auto power spectrum, finding a $2\sigma$ measurement of the amplitude without using the cross-spectrum information. We use this to constrain $b_v$ independently of the cross-spectrum to $b_v=0.26_{-0.05}^{+0.11}$, which is consistent with the measurement from the cross; we combine them to find $b_v=0.33\pm0.03$, an $11\sigma$ measurement.

Finally, we compare our $C_L^{vv}$ measurement to a $C_L^{vg}$ measurement, which we make by cross-correlating the kSZ-estimated velocity field with the galaxy overdensity directly, and perform several tests for foreground bias, finding hints of foreground bias in some parts of the $C_L^{vg}$ signal.

 This paper is organized as follows. In Section~\ref{sec:data} we describe the datasets that we use. In Section~\ref{sec:pipeline} we describe our pipeline. We present our theory modelling in Section~\ref{sec:theory}. We show a demonstration on simulations in Section~\ref{sec:simdem}. In Section~\ref{sec:results} we present our results.   We find a good fit for our scale-independent model in $C_L^{vv}$, and we use this measurement to measure the scale-dependent bias induced by $f_{\mathrm{NL}}$. We show tests for foreground contamination in Sec.~\ref{sec:fg_contamination}. We conclude in Section~\ref{sec:conclusion}.

\section{Data}\label{sec:data}

We use three datasets: the small-scale ACT DR6+\textit{Planck} component-separated temperature maps~\citep{2025arXiv250314451N,2024PhRvD.109f3530C}, the small-scale DESI DR9 extended photometric luminous red galaxy (LRG) sample~\citep{2023JCAP...11..097Z,2019AJ....157..168D}, and the large-scale DESI DR10 extended LRGs with and without velocity reconstruction performed~\citep{2023JCAP...11..097Z,2023AJ....165...58Z}. We describe these datasets in this Section.

\subsection{Small-scale temperature}\label{sec:acT_description}
We use the public ACT DR6+\textit{Planck} temperature map. This is a needlet internal linear combination (NILC)~\citep{1992ApJ...396L...7B,2009A&A...493..835D} estimation of the blackbody component of the multi-frequency ACT DR6 data~\citep{2025arXiv250314451N} and \textit{Planck} NPIPE data~\citep{2020A&A...643A..42P}, described in~\cite{2024PhRvD.109f3530C}. The blackbody components in the sky are the primary CMB anisotropies as well as the kSZ anisotropies, so we refer to this as a CMB+kSZ map. As the kSZ is a small-scale signal, the high-resolution ACT data is instrumental to this measurement. 

NILC provides a minimum-variance estimate of the signal of interest, but the map {contains} residual foregrounds, notably the thermal Sunyaev--Zel'dovich (tSZ) effect and the cosmic infrared background (CIB). The tSZ effect can be isolated in frequency space by its well-known $y$  distortion~\citep{1970Ap&SS...7....3S}; the CIB is harder to remove due to its less exact frequency dependence.  In order to test our signal for foreground contamination, we will use  other public ACT+\textit{Planck} products described in~\cite{2024PhRvD.109f3530C}, in particular the blackbody-deprojected Compton-$y$ map and the $y$-deprojected blackbody map. Both of these maps were created using constrained NILC~\citep{2009ApJ...694..222C,2011MNRAS.410.2481R}, which can exactly remove a foreground with known frequency dependence from an ILC map. The blackbody-deprojected $y$-distortion map will be used as a generic ``foreground'' map on which we expect a null signal; the $y$-deprojected blackbody map will be used as an alternative input to the kSZ estimator. This has different foregrounds than the unconstrained NILC map, though larger variance, leading to a lower signal-to-noise velocity reconstruction.

All of the ACT+\textit{Planck} maps have a beam which we approximate as Gaussian with full width at half maximum (FWHM) of $1.6^\prime$, and cover 30\% of the sky. In all cases, we project from the native Plate-Carr\'{e}e (CAR) coordinate system to a HEALPix~\citep{2005ApJ...622..759G} coordinate system with $N_{\mathrm{side}}=4096$, by first projecting to spherical harmonics with $\ell_{\mathrm{max}}=3\times4096$ and then to HEALPix.

\subsubsection*{Sky area}

The ACT survey mask that we use is shown in Fig.~\ref{fig:masks}. It leaves $30.1\%$ of the sky uncovered. In addition to this mask, we mask tSZ clusters by constructing a mask of circles of radius $5^\prime$ centered on each of the 4195 objects in the ACT DR5 tSZ cluster catalogue~\citep{2021ApJS..253....3H}.\footnote{We note that since the beginning of this work, the ACT DR6 cluster catalogue, which contains 9977 objects, has been completed and released~\citep{2025arXiv250721459A}. Switching to a mask defined with this catalogue did not affect any of our results.} We apodize this mask with a 5 arcminute apodization scale using the ``C1'' apodization routine of \texttt{pymaster}. The total area of sky lost is 0.46\% (of the whole sky); this is 1.4\% of the total ACT footprint, such that the final sky area available is $29.7\%$. We will explore the impact of masking the clusters on the analysis in Section~\ref{sec:autospectra}.

\subsubsection*{Power spectra}

\begin{figure}
\includegraphics[width=\columnwidth]{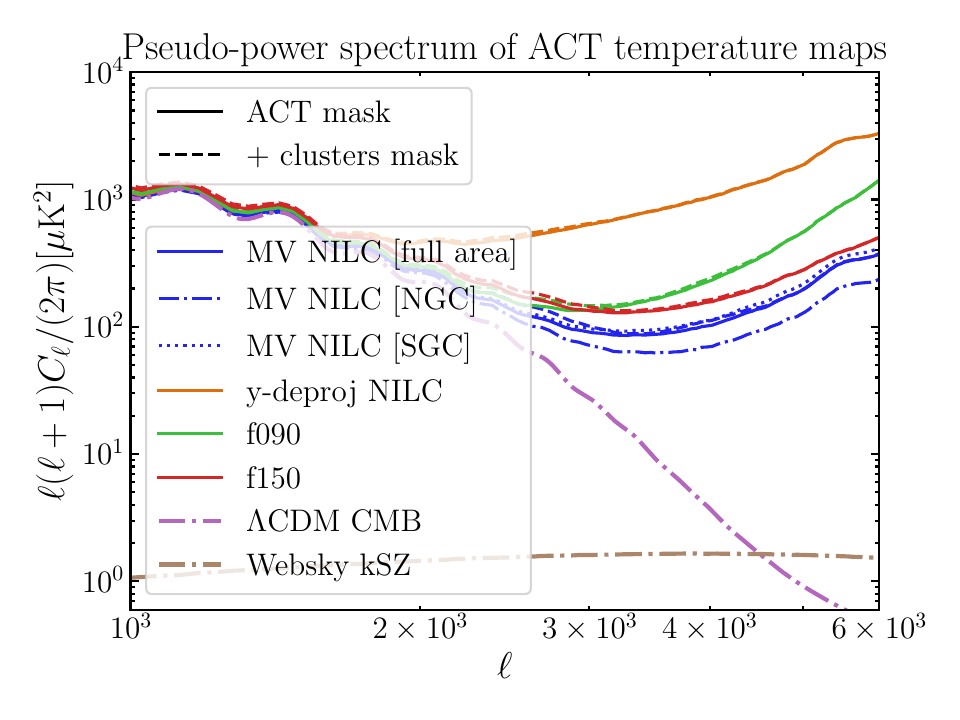}

\caption{An estimation of the power spectra of the ACT temperature maps we use, on the full ACT region (`` ACT mask'') as well as the ACT region with clusters masked (``ACT + clusters masked''). We show this both for minimum-variance NILC map (``MV NILC'') as well as for the tSZ-deprojected NILC map (``y-deproj NILC''), and the single-frequency maps. These are pseudo-$C_\ell$ estimates, estimated by measuring the raw power spectra of the masked map and dividing by the sky area available. The impact of deprojecting the tSZ is large; the impact of masking the tSZ clusters is much smaller. The variance on the NGC area is noticably lower than on the SGC area. In all cases we have divided by an estimate of the beam. Unless otherwise indicated, all measurements are on the full (NGC+SGC) area.  We include an estimate of the kSZ signal from the Websky simulations~\protect\citep{2020JCAP...10..012S}.
}\label{fig:ACTpower}
\end{figure}

We show the power spectra of the temperature maps we use in Figure~\ref{fig:ACTpower}. These are estimated simply by measuring the pseudo-$C_\ell$s on the available sky area and dividing by the sky area available.\footnote{We have checked that using filters computed from binned mask-decoupled spectra estimated with \texttt{pymaster} does not change our results.} 

\subsection{Small-scale galaxies}\label{sec:smallscaleg}

\begin{figure}
\includegraphics[width=\columnwidth]{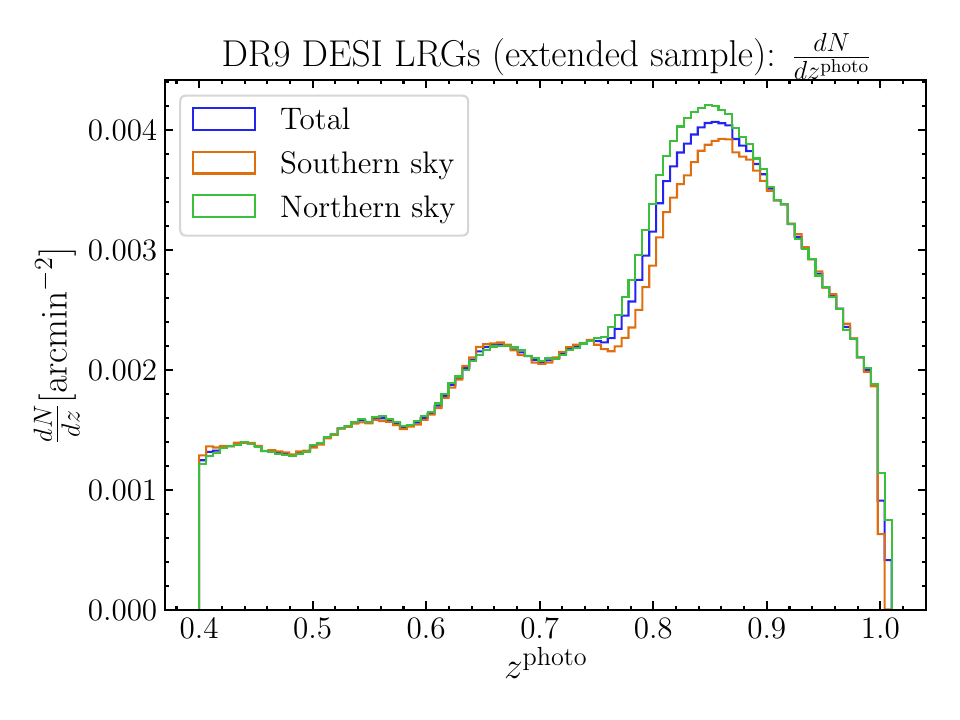}
\caption{The distribution of the photometric redshifts $z^{\mathrm{photo}}$ of the DR9 DESI LRGs that we use for the small-scale galaxy sample for kSZ velocity reconstruction. There are 27,253,833 objects in total, over 44.1\% of the sky.}\label{fig:dndz_dr9}
\end{figure}

For the small-scale galaxy leg of our $\left<T^S g^S g^L\right>$ bispectrum, we use the DESI LRGs with photometric redshifts described in~\cite{2023JCAP...11..097Z}. We use the extended sample, as we are interested in maximizing number density. These objects were selected from the imaging data from the DESI Legacy Survey DR9~\citep{2019AJ....157..168D,2023AJ....165...58Z}. We use galaxies with photometric redshift $z^{\mathrm{photo}}$ in the full range $0.4<z^{\mathrm{photo}}<1.01$.  There are 33,735,219 objects in total, covering {44.1\%} of the sky. They have photo-$z$ uncertainty $\frac{\sigma_z}{(1+z)}\sim{0.026}$. We use only the 27,253,833 objects that are assigned to one of the four tomographic redshift bins that were spectroscopically calibrated in~\cite{2023JCAP...11..097Z}; these include all of the objects with $0.4<z^{\mathrm{photo}}<1$. We separate the galaxies into a North Galactic Cap (NGC) and South Galactic Cap (SGC) region. The overall number density is ${0.22638}\,\mathrm{arcmin^{-2}}$, with ${0.22149}\,\mathrm{arcmin^{-2}}$ in the SGC and ${0.23038}\,\mathrm{arcmin^{-2}}$ in the NGC. The distributions of the photometric redshifts of these galaxies are shown in Fig.~\ref{fig:dndz_dr9}{, in $\mathrm{arcmin^{-2}}$}.

In all cases, the sky mask we use for the small-scale galaxies is the product of the four masks provided with the extended galaxies (one for each redshift bin). This mask is shown in Figure~\ref{fig:masks}, on the first row (right column).

We split the galaxies into redshift bins before performing kSZ velocity reconstruction. When we do so, we start with the pre-defined tomographic bins, which are defined by redshift boundaries 
$[0.4,0.545,0.713,0.885,1.01]$.  There are 3,825,983, 6,404,291, 8,409,482, and 8,614,077 objects in each of the 4 bins (in order of increasing redshift).  There are spectroscopically calibrated true redshift distributions $\frac{dN}{dz^{\mathrm{true}}}$ available for these bins as part of the release.\footnote{These binned dataproducts and spectroscopically-calibrated $\frac{dN}{dz}$ are available at \url{https://data.desi.lbl.gov/public/papers/c3/lrg\_xcorr_2023/}.}

\begin{table*}
\begin{tabular}{cl}
\hline
number of bins $N$ & Boundaries (in photo-$z$) \\
\hline
4 & 0.4, 0.545, 0.714, 0.854, 1 \\
8 & 0.4, 0.472, 0.545, 0.639, 0.713, 0.803, 0.854, 0.917, 1.01 \\
16 & 0.4, 0.436, 0.472, 0.509, 0.545, 0.593, 0.639, 0.676, 0.713, 
    0.763, 0.803, 0.832, 0.854, 0.885, 0.917, 0.953, 1.01 \\
\hline
\end{tabular}
\caption{The boundaries used in redshift space to construct the tomographic galaxy overdensity samples}\label{tab:redshiftboundaries}
\end{table*}

\begin{figure*}
\includegraphics[width=0.49\textwidth]{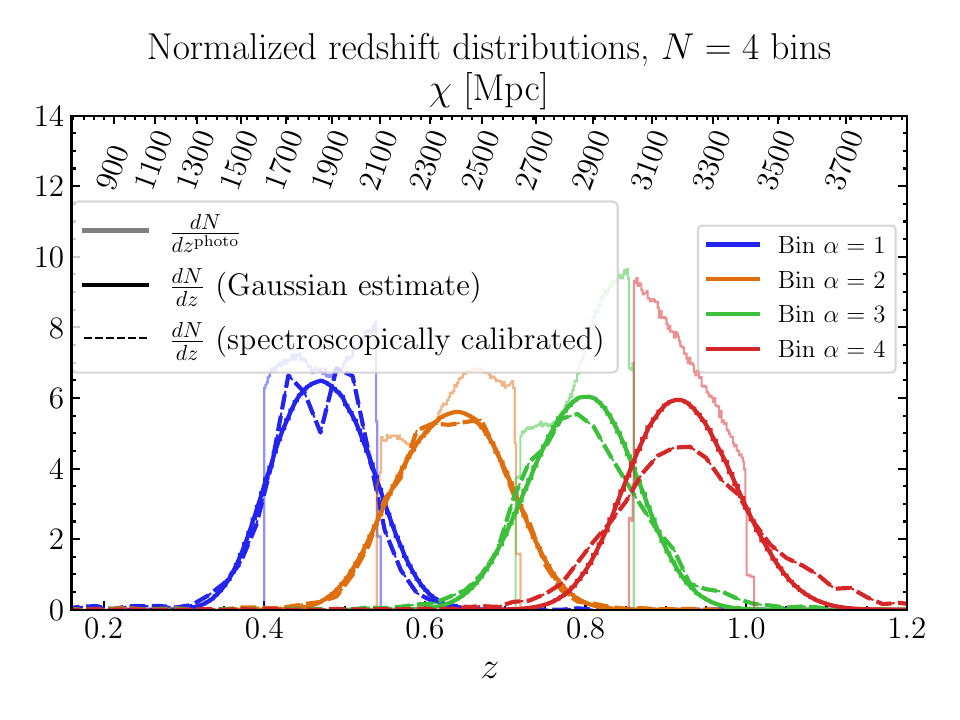}
\includegraphics[width=0.49\textwidth]{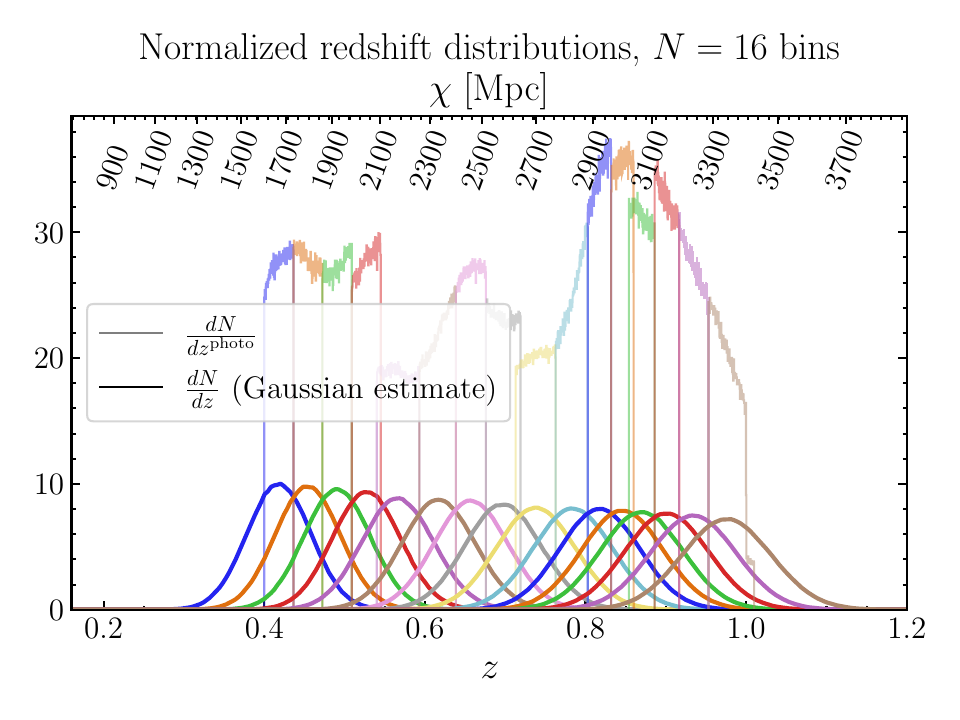}
\caption{The redshift distributions. The $\frac{dN}{dz^{\mathrm{photo}}}$, which we estimate just by histogramming the $z^{\mathrm{photo}}$ provided for each object, is in a faded line; the Gaussian-estimated $\frac{dN}{dz^{\mathrm{true}}}$ is shown in the solid line. The spectroscopically calibrated $\frac{dN}{dz^{\mathrm{true}}}$, provided by~\protect\cite{2023JCAP...11..097Z}, is also shown in the dashed lines. We use the spectroscopic estimates in our $N=4$ analysis, and the Gaussian estimates elsewhere. The coherence length for the velocity is $\sim$200 Mpc; we indicate the moving distance on the upper $x$ axis in order to compare to this number. In our 16-bin case, we approach this, but with broad redshift bins. 
}\label{fig:dndzestimate}
\end{figure*}

The kSZ effect is sensitive only to the radial velocity mode, which can cancel out along the line-of-sight if large  redshift bins are used. To avoid this, redshift bins smaller than the velocity correlation length ($\sim 200 \mathrm{Mpc}$) should be used.   Thus,  to maximize our sensitivity, we further bin the galaxies by further dividing each of the four previously defined redshift bins into two and four bins, resulting overall $N=8$ bin and  $N=16$ bin cases. When we do this, we choose boundaries in $z^{\mathrm{photo}}$ such that each bin is split into bins with an equal number of galaxies. The boundaries are shown in Table~\ref{tab:redshiftboundaries}. We estimate the true $\frac{dN}{dz}$ by convolving the $\frac{dN}{dz^{\mathrm{photo}}}$ (estimated by histogramming the data) with Gaussians of width $0.027\times (1+z)$ and show these quantities in Fig.~\ref{fig:dndzestimate}.

\subsubsection{Projection to 2+1 dimensions}\label{sec:projection_gals}
We convert each photo-$z$ bin into number density and then overdensity by projecting the galaxy number counts into $N_{\mathrm{side}}=4096$ HEALPix maps $g_\alpha(\hat n)$ where the value of each pixel is simply the number of objects in that pixel.
We convert from number counts $g_\alpha(\hat n)$ to overdensity according to
\begin{align}
\delta^{g}_{\alpha}(\hat n) = \frac{g_\alpha(\hat n) - \bar n_\alpha}{\bar n_\alpha}\label{numb_to_over}
\end{align}
where $\bar n_\alpha$ is the mean galaxy number density of  $g_\alpha(\hat n)$; we estimate this directly from the projected number counts maps by dividing the number of objects by the observed sky area.

\begin{figure*}
\includegraphics[width=0.49\textwidth]{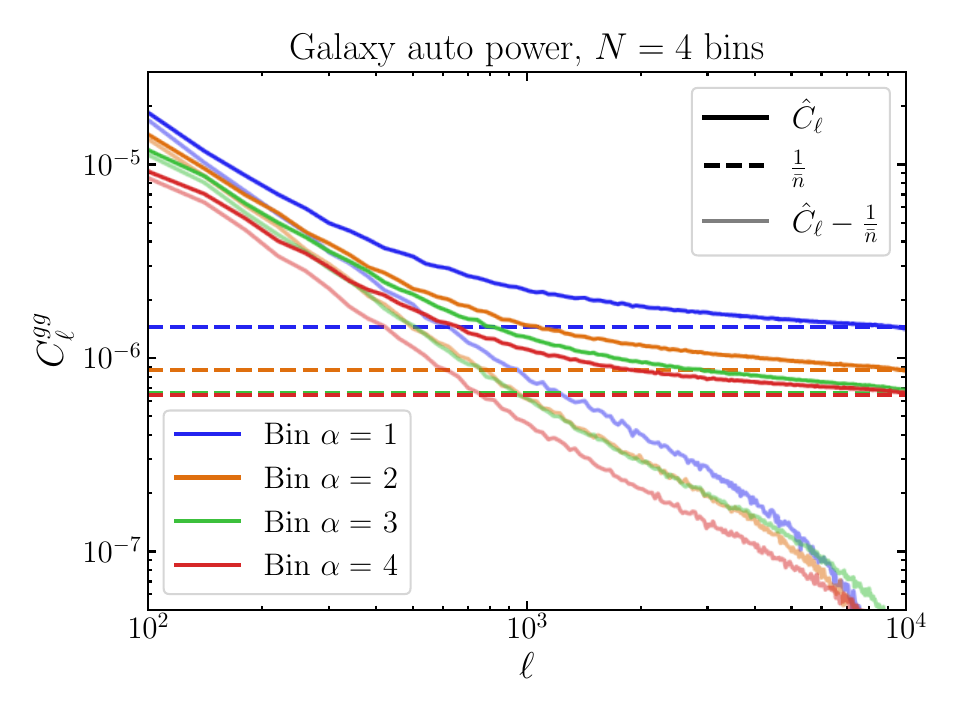}
\includegraphics[width=0.49\textwidth]{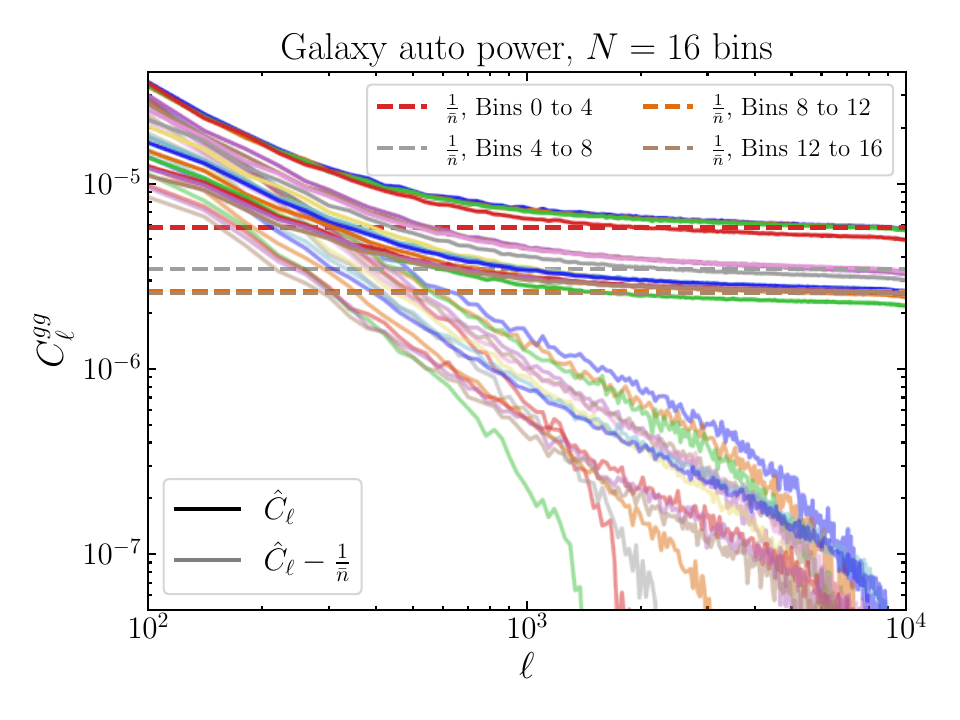}
\caption{The auto power spectra of the binned galaxy overdensity  maps for the $N=4,16$ cases. 
We also show an estimate of the shot noise, which is simply $\frac{1}{\bar n}$ with $\bar n$ the mean number density, and the measured power spectra with this subtracted, to indicate on what scales the measurements are shot-noise dominated. We have binned the spectra in  bins of width $\Delta\ell=40$ order to smooth them. Note that in the 16-bin case we define each group of 4 bins by dividing each of the bins in the 4-bin case into bins with an equal number of galaxies, so that each of these have equal galaxy number density (and hence shot noise estimates). 
}\label{fig:powergals}
\end{figure*}

 We show in Fig.~\ref{fig:powergals} estimates of the power spectra of these maps; these are estimated by simply measuring the pseudo $C_\ell$s  and dividing by the area of sky available. Due to the overlap in redshift distributions of the galaxies, the bins are correlated; we show the correlation coefficients arising from the inter-bin spectra in Appendix~\ref{sec:corrstructure_smallscale}. {This is relevant for our analysis as the signal in neighbouring redshift bins is correlated due to this overlap; we will include this correlation in our covariance matrix calculation.}

\subsection{Large-scale galaxies}

For the galaxies that we use for our large-scale tracer we use the extended photometric sample of LRGs from DESI Data Release 10 (DR10) \citep{2023JCAP...11..097Z,2023AJ....165...58Z}. We use these to make estimates both of the large-scale galaxy overdensity field and of the large-scale continuity-equation radial velocity field $v_r^{\mathrm{cont}}$ (hereafter we will drop the subscript $r$ as all scalar velocities that we refer to will be radial velocities). We will then use these fields to measure two large-scale signals: $C_L^{v^{\mathrm{kSZ}}v^{\mathrm{cont}}}$ and  $C_L^{v^{\mathrm{kSZ}}g}$, where  $C_L^{v^{\mathrm{kSZ}}X}$ is the large-scale power spectrum of the kSZ-estimated velocity with the field $X$ (for notational simplicity we use $g$ instead of $\delta^g$ in the superscripts). 

The galaxies are selected from the DESI Legacy Imaging Surveys~\citep{2019AJ....157..168D} that combine data from DECaLS on Blanco, MzLS on Mayall, and BASS on Bok. For this analysis we adopt the $i$-band based photometric redshifts (\texttt{Z\_PHOT\_MEDIAN\_I}), which have been shown to provide more accurate estimates. The overlap of this extended sample with the ACT footprint contains approximately 6.3 million galaxies.

Following \citet{2023JCAP...11..097Z}, we require the redshift error to satisfy $\sigma_z/(1+z) < 0.025$, which removes roughly $5\%$ of objects with poorly constrained redshifts. This cut ensures that the reconstructed density and velocity fields are not biased by outliers, while retaining the vast majority of the high-quality sample. 

We  describe our process of estimating the velocity and large-scale overdensities in Sections~\ref{sec:contvelrecpipeline} and~\ref{sec:largescaleg} respectively.

\subsubsection{Continuity equation velocity estimation}\label{sec:contvelrecpipeline}

We estimate the large-scale velocity field from the large-scale photometric galaxy catalog using the continuity equation. This photometric continuity equation reconstruction was explored in detail in~\cite{2024PhRvD.109j3533R} and~\cite{2024PhRvD.109j3534H} and applied to the DR9+DR10 DESI LRG sample in~\cite{2024arXiv240707152H}.  For this reconstrution, we treat the large-scale galaxy overdensity field as a fully 3-dimensional object, treating the photometric redshift as the true redshift. In Section~\ref{sec:simdem} we will explore the effect of such a reconstruction on simulations where we know the truth, to understand the behaviour of cosmological observables such as the angular galaxy velocity function $C_L^{vv}$.

The continuity equation is
\begin{equation}
 \vec \nabla \cdot \vec v^{\mathrm{cont}} + \frac{f}{b_g} \vec \nabla \cdot \left[  \left( \vec v^{\mathrm{cont}} \cdot \hat n \right)\hat n \right] = - a H f \frac{\delta^g}{b_g},\label{cont}
\end{equation}
where $f$ is the growth rate and $b_g$ is the galaxy bias. To solve this equation and estimate the peculiar velocity of each galaxy, we use the standard BAO reconstruction technique of \citet{ESSS07}, implemented in the \texttt{pyrecon}\footnote{\url{https://github.com/cosmodesi/pyrecon}} package \citep{2015MNRAS.450.3822W}, adopting the `RecSym' convention with the MultiGrid solver. This procedure provides an estimate of the first-order galaxy displacement field, which can then be converted into a velocity estimate. 

We split the galaxy catalog into four separate redshift bins, [0.4, 0.54], [0.54 0.713], [0.713, 0.86] and [0.86 1.024], on which we perform the reconstruction separately.\footnote{We have explored performing the continuity-equation reconstruction on all bins together, but found poorer results on data than expected compared to simulations. This is at the expense of some signal-to-noise, but we defer more in-depth exploration to future work.} Additionally, we split the galaxies into two regions: north galactic cap (NGC) and south galactic cap (SGC) to reduce our computational expenses. We adopt a cubic grid of size 512$^3$, and a smoothing radius of 12.5 ${\rm Mpc}/h$ following~\cite{2024PhRvD.109j3533R}. Throughout we adopt a constant linear bias of $b = 2.2$ and $Planck$ 2018 cosmology to estimate $f$ at the mean effective redshift of $\sim$0.8.

Once the three-dimensional velocity field has been reconstructed we evaluate it at the positions of the input DESI LRG galaxies.  We then create an estimate of the radial velocity field $v_r\equiv \vec v \cdot \hat n$, and project this to two-dimensional bins. We choose to make $N$ projections, where $N$ is the number of redshift bins, and we weight the projections by the galaxy distributions we use for kSZ velocity reconstruction (described in Section~\ref{sec:smallscaleg}) $\frac{dN}{dz^{\mathrm{true}}}$. The projected velocity in each pixel is
\begin{equation}
v_r^{\mathrm{cont},\alpha}(\hat n) = \frac{\sum_{i \in \hat n} v^{\mathrm{cont}}_{r,i}\frac{dN^\alpha}{dz^{\mathrm{true}}}( z_i^{\mathrm{photo}})}{\int \frac{dN}{dz^{\mathrm{true}}}(z)dz},\label{def_vcontprojection}
\end{equation}
where the sum is over the continuity-equation-estimated radial velocities $v_{r,i}^{\mathrm{cont}}$ of all galaxies $i$ in the pixel at $\hat n$, and we weight by the redshift kernel we wish to estimate the mean velocity over: $\frac{dN^\alpha}{dz^{\mathrm{true}}}$, where $\alpha$ labels the redshift bin.  Note that these are evaluated at $z_i^{\mathrm{photo}}$, which is the redshift at which we evaluated the three-dimensional velocity field. The quantity in the denominator is the integral over the $\frac{dN}{dz}$ of the whole dataset and ensures normalization.

\subsubsection{Large scale galaxy overdensity estimation}\label{sec:largescaleg}

We additionally project the large-scale galaxies into tomographically binned overdensity bins. To do this, we separate them into the photo-$z$ bins defined in Table~\ref{tab:redshiftboundaries} and project to 2+1 dimensions as described in Sec.~\ref{sec:projection_gals}.

\section{Pipeline}\label{sec:pipeline}

Our pipeline is as follows. We: 
\begin{enumerate}
\item Split the small-scale LRG galaxies into $N$ redshift bins, where $N$ is some reasonable number that should approach the maximum signal-to-noise while remaining computationally tractable. We explore $N=4,8,16$, and define the redshift bins using discrete cuts on the photometric redshift estimate $z^{\mathrm{photo}}$;
\item Perform kSZ velocity reconstruction with the $N$ small-scale LRG samples and the kSZ map using a quadratic estimator to get estimates of the radial velocity field $v^{\mathrm{kSZ}}_{\alpha}$ where $\alpha=1,\cdots,N$;
\item Perform continuity-equation velocity reconstruction on the large-scale LRG galaxies to get a three-dimensional velocity estimate $\vec v^{\mathrm{cont}}(z, \hat n)$;
\item Project the radial component of the three-dimensional continuity velocity field into $N$ two-dimensional bins $v^{\mathrm{cont}}_{\alpha}$;
\item Measure the angular cross-correlation $\hat C_L^{v^{\mathrm{kSZ}}_{\alpha}v^{\mathrm{cont}}_{\beta}}$ and estimate its covariance matrix;
\item Compare $\hat C_L^{v^{\mathrm{kSZ}}_{\alpha}v^{\mathrm{cont}}_{\beta}}$ with theory.
\end{enumerate}
We also compare with an alternate pipeline where, instead of performing continuity-equation  velocity estimation, we: 
\begin{enumerate}[label=(\roman*-b), start=3]
\item  Project the three-dimensional photometric galaxy catalogue  into $N$ two-dimensional bins $\delta^g_\alpha$ ;
\item Measure the angular cross-correlation $\hat C_L^{v^{\mathrm{kSZ}}_{\alpha}{g}_{\beta}}$  and estimate its covariance matrix.

\end{enumerate}

We perform the pipeline separately for the north galactic cap (NGC) and south galactic cap (SGC), and combine the final results assuming zero covariance.

We describe the different steps in more detail in several subsections below. Step (i) was described in Section~\ref{sec:projection_gals}. Step  (ii) is described in Section~\ref{sec:kszvelrecpipeline}. Steps (iii) and (iv) were described in Section~\ref{sec:contvelrecpipeline}, and step (iii-b) was described in Section~\ref{sec:largescaleg}. Steps  (v) and (iv-b), are  described in Sections~\ref{sec:xcorr} and~\ref{sec:covariance}. Step (vi) is described in Section~\ref{sec:likelihood}.

\subsubsection*{Preliminaries and notation}

Throughout this section, we use the notation $X_{\ell m}$ for the multipole moments of a real-space field $X$, which are related to the real-space field $X(\hat n)$ by:
\begin{align}
X_{\ell m} = \int X(\hat n) Y^*_{\ell m}(\hat n) d \Omega
\end{align}
where $Y_{\ell m}(\hat n)$ are the spherical harmonic functions (with the star indicating their complex conjugate) and $d\Omega$ is the integration measure on the 2-sphere. The inverse of this transform is
\begin{align}
X(\hat n) = \sum _{\ell=0}^\infty\sum _{m=-\ell}^{\ell}X_{\ell m}Y_{\ell m}(\hat n).\label{sphth}
\end{align}
We use upper-case $L,M$ for the multipole moments when we refer to a field of which we are interested in the large scales, and lower-case $\ell, m$ when we refer to a field of which we are interested in the small scales.

In practice the upper-limit of the sum over $\ell$ in Equation~\eqref{sphth} is cut off at some maximum $\ell$ which is set by the angular resolution. We use $\ell=3\times4096$ for our maximum $\ell$ for small-scale fields, as we are working in real-space with HEALPix maps of $\texttt{Nside}=4096$.

The (co-)variance in a map $X$ (or maps $X,Z$) as a function of scale is captured in the power spectrum $C^{XZ}_\ell$, which can be estimated as follows:
\begin{align}
C^{XZ}_\ell=\frac{1}{2\ell+1}\sum_{m=-\ell}^m X_{\ell m}Z^*_{\ell m}.
\end{align}

We will use the notation $\tilde X$ to refer to a field created with inverse-variance filtering:
\begin{align}
\tilde X_{\ell m} \equiv \frac{X_{\ell m}}{C_\ell^{XX}}.
\end{align}
In creating the inverse-variance filtered galaxy-overdensity maps $\tilde {\delta^{g}_\alpha}$ and temperature map $\tilde T$ we  measure the power spectra $C_\ell^{g_\alpha g_\alpha}$  and $C_\ell^{TT}$  directly from the data; see Figs~\ref{fig:ACTpower} and~\ref{fig:powergals}.

\subsection{kSZ velocity reconstruction pipeline}~\label{sec:kszvelrecpipeline}

\noindent We estimate the tomographically-binned projected large-scale velocity field from the small-scale temperature described in~\ref{sec:acT_description}  and the small-scale galaxy overdensity measurement described in Section~\ref{sec:smallscaleg}. Our pipeline is 2+1 dimensional in the sense that we work in two continuous angular dimensions on the sphere and one tomographically binned redshift dimension (labelled by Greek indices $\alpha$), as opposed to 3 continuous dimensions. The tomographic kSZ velocity field estimator requires a small-scale tomographic galaxy overdensity measurement $\delta ^{g}_\alpha$ as well as the small-scale kSZ temperature measurement.

Aside from different choices about galaxy binning to be described below, our kSZ velocity reconstruction pipeline is described in~\cite{2025JCAP...05..057M}. It is based on the \texttt{ReCCo} pipeline of~\cite{2023JCAP...02..051C}, which implements real-space version of a quadratic estimator (QE) introduced in~\cite{2018PhRvD..98l3501D}. The QE itself followed closely the formalism of~\cite{2003PhRvD..67h3002O}.

The theory of kSZ velocity reconstruction is described in Appendix~\ref{sec:kzsvelrectheory}. While we refer the reader to Section 3.1 of~\cite{2025JCAP...05..057M} for further details, the estimator is

\begin{equation}
 (\hat v^{\mathrm{kSZ}}_\alpha)_{LM}= A^\alpha_L\left(\tilde T(\hat n) \zeta^\alpha(\hat n)\right)_{LM}\label{vkSZ_recon}.
\end{equation}
Here, $\zeta^\alpha(\hat n)$ is the galaxy overdensity map refiltered such that it approximates the expected optical depth distribution in bin $\alpha$:
\begin{equation}
\zeta^\alpha_{\ell m} = C_\ell^{\tau {g_\alpha}} (\tilde {\delta^{g}_{\alpha}})_{\ell m}
\end{equation}
where $ C_\ell^{\tau g_\alpha} $ is a model for the galaxy-electron cross correlation in bin $\alpha$; we describe our model in Section~\ref{sec:gemodel}. Importantly, an incorrect model will lead to a biased velocity estimate; this bias is scale-independent and quantified by a quantity $b_v$ referred to as the ``velocity bias'' that will be marginalized over in our analysis.

The normalization $A^\alpha_L$ is given by
\begin{align}
A^\alpha_L = (2L+1)\left(\sum_{\ell; \ell^\prime}\frac{\left(f_{\ell \ell^\prime L }C_{\ell^\prime}^{\tau g_\alpha }\right)^2
}{C_\ell^{TT}C_{\ell^\prime}^{g_\alpha g_\alpha }}\right)^{-1}\label{a_normalization}
\end{align}
where
\begin{equation}
f_{\ell  \ell^\prime L} =  \sqrt{\frac{(2\ell+1)(2\ell^\prime+1)(2L+1)}{4\pi}}\wignerJ{\ell}{\ell^\prime}{L}{0}{0}{0}
\end{equation}
with $\wignerJ{\ell_1}{\ell_2}{\ell_3}{m_1}{m_2}{m_3}$ the Wigner 3-J symbols.
In practice, $A^\alpha_L$ is essentially scale-independent at the low $L$s that we are interested in.

The auto power spectra of the reconstructed velocity estimates have a noise bias $N^{(0)}_L$. Assuming the filters used in the inverse-variance filtering accurately describe the two-point power in the maps, this is given in the ensemble average by
\begin{align}
\left<\hat v_{\alpha,LM}^\mathrm{kSZ} \hat v_{\beta,L^\prime M\prime }^\mathrm{kSZ} \right>
&=\delta_{L L^\prime}\delta_{M M^\prime}\frac{A_L^\alpha A_L^\beta}{2L+1}\sum_{\ell;\ell^\prime}\frac{(f_{\ell \ell^\prime L }){}^2 C_{\ell^\prime}^{\tau g_\alpha}C_{\ell^\prime}^{\tau g_\beta}C_{\ell^\prime}^{g_\alpha g_\beta}}{C_\ell^{TT} C_{\ell^{\prime}}^{g_\alpha g_\alpha}C_{\ell^{\prime}}^{g_\alpha g_\beta}}\\
&\equiv N_L^{(0),\alpha\beta}.
\end{align}
Note that for the equal-redshift auto correlation we have
\begin{equation}
N_L^{(0),\alpha\beta}=A_L^\alpha\equiv N_L^{(0),\alpha}.
\end{equation}

We perform the velocity reconstruction separately in the north and south galactic caps. As we use the measured power spectra for the filters, this means that we measure them separately on each region (see Figs~\ref{fig:ACTpower} and~\ref{fig:dndz_dr9}). For homogeneous surveys, there would be no difference; however,  the inhomogeneous noise properties of the ACT map and the slightly different $\frac{dN}{dz}$ of the north and south galaxies results in filters that are  different on each region. In practice, this results in different $N^{(0)}_L$ in the different regions.

\subsection{Angular cross correlation}\label{sec:xcorr}

We measure the angular cross-correlations $\hat C_L^{v^{\mathrm{kSZ}}_{\alpha}v^{\mathrm{cont}}_\beta}$ and $\hat C_L^{v^{\mathrm{kSZ}}_{\alpha}g_{\beta}}$ ,  using a \texttt{pymaster}~\citep{2019MNRAS.484.4127A} pipeline. \texttt{pymaster} is a python implementation of the  MASTER algorithm~\citep{2002ApJ...567....2H}, which estimates a multipole-binned mask-decoupled power spectrum of two masked Gaussian fields. In~\cite{2025arXiv250621684L}, a more optimal quadratic maximum likelihood~\citep{2001PhRvD..64f3001T,2025arXiv251005215K} pipeline was implemented, leading to significant improvements in signal-to-noise (by an overall factor of $\sim2$). We leave such an implementation and a detailed further study of this improvement to future work.

We measure the mask-decoupled $\hat C_L$s in multipole bins with bandwidths $\Delta L = 12$, with two multipole bins with $\Delta L=3$ at very low $L$, i.e. bins with minimum $Ls$ defined by  $\left[0,3,6,18,24,\cdots\right]$. We do not use the lowest two bins in any analysis, taking an $L_{\mathrm{min}}$ of 6. Our maximum multipole is the one with boundaries $[42,53]$, where our signal-to-noise is saturated.  We have explored the stability of our signal on the size of $\Delta L$ and  the minimum multipole analyzed. 

\subsection{Covariance estimation}\label{sec:covariance}

We estimate the covariance between the $\hat C_L$s analytically using the \texttt{pymaster} implementation of the mask-decoupling of the simple ``Knox formula''  for the Gaussian covariance of the $C_L$s. This method requires an estimation of the true power in the maps. For this we use the measured power for the continuity equation velocity and galaxy  maps, and the theoretical noise power $N_L^{(0),\alpha\beta}$ for the kSZ velocity maps (we find that, after masking tSZ clusters, the theoretical power agrees well with the observed power). We have checked that adding a signal covariance matrix to the theoretical noise estimation results in a negligible change in the covariance. We include all inter-bin covariances. In general, before mask decoupling,
\begin{align}
\mathrm{Cov}(C_L^{v^{\mathrm {kSZ}}_{\alpha} v^{\mathrm {cont}}_{\beta}},C_{L^\prime}^{v^{\mathrm {kSZ}}_{\gamma} v^{\mathrm {cont}}_{\delta}})=&\frac{\delta _{L L^\prime}}{(2L+1) f_{\mathrm{sky}}}\nonumber\\
&\times\bigg{(}N_L^{(0),\alpha\gamma}C_L^{v_\beta^{\mathrm{cont}}v_\delta^{\mathrm{cont}}}\\\nonumber
&\hspace{2em}+C_L^{v_\alpha^{\mathrm{kSZ}}v_\delta^{\mathrm{cont}}}C_L^{v_\beta^{\mathrm{cont}}v_\gamma^{\mathrm{kSZ}}}\bigg{)}.
\end{align}
The second term in the brackets (the ``cosmic variance  term'' depends on the signal; we have run our pipeline with and without a best-fit signal included in the covariance, and find that it makes negligible difference. The covariance of $C_L^{v_\alpha^{\mathrm{kSZ}}g_\beta}$ is computed similarly, but with $C_L^{v_\beta^{\mathrm{cont}}v_\delta^{\mathrm{cont}}}$ replaced with $C_L^{g_\beta g_\delta}$. 

\begin{table}
\begin{tabular}{c|c|c|}
spec & $\chi^2$/ndof&PTE\\\hline\hline
$C_L^{v^{\mathrm{kSZ}}v^{\mathrm{cont}}}$ (diagonal, NGC)   &  55.013188 / 64  &  0.781 \\
$C_L^{v^{\mathrm{kSZ}}v^{\mathrm{cont}}}$ (diagonal, SGC)   &  58.408817 / 64  &  0.674 \\\hline
$C_L^{v^{\mathrm{kSZ}}v^{\mathrm{cont}}}$ (full, NGC)   &  994.06634 / 1024  &  0.743 \\
$C_L^{v^{\mathrm{kSZ}}v^{\mathrm{cont}}}$ (full, SGC)  &  994.62203 / 1024  &  0.739 \\\hline
$C_L^{v^{\mathrm{kSZ}}g}$ (first off-diagonal, NGC) &   111.72567 / 128  &  0.847 \\
$C_L^{v^{\mathrm{kSZ}}g}$ (first off-diagonal, SGC) &   110.8865 / 128  &  0.86 \\\hline
$C_L^{v^{\mathrm{kSZ}}g}$ (full, NGC) & 1043.4999 / 1024  &  0.329 \\
$C_L^{v^{\mathrm{kSZ}}g}$ (full, SGC)   &  1031.0789 / 1024  &  0.432 
\\\hline
\end{tabular}\caption{The PTEs of the $\chi^2$-to-zero for a nulled version of the dataset, where the redshifts of the small-scale galaxies are randomly shuffled. We list the $\chi^2$-to-zero over the number of degrees of freedom ndof, which we convert to a PTE assuming an analytic $\chi^2$ distribution. In all cases there are 4 bandpowers in each spectrum, with an $L_{\mathrm{min}}$ of 6 and an $L_{\mathrm{max}}$ of 53. We list the PTE for our cross-correlation with the continuity equation velocity ($C_L^{v^{\mathrm{kSZ}}v^{\mathrm{cont}}}$) as well as with the galaxy overdensity ($C_L^{v^{\mathrm{kSZ}}g}$), for the full inter-redshift-bin measurement (``full'') as well as case when we look at just the redshift-space diagonals (\textit{ie}, only the equal-redshift-bin correlations) for $C_L^{v^{\mathrm{kSZ}}v^{\mathrm{cont}}}$ and neighbouring bin cross spectra for $C_L^{v^{\mathrm{kSZ}}g}$ (``first off-diagonal''). All PTEs are acceptable, indicating that our covariance matrix correctly describes the data according to this test.}\label{tab:nullptes}
\end{table}

When we include the auto power spectrum of the kSZ velocity in the likelihood, we use
\begin{align}
\mathrm{Cov}(C_L^{v^{\mathrm {kSZ}}_{\alpha} v^{\mathrm {kSZ}}_{\beta}},C_{L^\prime}^{v^{\mathrm {kSZ}}_{\gamma} v^{\mathrm {kSZ}}_{\delta}})=&\frac{\delta _{L L^\prime}}{(2L+1) f_{\mathrm{sky}}}\nonumber\\
&\times\bigg{(}N_L^{(0),\alpha\gamma}N_L^{(0),\beta\delta}\\\nonumber
&\hspace{2em}+N_L^{(0),\alpha\delta}N_L^{(0),\beta\gamma}\bigg{)}.
\end{align}
The expression for the covariance of the auto power spectrum with the cross power spectrum is linear in the cosmic variance term:
\begin{align}
\mathrm{Cov}(C_L^{v^{\mathrm {kSZ}}_{\alpha} v^{\mathrm {kSZ}}_{\beta}},C_{L^\prime}^{v^{\mathrm {kSZ}}_{\gamma} v^{\mathrm {cont}}_{\delta}})=&\frac{\delta _{L L^\prime}}{(2L+1) f_{\mathrm{sky}}}\nonumber\\
&\times\bigg{(}N_L^{(0),\alpha\gamma}C_L^{v_\beta^{\mathrm{kSZ}}v_\delta^{\mathrm{cont}}}\\\nonumber
&\hspace{2em}+N_L^{(0),\beta\gamma}C_L^{v_\alpha^{\mathrm{kSZ}}v_\delta^{\mathrm{cont}}}\bigg{)}.
\end{align}
Again, we find that inclusion of this term is a negligible effect on our analysis.

{We note that an analytical covariance with a theoretical power spectrum instead of the measured power may be preferable, or alternatively a covariance measured from simulations. However, given the requirement of many simulations for convergence, we defer such a study to future work, although we note that the converged simulation-based covariance used in~\cite{2025JCAP...05..057M} was very close to the analogous theoretical estimate.

We test the covariance matrix by performing null tests. In particular, we run our pipeline on our dataset with the redshifts of the small-scale galaxies randomly shuffled; in this case we should detect a signal consistent with zero. We calculate the overall $\chi^2$ with respect to zero, and convert this into a probability-to-exceed (PTE) assuming an analytical $\chi^2$ distribution with number of degrees of freedom given by the number of datapoints. The PTEs are perfectly acceptable in all cases; we list them in Table~\ref{tab:nullptes}.}

\subsection{Likelihood and comparison with theory}\label{sec:likelihood}

We compare our measured bandpowers $\hat C_L^{v^{\mathrm{kSZ}}_\alpha v^{\mathrm{cont}}_\beta}$ to theory $C_L^{v^{\mathrm{kSZ}}_\alpha v^{\mathrm{cont}}_\beta}(\theta)$ (where $\theta$ is a parameter vector) using a Gaussian likelihood for the bandpowers 
\begin{align}
-2\ln \mathcal {L} (\hat C_L^{v^{\mathrm{kSZ}}_\alpha v^{\mathrm{cont}}_\beta}, \theta) &= \left(\hat C_L^{v^{\mathrm{kSZ}}_\alpha v^{\mathrm{cont}}_\beta}- C_L^{v^{\mathrm{kSZ}}_\alpha v^{\mathrm{cont}}_\beta}(\theta) \right ) \mathbb{C}^{-1} \nonumber\\
&\hspace{2em}\cdot\left(\hat C_L^{v^{\mathrm{kSZ}}_\alpha v^{\mathrm{cont}}_\beta}- C_L^{v^{\mathrm{kSZ}}_\alpha v^{\mathrm{cont}}_\beta}(\theta) \right ) \label{likelihood_def}\\
&\equiv \chi^2\label{chi2_def},
\end{align}
where $\mathbb C$ is the covariance matrix of $\hat C$.
The theory is described in Section~\ref{sec:theory}. The parameter vector contains parameters encompassing the velocity bias $b_v$, the galaxy bias $b_g$, and cosmology $f_{\mathrm{NL}}$ (with all other cosmological parameters assumed to be degenerate with $b_v$). In practice, we look at four models: two fixed-cosmology model where $f_{\mathrm{NL}}=0$, and two varying cosmology models where $f_{\mathrm{NL}}$ is freed. In each case, we consider one case where there is one $b_v$ parameter that is the same in all redshift bins, and one case where the velocity bias is allowed to vary independently in each redshift bin. In the latter case there are $N$ velocity bias parameters which we label as $b_v^\alpha$ (recall that $N$ is the number of redshift bins).

We always allow $N$ independent galaxy bias parameters $b_g^\alpha$. The galaxy bias is well-known, and so we take tight priors $b_g=2.1\pm0.1,2.1\pm0.1,2.3\pm0.1,2.4\pm0.1$ in to the redshift ranges $0.4<z^{\mathrm{photo}}<0.545$, $0.545<z^{\mathrm{photo}}<0.713$, $0.713<z^{\mathrm{photo}}<0.885$, $0.885<z^{\mathrm{photo}}<0.101$ respectively, corresponding to the bins and measurements of~\cite{2024JCAP...12..022K} and~\cite{2025JCAP...06..008S}, which analyzed the cross-correlation of the clustering of these galaxies with \textit{Planck} and ACT CMB lensing~\citep{2022JCAP...09..039C,2024ApJ...962..112Q,2024ApJ...962..113M,2024ApJ...966..138M}. Explicitly, in the 16-bin case, we use the first prior in the first four bins (which are created from rebinning the  $0.4<z^{\mathrm{photo}}<0.545$ bin, and the second prior in the second four bins, etc.

We perform a Markov chain Monte--Carlo (MCMC) sampling, using  \texttt{Cobaya}~\citep{2019ascl.soft10019T,2021JCAP...05..057T}. We minimize the likelihood using the BOBYQA minimizer~\citep{bobyqa,2018arXiv180400154C,2018arXiv181211343C}  implemented in \texttt{Cobaya} to find the best-fit parameters. We visualize our posteriors and compute summary statistics with \texttt{getDist}\footnote{\url{https://getdist.readthedocs.io/en/latest/}}~\citep{Lewis:2019xzd}.

 When we quote signal-to-noise ratios, we use the mean $b_v$ from the one-parameter model  from the chains divided by the errorbar estimated from its covariance: $\mathrm{SNR}=\frac{\bar b_v}{\sigma(b_v)}$. We have checked in many cases that this agrees with the SNR estimated of~\cite{2025arXiv250621684L}, where the separate estimates of $b_v^\alpha$ are co-added using their covariance matrix, and also that this agrees with the SNR quoted in~\cite{2025JCAP...05..057M}, which computed the improvement in $\chi^2$ between the best-fit point and the null case and converted the equivalent probability-to-exceed (PTE) to a  Gaussian number of sigma with the same PTE.

\section{Theory}{\label{sec:theory}}

In this Section we describe the theory quantities that we require to interpret our observables. In all cosmological calculations, we use the \textit{Planck} 2018 $\Lambda$CDM cosmology~\citep{2020A&A...641A...6P}, where $H_0\equiv 100\,h\, \mathrm{km/s/Mpc}=67.66\,  \mathrm{km/s/Mpc}$ is the Hubble parameter; $\Omega_b h^2=0.02242$ is the physical density of baryons today; $\Omega_c h^2=0.11933$ is the physical density of cold dark matter today;  and $\ln(10^{10}A_s)=3.047$ and $n_s=0.9665$, with  and $A_s$ and $n_s$ the amplitude and spectral index of scalar fluctuations respectively (at a pivot scale of $0.05\, \mathrm{Mpc}^{-1}$).\footnote{We note that since the beginning of this work, the most precise cosmological parameters have been superseded by the ACT DR6+\textit{Planck} cosmology~\citep{2025arXiv250314452L}. The small change in the cosmology would have negligible impact in our work.}

Two distinct aspects of theory modelling are required for this measurement: an astrophsyical model for the small-scale $C_\ell^{g\tau}$ that we use to create the kSZ velocity estimate, and cosmological models for the large scale observables $C_L^{v^{\mathrm{kSZ}}g}$ and $C_L^{v^{\mathrm{kSZ}}v^{\mathrm{cont}}}$. We discuss the small-scale $C_\ell^{g\tau}$ model in Section~\ref{sec:gemodel} and the large-scale cosmological modeling in Section~\ref{sec:largescale_theory}. Finally, we present our  model for scale-dependent bias induced by non-Gaussianity in Section~\ref{sec:fnl_theory}.

\subsection{Small-scale theory: galaxy-electron model}\label{sec:gemodel}

\begin{figure*}
\includegraphics[width=0.49\textwidth]{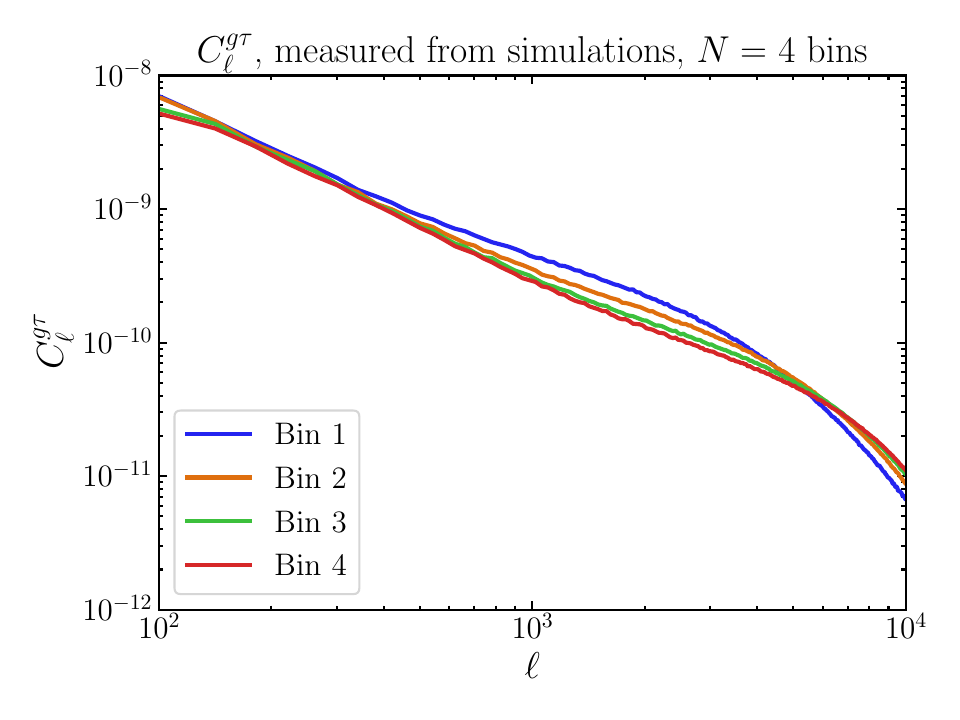}
\includegraphics[width=0.49\textwidth]{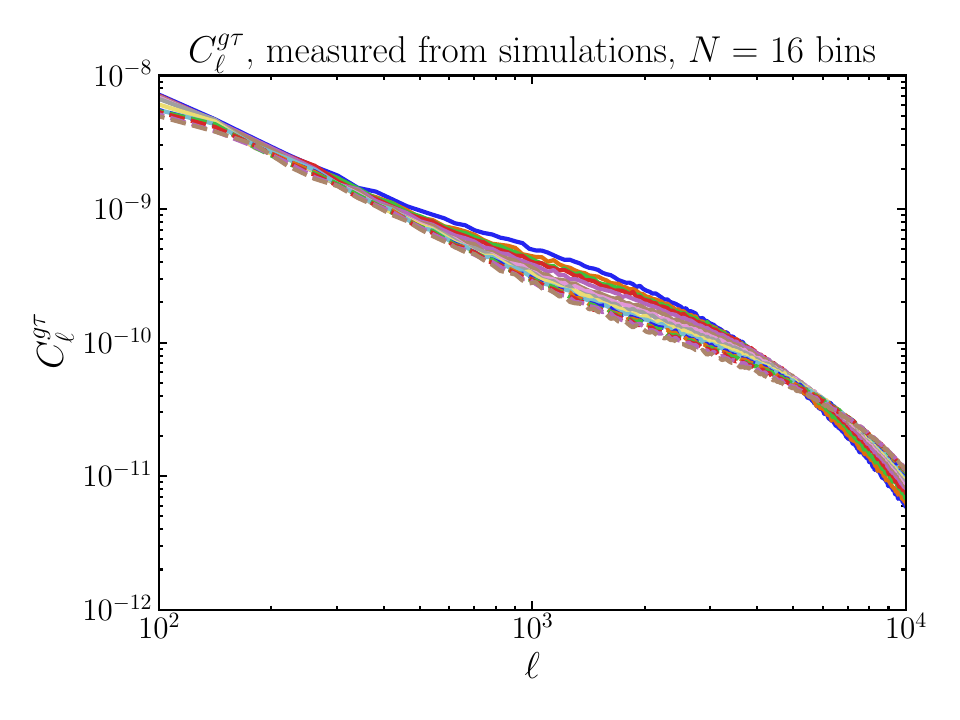}
\caption{The $C_\ell^{g\tau}$ model we use in our filters. These are measured from the \texttt{abacus} HOD and $\tau$ simulations. We measure this at every $\ell$, but for visualization purposes, we have binned them in $\ell$ space with a width $\Delta\ell=400$.
}\label{fig:filters}
\end{figure*}

\subsubsection{Galaxy-electron model}\label{sec:gemodeldescription}

kSZ velocity reconstruction requires a model for $C_\ell^{\tau g}$, to compute the normalization $A^\alpha_L$ (see Equation~\eqref{a_normalization}). In previous work, we modeled this using a halo model (for a review, see~\cite{2002PhR...372....1C}). In this work, we bypass the halo-model calculation and use filters measured  directly from $C_\ell^{g\tau}$ in simulations created from $N$-body simulations populated with LRG-like galaxies using a halo occupation distribution (HOD) and with a $\tau$ signal matched to hydrodynamic simulations. We create galaxy overdensity and $\tau$ maps from the simulations and cross-correlate them to directly estimate $C_\ell^{g\tau}$.  

The N-body simulations we use are the \texttt{AbacusSummit} light cone simulations~\citep{2021MNRAS.508.4017M,2022MNRAS.509.2194H}; we describe these, and our creation of galaxy and tau maps from them, in Appendix~\ref{sec:abacus_description}. 

We show the  $C_\ell^{g\tau}$ in Fig.~\ref{fig:filters}.

\subsubsection{Optical depth bias}\label{sec:optbias}

The assumption of a model for $C_\ell^{ \tau g_\alpha}$ results in a degeneracy: any parameter constraints we infer from $v^{\mathrm{kSZ}}_{\alpha}$ are degenerate with mismodelling of this quantity. While there is hope in the future of calibrating this quantity with direct tracers of the electrons such as fast radio bursts~\citep{2019PhRvD.100j3532M} or anisotropic CMB screening~\citep{2024arXiv240113033C,2024PhRvD.109j3539S,2025arXiv250617379H}, for now we use the fact that the optical depth bias is a scale-independent multiplicative quantity on large scales to define  $N$ optical depth bias parameters $b_v^\alpha$ (where $\alpha$ runs over 1 to the number of redshift bins $N$).\footnote{The scale-dependence of $b_v^\alpha$ has been tested on simulations~\citep{2022JCAP...09..028G,2023JCAP...02..051C}, with~\citep{2023JCAP...02..051C} finding that $b_v^\alpha$ was scale-independent at multipoles $L<200$.} If the data are not sufficiently constraining, these $N$ parameters can be condensed to one parameter $b_v$. In our previous work~\citep{2025JCAP...05..057M} we quoted an ``amplitude parameter'' $A$; this is simply different notation for $b_v$ and in this work we update to $b_v$ and $b_v^\alpha$ for consistency with other works in the literature.

The expected value of $b_v^\alpha$ given the true underlying galaxy-optical depth correlation $\left[C_\ell^{\tau g_\alpha}\right]^{\rm true}$ is
\begin{eqnarray}
 b_v^\alpha \simeq \frac{\left(\sum_{\ell} \frac{(2\ell+1) C_{\ell}^{\tau g_\alpha} \left[C_\ell^{\tau g_\alpha}\right]^{\rm true}}
{C_\ell^{TT}C_{\ell}^{g_\alpha g_\alpha }}\right)}{
\left(\sum_{\ell} \frac{(2\ell+1) \left(C_{\ell}^{\tau g_\alpha}\right)^2 }
{C_\ell^{TT}C_{\ell}^{g_\alpha g_\alpha }}\right)}.
\end{eqnarray}
It is important to note that this depends on the details of the CMB dataset through $C_\ell^{TT}$ (containing foregrounds, instrumental noise, etc) and the details of the galaxy survey through the definition of the redshift-binning as well as through $C_{\ell}^{g_\alpha g_\alpha }$ (containing shot noise, galaxy bias, etc). Care should therefore be taken when comparing the measurement of $b_v$ from different data combinations and analysis choices.

Any cosmological constraints from kSZ velocity reconstruction should be marginalized over $b_v$. Thus, {with low-precision measurements of $b_v$}, it is difficult to use the kSZ-reconstructed velocity to place constraints on scale-independent quantities like growth; much of the interest in the signal will be for parameters which induce a scale-dependence like local primordial non-Gaussianity~\citep{2008PhRvD..77l3514D,2019PhRvD.100h3508M}.

\subsection{Large-scale theory: velocity and galaxy power spectra}\label{sec:largescale_theory}

\subsubsection{Velocity-velocity power specrum}

We measure the cross-correlation between the kSZ-reconstructed radial velocity $\hat v^{\mathrm{kSZ}}_{\alpha}$ and the continuity-equation reconstructed radial velocity $\hat v^{\mathrm{cont}}_{\alpha}$. Each of these velocity fields is a projection of the true three-dimensional radial velocity field $\vec v(\chi, \hat n)$ with $\hat v^{\mathrm{kSZ}}_{\alpha}$ following the projection kernels stated in Equations~\eqref{ksz_proj} and $\hat v^{\mathrm{cont}}_\alpha$ defined in~\eqref{def_vcontprojection}. Assuming perfect reconstruction in both cases,  we can model both as a projection of the underlying radial velocity field with the same redshift kernel:
\begin{align}
v_{\alpha}(\hat n) = \int d\chi W^\alpha(\chi) v_r(\chi, \hat n),
\end{align}
where $W^\alpha(\chi)$ is the normalized redshift distribution of the galaxies in the bin $\alpha$:
\begin{equation}
W^\alpha(\chi) = \frac{\frac{dN^\alpha}{d\chi}}{\int \frac{dN^\alpha}{d\chi}d\chi}.
\end{equation}

\begin{figure*}
\includegraphics[width=\textwidth]{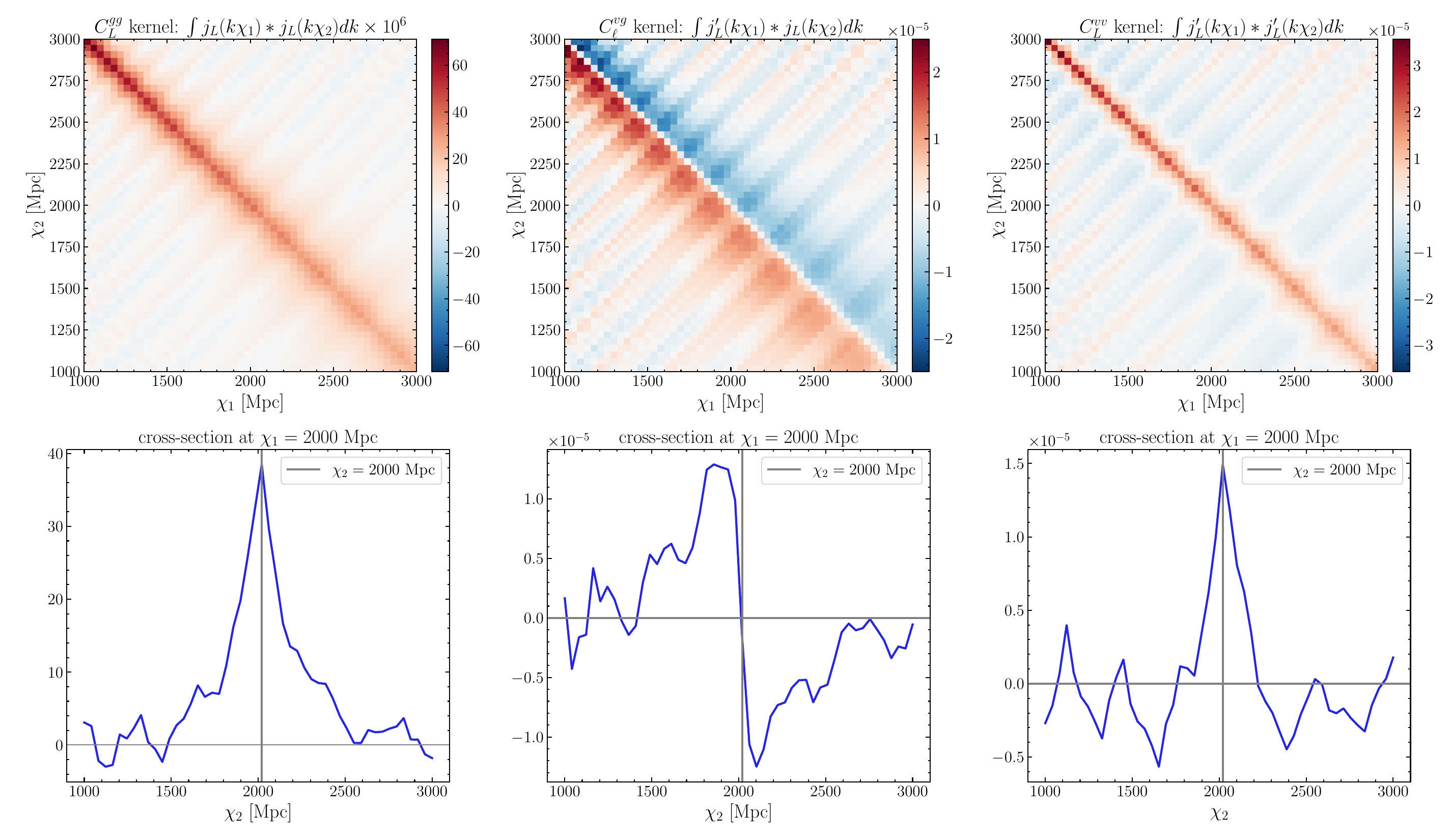}
\caption{The structure of the $(\chi_1,\chi_2)$ kernels used to evaluate $C_\ell^{gg}$, $C_\ell^{gv}$, and $C_\ell^{vv}$. The $C_\ell^{gv}$ kernel displays a clear dipolar nature, peaking at non-local regions in the ($\chi_1,\chi_2)$ plane. The $C_\ell^{gg}$ and $C_\ell^{vv}$ kernels are local and peak in the equal-redshift $\chi_1=\chi_2$.}\label{fig:structure}
\end{figure*}

The angular two-point correlation function of   the projected radial velocities is modelled as
\begin{align}
C_L^{v_\alpha v_\beta} =& \int d\chi_1 d\chi_2 W^\alpha(\chi_1) W^\beta(\chi_2)\nonumber\\
&\times\int \frac{k^2 dk}{(2\pi)^3}\mathcal K_L^v(\chi_1, k)\mathcal  K_L^v(\chi_2, k)P_{\mathrm{lin}}(\chi_1,\chi_2,k),\label{general_clvv}
\end{align}
where $P_{\mathrm{lin}}(\chi_1,\chi_2,k)$ is the linear matter power spectrum and the velocity kernel $\mathcal K^v_L(\chi, k)$ is given by
\begin{align}
\mathcal K_L^v(\chi, k) = 4 \pi i^L\frac{f(\chi) H(\chi) a(\chi)}{(2L+1)k}(L j_{L-1}(k\chi)  \nonumber\\
-(L+1) j_{L+1}(k\chi)),\label{velkernel}
\end{align}
where $j_L(x)$ are the Bessel functions of degree $L$; $f(\chi)\equiv\frac{d\ln \delta}{d\ln a}$ (with $\delta$ the dark matter overdensity and $a$ the scale factor) is the growth rate; and $H(\chi)$ is the Hubble parameter at $\chi$. 

\subsubsection{Velocity-continuity equation velocity power spectrum}

To estimate the velocity field from the large-scale galaxy distribution, we solve the redshift-space continuity equation (Equation~\eqref{cont}), which we restate here:
\begin{equation}
 \vec \nabla \cdot \vec v^{\mathrm{cont}} + \frac{f}{b_g} \vec \nabla \cdot \left[  \left( \vec v^{\mathrm{cont}} \cdot \hat n \right)\hat n \right] = - a H f \frac{\delta^g}{b_g},\label{cont}
\end{equation}
 where $b_g$ is an estimate of the galaxy bias. Thus the continuity equation velocity which we estimate is dependent on this galaxy bias estimate, and contains galaxy bias information, which is often suppressed when writing $v^{\mathrm{cont}}$. Restoring this, and neglecting the redshift space distortion (RSD) term (which is the second term on the left hand side of equation~\eqref{cont}), we see that the continuity equation velocity is related to the true velocity according to 
\begin{align}
\vec v^\mathrm{cont} = \frac{b_g(\chi, k)}{b_g^{\mathrm{fid}}} \vec v^{\mathrm{true}},
\end{align}
where $ \vec v^{\mathrm{true}}$ is the true underlying velocity field, $b_g^{\mathrm{fid}}$ is the fiducial bias chosen for the velocity reconstruction, and $b_g(\chi, k)$ is the true bias of the galaxies, which could have both redshift and scale dependence. Thus, the velocity-continuity equation velocity power spectrum is in fact
\begin{align}
C_L^{v_\alpha v^{\mathrm{cont}}_\beta} =& \int d\chi_1 d\chi_2 W^\alpha(\chi_1) W^\beta(\chi_2)\int \frac{k^2 dk}{(2\pi)^3}\nonumber\\
&\times\frac{b_g(\chi_2,k)}{b_g^{\mathrm{fid}}}\mathcal K_L^v(\chi_1, k)\mathcal  K_L^v(\chi_2, k)P_{\mathrm{lin}}(\chi_1,\chi_2,k).\label{clvv_fnl}
\end{align}

This modelling assumes perfect continuity-equation velocity reconstruction. However, as our pipeline is performed on imperfect data---with a notable problem being the photo-$z$ uncertainty of the data---further modelling is required. In order to incorporate effects of such properties of the data on the signal, we use simulations to calibrate a transfer function $T_L$ such that our final model is 
\begin{align}
C_L^{v_\alpha \hat v^{\mathrm{cont}}_\beta} =& T_L^{\alpha\beta}C_L^{v_\alpha  v^{\mathrm{cont}}_{\beta}}.
\end{align}
We present the simulations and the transfer function in Section~\eqref{sec:simdem}.

\subsubsection{kSZ velocity-continuity equation velocity power spectrum}\label{sec:clvvcont}

The kSZ velocity is modeled as scale-independent bias times the true velocity:
\begin{align}
v^{\mathrm{kSZ}}=b_v^\alpha v^{\mathrm{true}},
\end{align}
where $b_v^\alpha$ is the kSZ velocity bias that arises due to the mismodelling of $C_\ell^{\tau g}$. Thus the full expression we use for the kSZ velocity-continuity equation velocity power spectrum is 
\begin{align}
C_L^{v^{\mathrm{kSZ}}_{\alpha}\hat v^{\mathrm{cont}}_{\beta}} = b_v^\alpha C_L^{v_\alpha \hat v^{\mathrm{cont}}_{\beta}}.
\end{align}
In practice, unless scale-dependence is being constrained, the mis-specification of the linear bias $b_g^{\mathrm{fid}}$ is degenerate with the kSZ velocity bias $b_v^\alpha$ at the level of $C_L^{v^{\mathrm{kSZ}}v^{\mathrm{cont}}} $.

\subsubsection{Velocity-galaxy power spectrum}

The cross power spectrum between galaxies and true underlying binned radial velocity is
\begin{align}
C_L^{v_\alpha g_\beta} =& \int d\chi_1 d\chi_2 W^\alpha(\chi_1) W^\beta(\chi_2)\nonumber\\
&\times\int \frac{k^2 dk}{(2\pi)^3}\mathcal K_L^v(\chi_1, k)j_L(k \chi_2) b_g^{\beta}(\chi_2,k)P_{\mathrm{lin}}(\chi_1,\chi_2,k),\label{general_clvg}
\end{align}
where $b_g^\alpha(\chi_2,k)$ is the linear galaxy bias of the galaxies in redshift bin $\alpha$ at $\chi_2$, which could possibly depend on scale $k$. 

\subsubsection{kSZ velocity-galaxy power spectrum}

The power spectrum between the kSZ-reconstructed velocity and the galaxies is 
\begin{align}
C_L^{v^{\mathrm{kSZ}}_{\alpha} g_\beta} =b_v^\alpha C_L^{v_\alpha g_\beta}.
\end{align}
Just as for $C_L^{v^{\mathrm{kSZ}}v^{\mathrm{cont}}}$, misspecification in galaxy bias is degenerate with the velocity bias at the level of $C_L^{v^{\mathrm{kSZ}}g}$.

\subsubsection{Implementation of the theory}

We calculate the large-scale theory observables with a modified version of \texttt{ReCCO}~\citep{2023JCAP...02..051C},~\footnote{\url{https://github.com/jcayuso/ReCCO}} which uses \texttt{CAMB}~\citep{2000ApJ...538..473L}\footnote{\url{https://camb.info}} to compute background cosmological quantities, the linear matter power spectrum, and perturbation  transfer functions. 

On the extremely {large} scales of interest, the widely-used Limber approximation~\citep{1953ApJ...117..134L} is extremely poor. We use the beyond-Limber corrections of~\cite{2020JCAP...05..010F} to compute the full expressions~\eqref{general_clvv} and~\eqref{general_clvg}; the implementation is described in detail in {Appendix A of}~\cite{2023JCAP...02..051C}.

\subsubsection{Redshift-structure of the theory}\label{sec:redshiftstructure}

Due to the integral over spherical Bessel functions and their derivatives\footnote{Note that the expression in brackets Equation~\eqref{velkernel} is proportional to the derivative of a $j_L(k\chi)$}, an interesting structure of $C_L^{vg}$ and $C_L^{vv}$ emerges in redshift space. The velocity-galaxy signal is exactly zero at $\chi_1=\chi_2$, and exhibits a dipolar nature. Thus the $C_L^{v_\alpha g_\beta}$ signal peaks on the off-diagonals $(\alpha\ne\beta)$. The $C_L^{v_\alpha v_\beta}$ and $C_L^{g_\alpha g_\beta}$ signals peak on the equal-redshift diagonal. This structure is illustrated in Fig.~\ref{fig:structure}. This structure is the direct result of the velocity and density modes being related by a derivative and thus out-of-phase with each other.

\subsection{Scale-dependent bias induced by primordial non-Gaussianity}\label{sec:fnl_theory}

\subsubsection{Introduction}
Local primordial non-Gaussianity is a non-Gaussianity in the initial conditions of the Universe parametrized by $f_{\mathrm{NL}}^{\mathrm{loc}}$ of the type
\begin{align}
\Phi(\vec{x}) = \phi(\vec{x}) +f_{\mathrm{NL}}^{\mathrm{loc}} \left(\phi(\vec{x}) - \left<\phi\right>\right)^2,
\end{align}
where $\Phi(\vec{x})$ is the Newtonian potential at position $\vec{x}$ and $\phi(\vec{x})$ is an underlying Gaussian field. Note that, as $\phi$ is of order $\mathcal{O}(10^{-5})$, a value of $f_{\mathrm{NL}}^{\mathrm{loc}}$ of 1 in fact corresponds to a highly Gaussian field (hereafter we will rever to $f_{\mathrm{NL}}^{\mathrm{loc}}$ as  $f_{\mathrm{NL}}$) . Constraints on primordial non-Gaussianity are of great interest, as they can give insight into the physics of the early Universe. 
Local-type primordial non-Gaussianity in particular is sensitive to whether the process driving inflation involved one or multiple scalar fields, with scenarios of multi-field inflation generically leading to $f_{\mathrm{NL}}\sim\mathcal{O}(1)$(see, \textit{eg}~\citealt{2017PhRvD..95l3507D}). All current measurements are consistent with Gaussian initial conditions, with the tightest constraints from measuring the bispectrum of the \textit{Planck} CMB~\citep{2020A&A...641A...9P}  giving $f_{NL}=-0.9\pm5.1$. Non-Gaussianity of other types is also consistent with zero.

Local-type non-Gaussianity leads to a distinct signature in the clustering galaxies on very large scales~\citep{2008PhRvD..77l3514D}, inducing a scale-dependent galaxy bias $b^g\rightarrow b^g + \frac{f_{\mathrm{NL}}}{k^2}(b^g-p)$, where $b^g$ is the galaxy bias with respect to dark matter in a Gaussian scenario (which is scale-independent), and $p$ is a parameter which depends on galaxy assembly history (for objects following a universal halo mass relation, $p=1$). This unique signature has led to a significant large-scale-structure programme to constrain $f_{\mathrm{NL}}$. Reaching the errorbars $\sigma(f_{\mathrm{NL}})\sim1$ required to be sensitive to multi-field inflation will be difficult for two reasons: the difficulty of making robust large-scale measurements, and the cosmic (or sample) variance limit inherent on large scales.  
Cross correlation measurements with unbiased\footnote{or more generally, several differently-biased} (i.e., $b^g=1$) tracers (such as the true velocity field or the CMB lensing potential) have been shown to be able to bypass the sample variance limit through sample variance cancellation~\citep{2009PhRvL.102b1302S}: heuristically, by simultaneously measuring the biased power spectrum $b P(k)$ and the unbiased power spectrum $P(k)$ on the same area of sky, one can measure the non-stochastic quantity $\frac{b P(k)}{P(k)}=b$ and avoid the cosmic variance inherent in a measurement of the stochastic quantity $P(k)$. This effect was shown to be very promising in the context of kSZ tomography in~\cite{2019PhRvD.100h3508M} and~\cite{2019JCAP...10..024C}; recently, forecasts were revisited in~\cite{2025arXiv251025821T}. It has also been realized that the large-scale measurements required to access very low $k$ have many systematics and a cross-correlation \textit{alone} measurement may be significantly more robust. This has led to many recent constraints from cross-correlation-alone measurements on $f_{\mathrm{NL}}$, with the Integrated Sachs Wolfe effect~\citep{2014MNRAS.441L..16G}, CMB lensing~\citep{2023PhRvD.108h3522M,2024JCAP...03..021K,2025A&A...698A.177B,2025arXiv250420992F}, and most recently the kSZ-reconstructed velocity~\citep{2024arXiv240805264K,2025PhRvL.134o1003L,2025arXiv250621657H}. 

The tightest constraints from large-scale-structure remain auto-power-spectrum constraints, with $f_{\mathrm{NL}}=-3.6^{+9.0}_{-9.1}$ from DESI DR1 LRGs and quasars~\citep{2025JCAP...06..029C}. The tightest constraints from the kSZ velocity-galaxy cross correlation come from the combination of ACT DR5 and DESI LRG data and give $f_{\mathrm{NL}}=-39^{+40}_{-33}$~\citep{2025arXiv250621657H}.

\subsubsection{Theory and modelling details}\label{sec:fnltheory}

The precise effect of $f_{\mathrm{NL}}$ on the Gaussian bias is
\begin{align}
b^g \rightarrow b^g + f_{\mathrm{NL}}\frac{3 \Omega_m H_0^2}{k^2 T(k) D(z)}\delta_c(b^g-p),
\end{align}
where $\Omega_m$ is the mean density of matter today; $H_0$ is the Hubble constant; $T(k)$ and $D(z)$ are the transfer and growth functions of the density field, respectively, with $T(k)$ normalized to 1 at low $k$ and $D(z)$ normalized such that $D(z) = \frac{1}{1+z}$ during matter domination; and $\delta_c$ is the critical overdensity above which objects undergo gravitational collapse. We take $\delta_c=1.686$ and $p=1$ (appropriate for objects which follow a Universal halo mass relation, like dark matter halos). To constrain $f_{\mathrm{NL}}$, we simply use this model for the bias in Equation~\eqref{clvv_fnl}. 

We note that there is complete degeneracy between $f_{\mathrm{NL}}$ and $p$ (or the $b_\phi$ parameter that quantifies the coupling of galaxies to the primordial potential)~\citep{2008JCAP...08..031S,2022JCAP...11..013B}. Exploration of this degeneracy is beyond the scope of this paper.

\section{Demonstration of our pipeline on simulations}\label{sec:simdem}

\begin{figure*}
\centering
\includegraphics[width=0.8\textwidth]{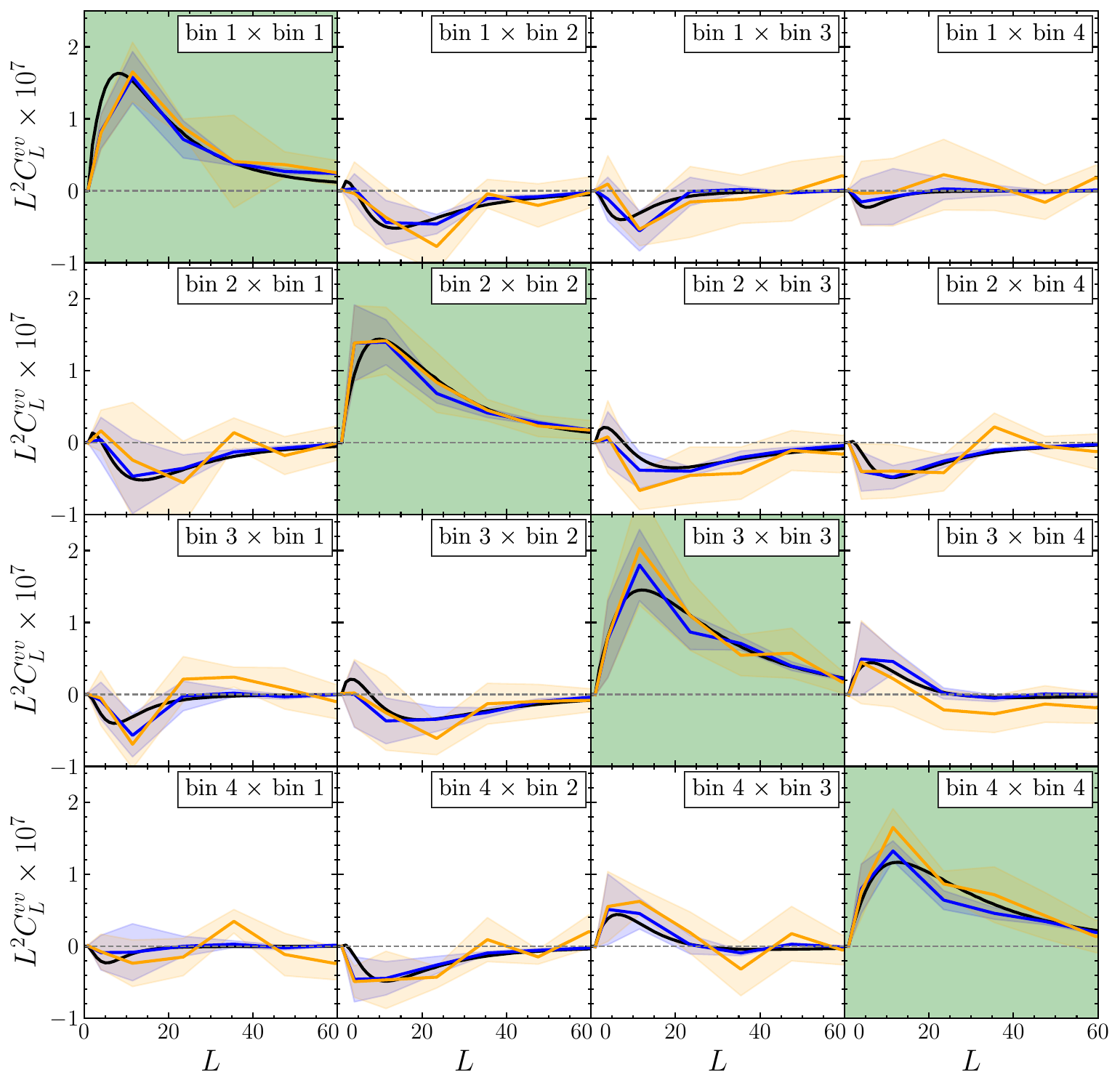}\\
\includegraphics[width=0.19\textwidth]{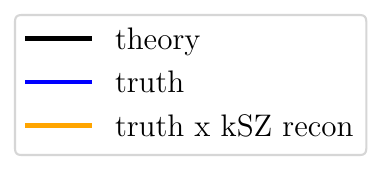}
\caption{The results  kSZ reconstruction  applied to simulations for the 4-bin case on the SGC geometry. The reconstructed power is consistent with the true power. We show the true power spectrum and the cross spectrum of the kSZ reconstruction with truth;  the kSZ reconstruction is noise dominated.  The shaded bands show the $1\sigma$ regions. The diagonal is highlighted in green to guide the eye. The mean power spectrum of the true velocity field (blue) matches the theoretical prediction (black), providing a validation of our theory code; additionally, the kSZ-reconstructed velocity follows this closely as well.}\label{fig:compareksztrue4bins}
\end{figure*}

\begin{figure*}
\centering
\includegraphics[width=0.8\textwidth]{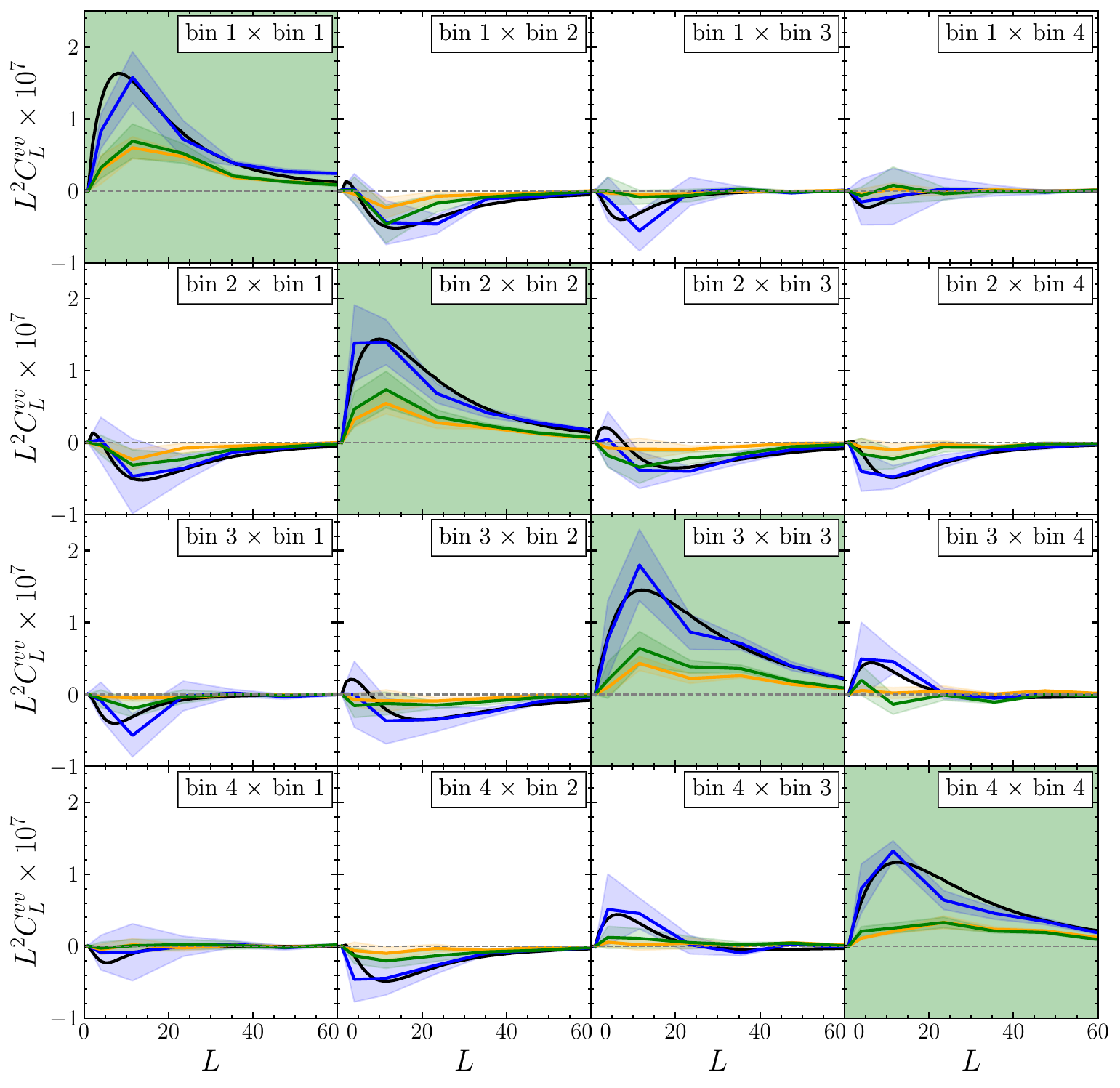}\\
\includegraphics[width=0.39\textwidth]{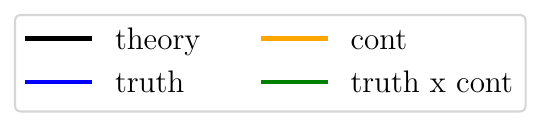}
\caption{ The $C_L^{vv}$ measured on the \texttt{AbacusSummit} lightcone simulations, for the $N=4$ tomographic bin case on the SGC, both before and after velocity reconstruction. The measured ``true'' power spectrum (\textit{ie}, the power spectrum of $v^\alpha_{\mathrm{true}}$) is shown in blue, with the mean over the six quasi-independent simulations in solid lines and the $1\sigma$ region in shaded lines.  The power spectrum of the continuity-equation reconstruction is shown in orange, and the cross-power spectrum of the truth and the reconstruction in green. {We show a $\Lambda$CDM theory prediction for each case in black; we find good agreement between the theory and the truth.}  The diagonal is highlighted in  green to guide the eye.}\label{fig:4bin_simspower}
\end{figure*}

We demonstrate and calibrate various aspects of our pipeline on the \texttt{AbacusSummit} light-cone simulations~\citep{2021MNRAS.508.4017M,2022MNRAS.509.2194H}.  We describe these simulations in Appendix~\ref{sec:abacus_description}.

We demonstrate the unbiased recovery of the true velocity (given the correct model for $C_\ell^{g\tau}$) in Section~\ref{sec:compareksztrue4bins}. As discussed in Section~\ref{sec:clvvcont}, the continuity-equation estimated velocity field is biased, and requires calibration on simulation; we discuss this in more detail and present the calibration in Section~\ref{sec:continuity_calibration}.

\subsection{kSZ reconstruction demonstration}\label{sec:compareksztrue4bins}

We treat our simulations like the dataset and run our kSZ velocity reconstruction pipeline (described in Section~\eqref{sec:kszvelrecpipeline}). We add uncorrelated Gaussian noise to the kSZ map beforehand, in order to match the two-point power in the data. The results are shown in Fig.~\ref{fig:compareksztrue4bins}, where very good agreement is seen between the kSZ reconstruction and the truth.  Thus we confirm that the QE makes an unbiased estimate of the velocity, in the case where $C_\ell^{g\tau}$ is perfectly known.

A validation of our theory code is also included in  Fig.~\ref{fig:compareksztrue4bins}, as we include the theoretical prediction for $C_L^{vv}$ as well as the measured truth, and find that the theory describes the truth well.

\subsection{Continuity equation velocity calibration}\label{sec:continuity_calibration}

\begin{figure*}
\centering
\includegraphics[width=0.8\textwidth]{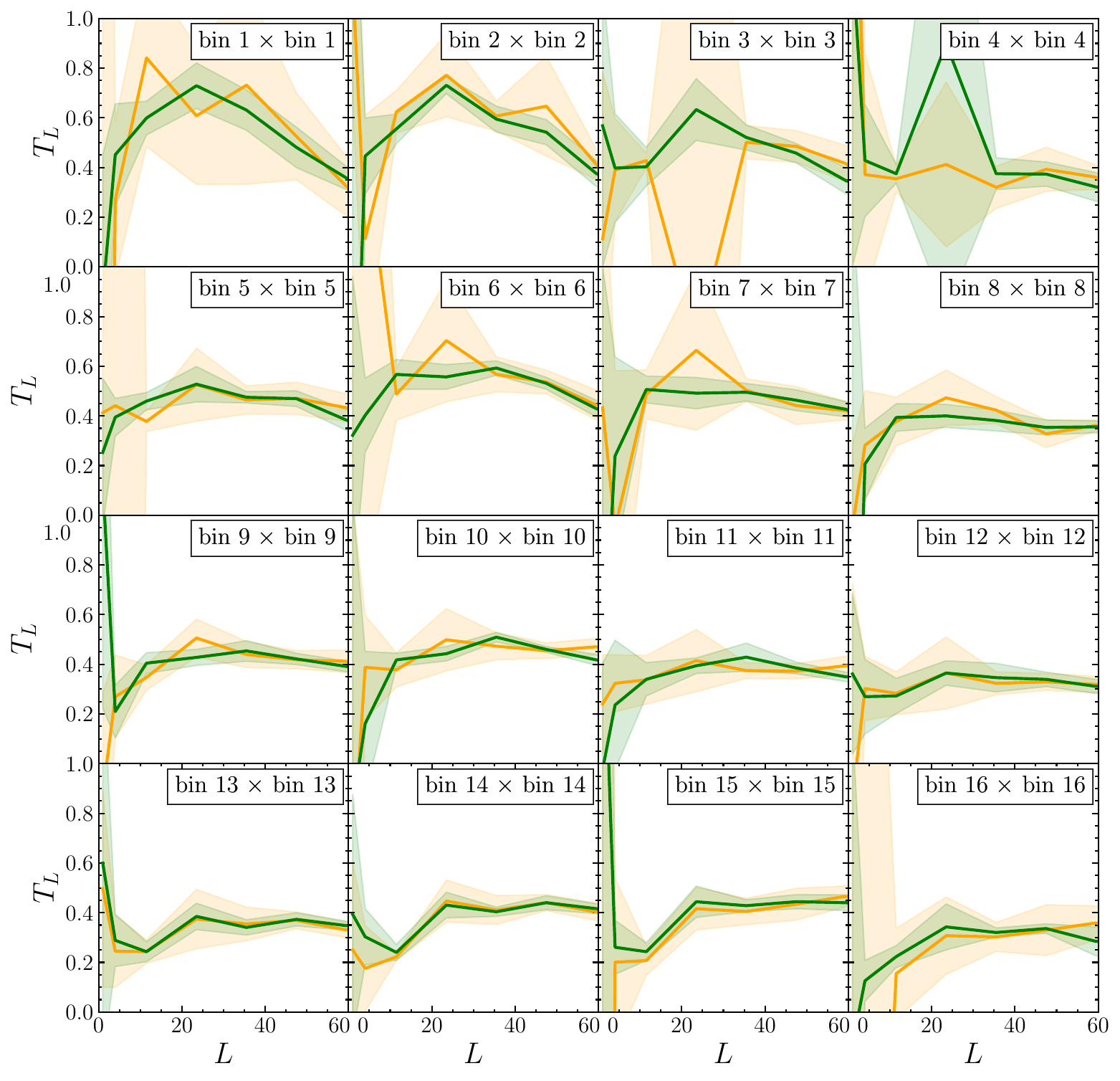}\\
\includegraphics[width=0.23\columnwidth]{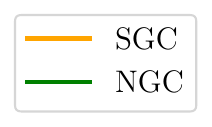}

\caption{The transfer functions measured from the simulations with $N=16$ tomographic bins. We show $\left<\frac{C_L^{v^{\mathrm{true}}v^{\mathrm{cont}}}}{C_L^{v^{\mathrm{true}}v^{\mathrm{true}}}}\right>$  In general, a transfer function $T_L\sim0.5$ is observed.  We plot the mean over the six quasi-independent simulations with the standard deviation indicated with a shaded band..}\label{fig:4bin_simtransfer}
\end{figure*}
\subsubsection{The need for simulation calibration of the theory}

For the continuity equation velocity reconstruction part of our pipeline, we must understand the fidelity of the velocity reconstruction. In~\cite{2024PhRvD.109j3533R,2024PhRvD.109j3534H}, it was shown that for a DESI LRG-like galaxy sample with $\frac{\sigma_z}{(1+z)}\sim0.02$, the velocity reconstruction has a correlation coefficient with the truth of $\sim0.3$. This correlation coefficient was defined as
\begin{align}
r^{v_{\mathrm{cont}},v_{\mathrm{true}}}=\frac{\left<v_{\mathrm{cont}}^i v_{\mathrm{true}}^i\right>}{\sqrt{\left<v_{\mathrm{cont}}^i v_{\mathrm{cont}}^i\right>\left<v_{\mathrm{true}}^i v_{\mathrm{true}}^i\right>}},
\end{align}
where the superscript $i$ indicates velocity evaluated at the position of the object labelled by $i$, $v_{\mathrm{true}}$ is the true halo velocity, and $v_{\mathrm{cont}}$ is the continuity-equation-estimated velocity. The angular brackets indicate the mean over all objects $i$. This quantity is an average over all scales; there was some exploration in~\cite{2024PhRvD.109j3534H} of the scale-dependence of $r$ (see Fig. 6 therein), although not for photometric samples.

The correlation coefficient  is not the quantity that we will need to quantify the performance of our continuity equation reconstruction. Our observable is $C_L^{v^{\mathrm{cont}}v^{\mathrm{kSZ}}}$, which we will for simplicity treat  as $C_L^{v^{\mathrm{cont}}v^{\mathrm{true}}}$ (having validated our kSZ-reconstruction estimation in the previous section). Under the assumption that the continuity equation reconstruction creates an unbiased estimate of the velocity field,~\textit{i.e.}
\begin{align}
v^{\mathrm{cont}} = v^{\mathrm{true}}+n,
\end{align}
where $n$ is some noise, then even with a low correlation coefficient we can recover an unbiased measurement of the cross-power. Conversely, if the continuity-equation reconstruction creates a biased but noiseless estimate of the velocity field,
\begin{align}
v^{\mathrm{cont}} = Tv^{\mathrm{true}},
\end{align}
with transfer function $T\ne1$, a high correlation coefficient can still result in an in correct cross-correlation with the truth. In reality, we model the picture as somewhat in between:
\begin{align}
v^{\mathrm{cont}} = Tv^{\mathrm{true}} + n,
\end{align}
where $T$  quantifies the quality of the velocity reconstruction and allows for missing modes in the continuity-equation reconstruction, and $n$ is all noise that is not correlated with $v^{\mathrm{true}}$.

\begin{figure*}
\includegraphics[width=\textwidth]{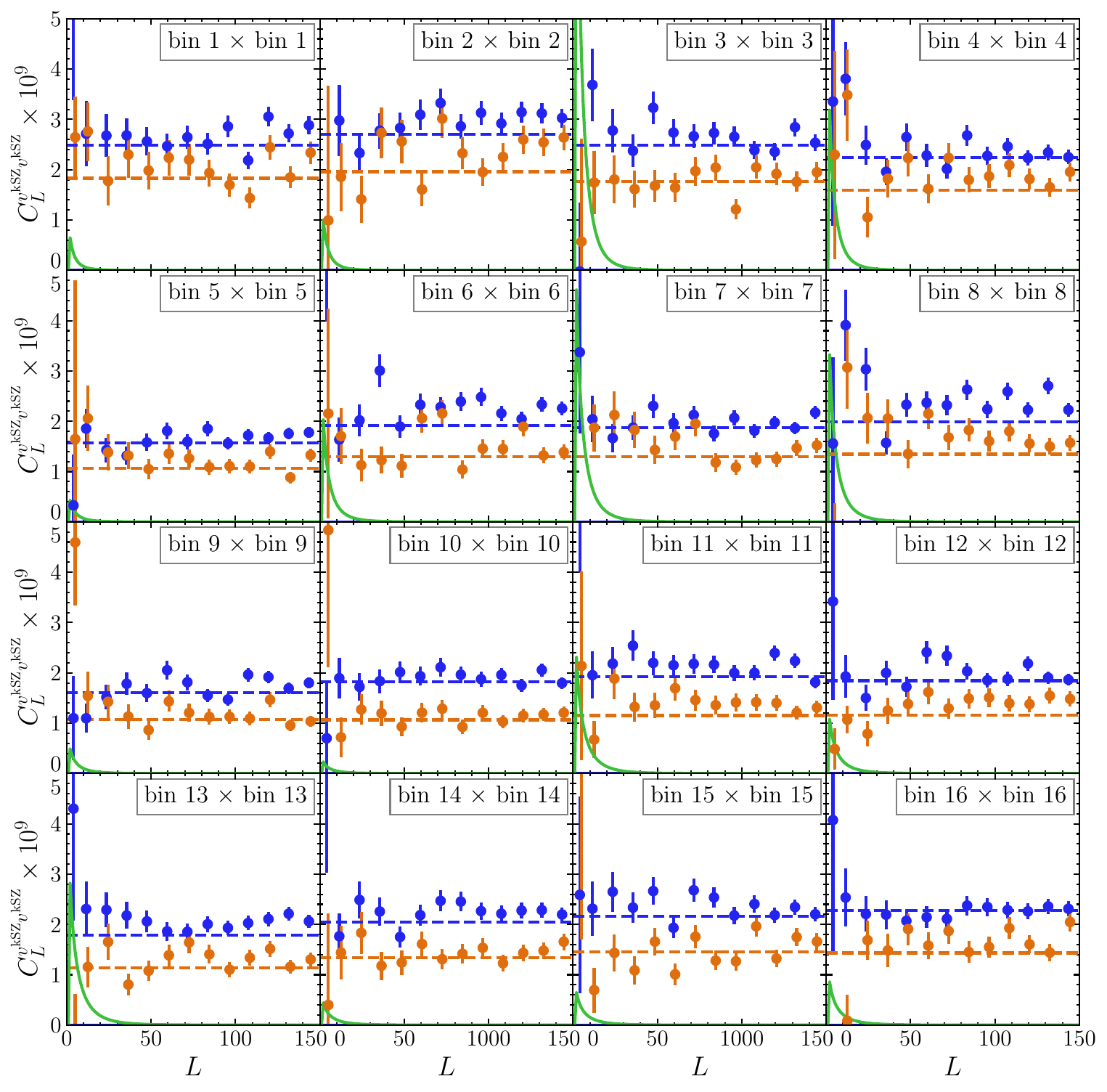}
\includegraphics[width=0.6\textwidth]{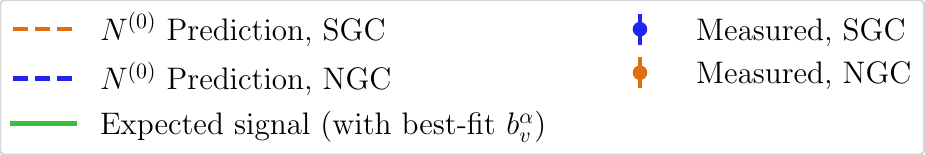}
\caption{
The measured auto power spectra of the kSZ-reconstructed velocity. The NGC region has lower $N^{(0)}$ bias due to the lower-variance  CMB measurement in this region, as well as slightly higher galaxy number density in some redshift bins. In general, we find good agreement between the measured power and the analytical $N^{(0)}_L$ prediction (when masking galaxy clusters in the CMB+kSZ map before performing kSZ velocity reconstruction). The best-fit signal indicated uses the $b_v^\alpha$ from the best-fit model {to the cross $C_L^{v^{\mathrm{kSZ}}v^{\mathrm{cont}}}$, as discussed in Section ~\ref{sec:crosspower}}. }\label{fig:autospectra}
\end{figure*}

In order to understand the quantity $C_L^{v^{\mathrm{cont}}v^{\mathrm{true}}}$ it will be helpful to ignore (for now) any anisotropy in $T$ (due to, \textit{e.g.}, relatively poor performance near the edges of the footprint), and write
\begin{equation}
v^{\mathrm{cont}}_L = T_Lv^{\mathrm{true}}_L + n_L.
\end{equation}
We can then write for the cross-power spectrum we are interested in
\begin{align}
C_L^{v^{\mathrm{cont}}v^{\mathrm{true}}} =& T_L C_L^{v^{\mathrm{true}}v^{\mathrm{true}}};
\end{align}
in order to interpret the $C_L^{v^{\mathrm{kSZ}}v^{\mathrm{cont}}}$ that we measure, it will be necessary (in the absence of deeper analytical understanding) to calibrate $T_L$ from  simulations.

Comparing the measured auto-correlation of the continuity-equation velocity reconstruction on data $C_L^{v^{\mathrm{cont}}v^{\mathrm{cont}}}$ to the measured auto correlations of the simulations will serve as a check that our $T_L$ is realistic. 

Note that by using the same survey geometry for the simulations as we do for the data, any positional dependence in $T$ should be appropriately encompassed into the average $T_L$ that we calibrate (although in practice we find little dependence on survey geometry, at least at the current level of sensitivity).

Note that we have suppressed the redshift indices $\alpha$ in the above discussion, although we will need to calibrate a $T_L^{\alpha\beta}$ for all inter-bin cross correlations, in order to allow for appropriate redshift dependence in $T_L$. We describe our calibration of $T_L$ in section~\ref{sec:conttl}.

\begin{figure*}
\includegraphics[width=\textwidth]{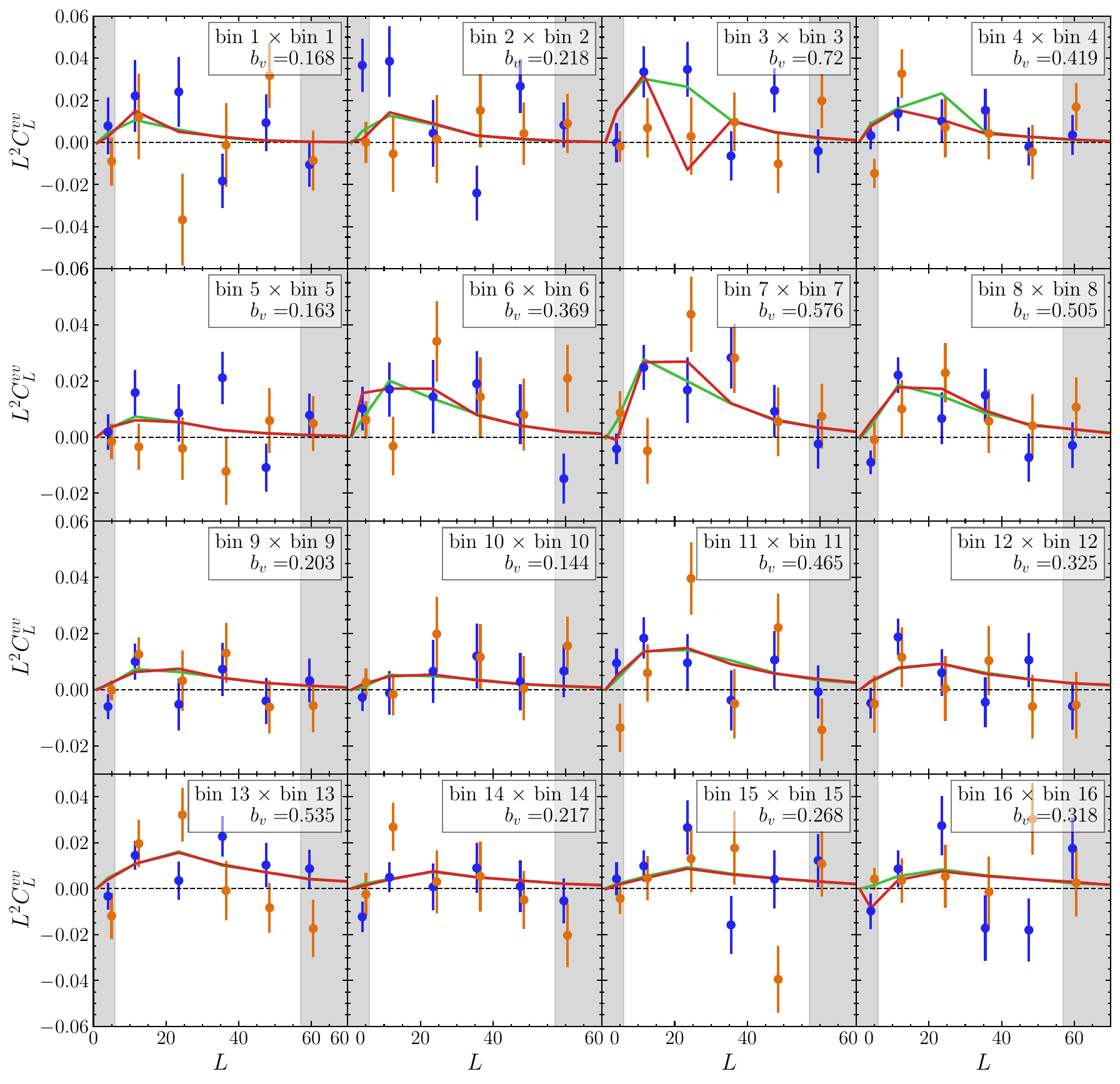}
\includegraphics[width=0.5\textwidth]{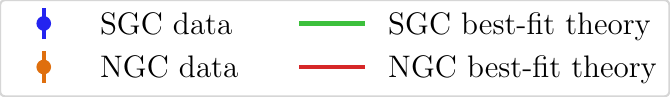}

\caption{The baseline datapoints and the best-fit model. The best-fit model has 16 parameters $b_v^\alpha$ and is jointly fit to all bins; the only difference in the model between the NGC and SGC cases is due to the different transfer functions. While there are a minority of cases where the transfer function looks unconverged, we have explored constraints dropping these bins and find negligible effect. We do not use the datapoints in the gray regions in any fits. For visualization purposes, we co-add these and display them in Fig.~\ref{fig:coadd}.}\label{fig:bestfit16bins}

\end{figure*}

\begin{figure*}
\includegraphics[width=0.8\textwidth]{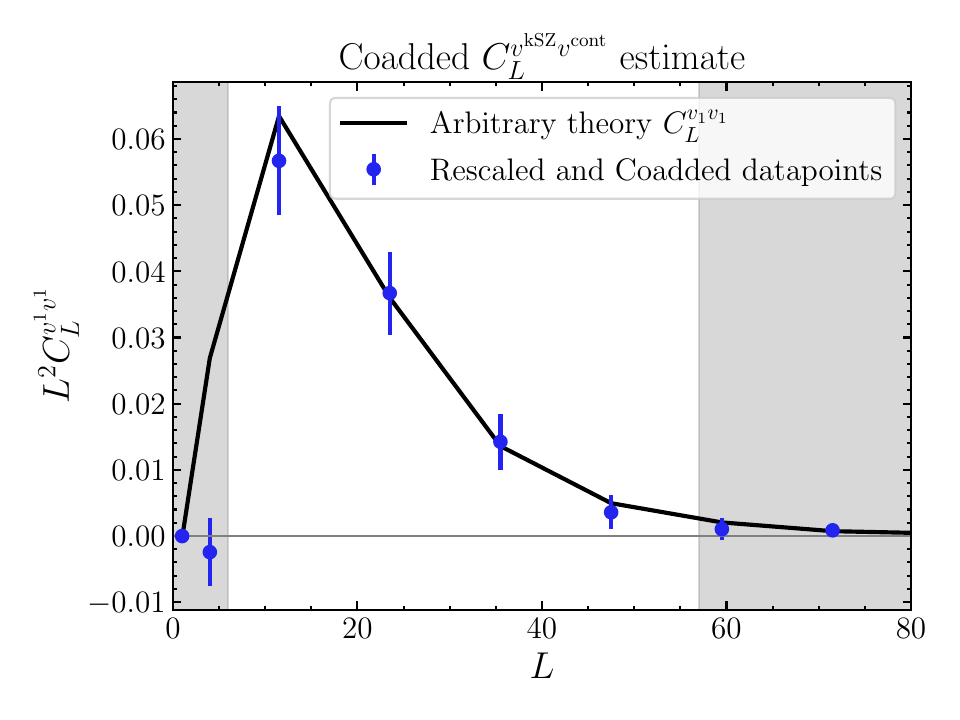}
\caption{A co-add of the datapoints rescaled to  estimate  $C_L^{v_1v_1}$ (arbitrarily chosen). The scales not used in the analysis are indicated in the gray regions. We do not use these points in the analysis; we analyze the points shown in Fig.~\ref{fig:bestfit16bins}.}\label{fig:coadd}
\end{figure*}
\subsubsection{Demonstration of continuity-equation reconstruction on simulations}\label{sec:conttl}

We perform our continuity-equation reconstruction pipeline on the DESI DR10 LRG simulations described in Appendix~\ref{sec:abacus_description}. We
 carry out the continuity-equation velocity reconstruction independently in each of the four larger redshift bins (with redshift boundaries shown in the first row of Table~\ref{tab:redshiftboundaries}).

Before and after performing the continuity-equation velocity reconstruction, we  project each bin to two dimensions and measure both  mean true velocity and reconstructed velocity in each $N_{\mathrm{side}}=64$ HEALPix pixel, resulting in $N$ $v^\alpha_{\mathrm{true}}$ and $v^\alpha_{\mathrm{cont}}$ maps respectively. We then measure $C_L^{v^{\mathrm{true}}_{\alpha}v^{\mathrm{true}}_{\beta}}$, $C_L^{v^{\mathrm{cont}}_{\alpha}v^{\mathrm{cont}}_{\beta}}$, and $C_L^{v^{\mathrm{true}}_{\alpha}v^{\mathrm{cont}}_{\beta}}$, with a \texttt{pymaster} mask-decoupled estimate corresponding to the same bandpowers we use in the analysis.

\begin{figure*}
\includegraphics[width=\columnwidth]{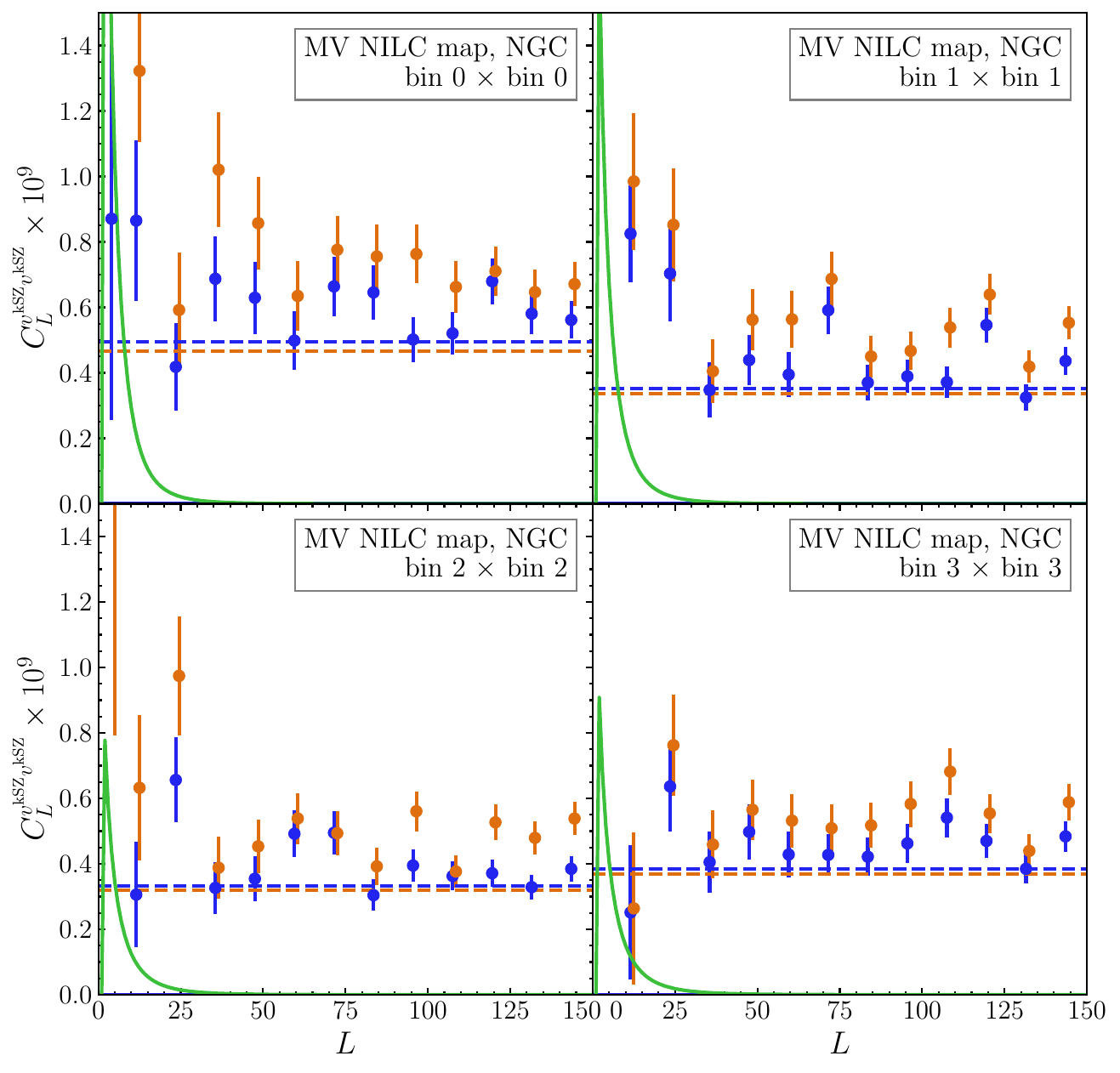}
\includegraphics[width=\columnwidth]{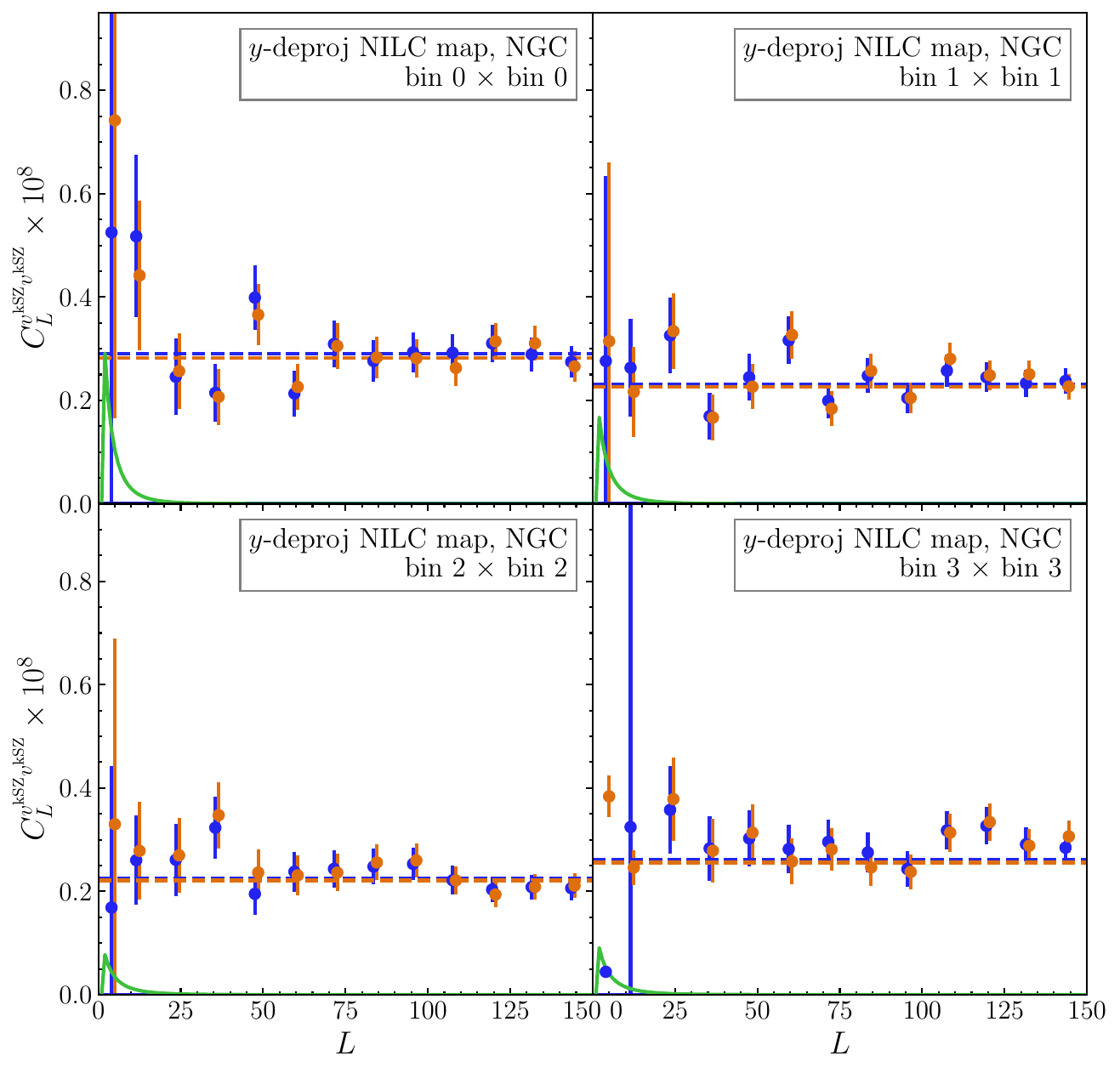}
\includegraphics[width=1.6\columnwidth]{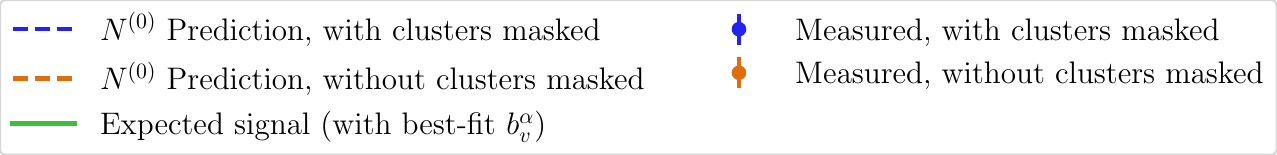}

\caption{Auto power spectra of the kSZ reconstructed velocity for the $N=4$ bin case on NGC. On the left we show the reconstruction with the unconstrained NILC, and on the right with the SZ-deprojected NILC; we show this only for NGC but conclusions are similar for SGC. We see that only for the unconstrained NILC when tSZ clusters are unmasked does the analytical prediction disagree with the measured power, indicating that removing these clusters (either through masking or deprojecting the tSZ effect) makes the analytical prediction much closer to the measurement. Note the significantly larger $y$-axis scale on the right; the SZ-deprojected velocity reconstructions have a factor of$\sim 5\times$ more power than the MV NILC measurement. Note also that we have slightly offset the orange ``measured, without clusters masked'' points for slightly better comparison with the other points. }\label{fig:auto_tsz}
\end{figure*}

The mean $C_L^{v_\alpha v_\beta}$ for the 4-bin case are shown in Fig.~\ref{fig:4bin_simspower}.  In all plots we include a theory estimate (see Section~\eqref{general_clvv}); we find good agreement between our theory and the observed true velocity power spectrum. The relevant transfer functions are shown in Fig.~\ref{fig:4bin_simtransfer}  for the  16-bin case, for the $\alpha=\beta$ diagonal of $C_L^{v_\alpha v_\beta}$ equal-redshift correlations only. 

We also attempt to calibrate transfer functions for the inter-bin ($a\ne\beta$) measurement, but found that these are unconverged for bins separated by a large distance (when the signal is zero). Thus we never include such bins in our analysis and restrict ourselves only to the diagonal ($\alpha=\beta$) for our nominal analysis, although we also perform a fit for the case when we also include the first off-diagonal ($\alpha=\{\beta+1,\beta-1\}$)). Most of the signal-to-noise is concentrated along or close to the diagonal.

\subsubsection{Comparison to data}

As we will use the $T_L$ in our analysis, it is an important check to compare the equivalent quantity for the auto-spectrum to the data products. In general, we find good  consistency between the auto power spectra of the data and the simulations; we show in App.~\ref{sec:boryanauto} the auto power spectra of the velocity field as measured on the data along with the mean measurement of the same signal from our simulations.

\section{Results}\label{sec:results}

In this Section we present, discuss, and analyze our measurement.  We make the measurement for an $N=4$, $N=8$, and $N=16$ tomographic bin case and compare these with each other for consistency. All plots in this section are for the baseline $N=16$ bin case, unless otherwise stated (note that sometimes we will plot examples for the $N=4$ bin case to improve readability). 

Our ``baseline''signal is $C_L^{v^{\mathrm{kSZ}}_\alpha v^{\mathrm{cont}}_\alpha}$   for a 16-bin case, where $\alpha$ runs from 1 to 16. For the larger more general $C_L^{v^{\mathrm{kSZ}}_\alpha v^{\mathrm{cont}}_\beta}$  signal there are  $16\times16=$256 inter-bin combinations; plots displaying these datapoints are difficult to read and interpret. Indeed, as most of the signal-to-noise is concentrated at equal-redshift ($\alpha=\beta$) diagonal, and because we find a good fit when we look at the diagonal-only case, we therefore primarily include only plots looking at the $\alpha=\beta$ equal-redshift diagonal. When we refer to our ``baseline'' dataset, we mean this $\alpha=\beta$ redshift diagonal. We defer a detailed analysis of the signal-to-noise penalty of removing the off-diagonal bins to future work.

In Section~\ref{sec:autospectra} we present the auto spectra of the kSZ-reconstructed velocities; these  are shown in Fig.~\ref{fig:autospectra}. In Section~\ref{sec:crosspower} we present and analyze the cross-power measurement between $\hat v^{\mathrm{kSZ}}$ and $\hat v^{\mathrm{cont}}$; our baseline datapoints, along with the best-fit model, are shown in Fig.~\ref{fig:bestfit16bins}; we show a co-add of these points in Fig.~\ref{fig:coadd}.  In Section~\ref{sec:scale_dep} we open up the scale dependence of the model and constrain $f_{\mathrm{NL}}$.  In Section~\ref{sec:auto} we  include the auto spectrum of  $\hat v^{\mathrm{kSZ}}$ in the analysis and report a $2.1\sigma$ measurement of the auto spectrum.

\subsection{Auto power spectra of the kSZ reconstructions}\label{sec:autospectra}

The auto power spectra of our kSZ-estimated velocities are shown in Figure~\ref{fig:autospectra}.

For reconstruction performed on Gaussian fields with zero signal, the kSZ velocity auto power-spectra contain only the reconstruction noise $N^{(0),\alpha}_L$. This is scale independent (on large scales) and, if the small-scale auto power of the fields matches the $C_\ell^{TT}$ and $C_\ell^{gg}$ filters used in the filtering and reconstruction normalization ({i.e.}, if the filters used in the inverse-variance filtering of the maps and the theoretical computation of the normalization in ~\eqref{a_normalization} truly represent the data), can be analytically calculated to be
\begin{align}
N_L^{(0),\alpha}=A^\alpha_L,
\end{align}
where $A^\alpha_L$ is the normalization in Equation~\eqref{a_normalization}.

The assumptions of this analytical calculation are that all fields are Gaussian, and that $C_\ell^{gT}=0$ ({i.e.}, that the temperature and galaxy fields are uncorrelated). Of course,  a non-zero $C_\ell^{gT}$ is expected due to the correlation between the galaxies and  foregrounds to the temperature field such as the CIB or tSZ effect; it is straightforward to include such an effect into the calculations both of the optimal estimator weights and of the resulting $N_L^{(0)}$ (the expressions are given in full in~\cite{2023JCAP...02..051C}); however, this  effect is subdominant as it is a correction of order $r_\ell\equiv\frac{C_\ell^{gT}}{\sqrt{C_\ell^{gg}C_\ell^{TT}}}$ which we have measured  to be everywhere less than $2\%$. Thus, we neglect this correction. The remaining effect that we expect to observe from the non-zero $C_\ell^{gT}$ is a contribution to the estimator \textit{monopole}. We  remove this contribution and suppress any variance sourced by it by subtracting the observed monopole (estimated on the sky area available) before making any auto- and cross-power spectrum measurements.

Previous work~\citep{2024arXiv240500809B,2025JCAP...05..057M,2025arXiv250621684L} found an offset of order $10-70\%$ between the theory prediction for noise and the measured high-$L$ power (where in the context of $C_L^{vv}$ high-$L$ now means $L>\sim50$, where no signal is expected, and the measured power was always higher than the prediction).\footnote{The disagreement  in \citep{2024arXiv240500809B} can be eliminated with a more accurate estimation of the Planck temperature power spectrum.} The noise amplitude had to be separately calibrated in the estimation of the covariance (in all of these works). For~\cite{2025JCAP...05..057M} we performed some tests replacing either the galaxy or the temperature data with simulations which matched the power spectra of the data; we found that in both cases, the measured auto spectra agreed much more closely with the analytical prediction. Given that the kSZ reconstruction auto power spectrum is a trispectrum of the form $\left<T^ST^S\delta^S\delta^S\right>$, our simulation tests suggested that this was caused by a bias  of the form $\left<\mathrm{FG}^S\mathrm{FG}^S \delta^S\delta^S\right>$, where $\mathrm{FG}^S$ is some smal-scale foreground to the temperature measurement that is correlated with the galaxies. As the Gaussian contribution from $C_\ell^{gT}$ is subdominant (as discussed above),  we expected this to be from the connected part of this trispectrum.

\begin{figure}
\includegraphics[width=\columnwidth]{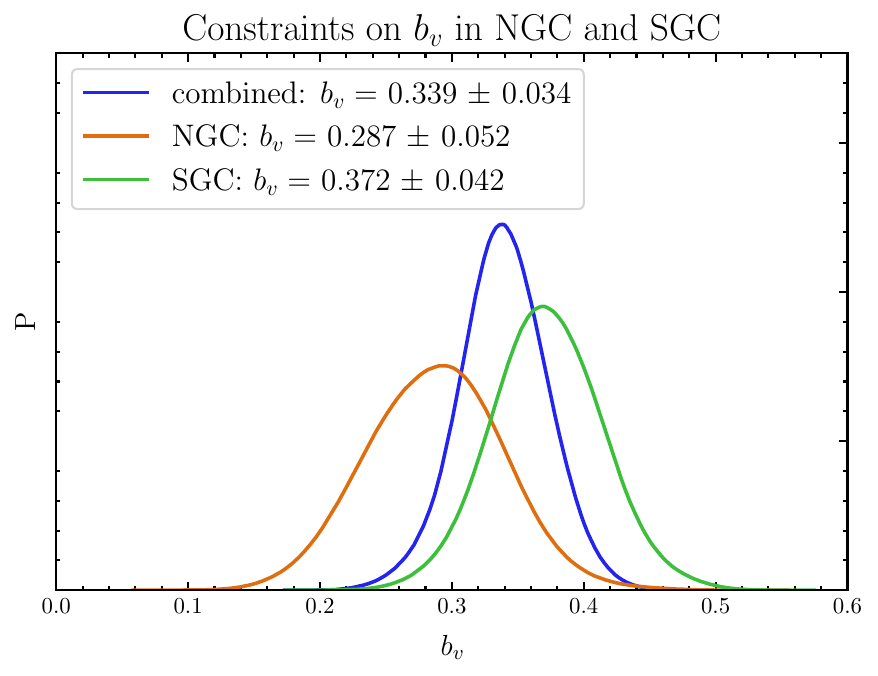}

\caption{ The velocity bias $b_v$ inferred on NGC, SGC, and their combination. The NGC and SGC results are consistent. Our combined measurement is $b_v=0.339\pm0.034$.}\label{fig:bvpercap}

\end{figure}

Interestingly, we find that this offset is significantly mitigated by the masking of tSZ clusters. We demonstrate this in Fig.~\ref{fig:auto_tsz} for the $N=4$ case in Appendix~\ref{sec:clustermasking_auto} for the $N=16$ case. We also show in Fig.~\ref{fig:auto_tsz} that deprojecting the tSZ also removes the offset between analytical and measured power, although the overall reconstruction power (both analytical and measured) is $\sim 5\times$ higher than for the unconstrained NILC. As the tSZ clusters are highly non-Gaussian and also correlated with the temperature, this is consistent with the offset being caused by a term of the form $\left<FG FG \delta\delta\right>$. 
It would be of interest to study this contribution on simulations or analytically; such a study is beyond the scope of this paper.

Going forward, all spectra that we show will be with the tSZ clusters masked. All of our results are derived from this case.

 We defer to future work a more careful analysis of these the auto power spectra of $v^{\mathrm{kSZ}}$, although we  show results where we include the datapoints in our Gaussian likelihood in Section~\ref{sec:auto}.

\subsection{Cross power spectra of the kSZ and continuity-equation velocity}\label{sec:crosspower}

\subsubsection{Measurements of $C_L^{v^{\mathrm{kSZ}}v^{\mathrm{cont}}}$}

We show the measured cross-correlation between the kSZ-reconstructed velocity and the continuity-equation-reconstructed velocity, and the best-fit model, in 
Fig.~\ref{fig:bestfit16bins} for the diagonals of the 16-bin case. These are our baseline datapoints. We perform tests for foreground contamination by using different signal maps in Section~\ref{sec:fg_contamination}; we find no evidence for foreground contamination in this signal. As the signal-to-noise is spread over 16 redshift bins, it is difficult to assess by eye the signal to noise.  To visually demonstrate this, we co-add our datapoints into an estimate of $C_L^{v_1v_1}$ as in~\cite{2025JCAP...05..057M}. We show this co-added plot in Fig.~\ref{fig:coadd}.

We fit two models to the baseline dataset: an $N+1$-dimensional model where we have one $b_v$ parameter and $N$ galaxy bias parameters, and a $2N$-dimensional model where we have $N$ galaxy bias parameters and $N$ velocity bias parameters $b_v^\alpha$. We summarize the constraints, signal-to-noise ratios, and goodness of fit  in Table~\ref{tab:compare_sgcngc}, on both SGC and NGC and for their combination. The SNR is calculated as the mean $b_v$ from the one-$b_v$ model divided by its errorbar. We find good fits with acceptable PTEs (note that we neglect the galaxy biases in our degrees of freedom counting, but we also do not count the contribution of the galaxy bias priors to the $\chi^2$) when calculating the PTE. We show the posteriors on $b_v$ for each case in Fig.~\ref{fig:bvpercap}. We prefer the multiple-$b_v^\alpha$ model over the single-$b_v$ model with a $\Delta\chi^2$ of 27 for 15 more parameters; this has a PTE 0.04\%, and indicates a $\sim2\sigma$ preference for the several-$b_v^\alpha$ model.

\begin{figure}
\includegraphics[width=\columnwidth]{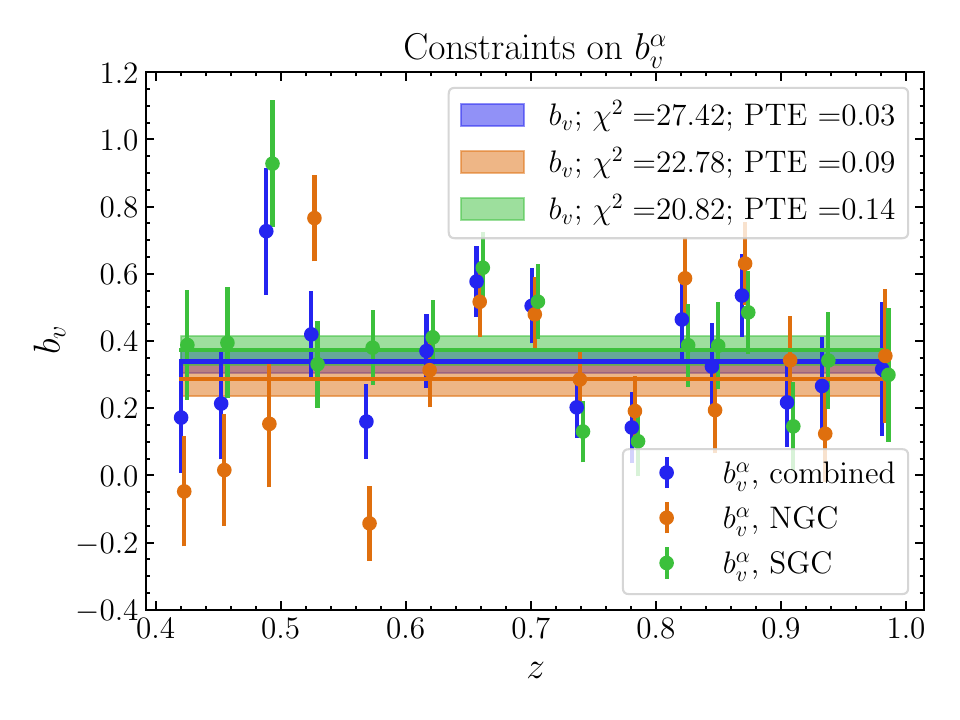}

\caption{The $b_v^\alpha$  measured in each redshift bin  as compared to the one-$b_v$ model where all are constrained to be the same. We compare the separate $b_v^\alpha$ points to the single $b_v$ point and quantify their difference by converting to a $\chi^2$ using the full covariance matrix from the chains. We convert this to a PTE assuming 16-1 degrees of freedom, and find a PTE of 0.03, indicating that the varying $b_v^\alpha$ model is $\sim2\sigma$ consistent with the single-$b_v$ model. }\label{fig:bvevolution}

\end{figure}

The  individually recovered $b_v^\alpha$ are shown in Fig.~\ref{fig:bvevolution}. We compare both models by defining a $\chi^2$ according to
\begin{align}
\chi^2 = \left(b_v^\alpha - b_v\right) C_{b_v^\alpha}^{-1} \left(b_v^\alpha - b_v\right)
\end{align}
where $C_{b_v^\alpha}$ is the full covariance matrix of the $b_v^\alpha$ parameters which we obtain from the MCMC chains. We convert this to a PTE assuming 15 degrees of freedom, and find  $\chi^2$s  with slightly low PTEs, similar to the PTE of the $\Delta\chi^2$ between the two model, again demonstrating the $2\sigma$ preference for the $b_v^\alpha$ model.

\begin{table*}
\begin{tabular}{|c|c|c|c|c|c|c|c|c|}
\hline
spec & constraint (combined) & SNR & $N_{\mathrm{points}}$ & $\chi_{\mathrm{min}}^2$ ($b_v^\alpha$)  & PTE ($b_v^\alpha$)& $\chi_{\mathrm{min}}^2$ ($b_v$)& PTE ($b_v$)  \\\hline\hline

$C_L^{v^{\mathrm{kSZ}}v^{\mathrm{cont}}}$(diagonal) (NGC) & 0.287 $\pm$ 0.052& 5.53 & 64 & 55.49 & 0.21 & 78.45 & 0.09  \\
$C_L^{v^{\mathrm{kSZ}}v^{\mathrm{cont}}}$(diagonal) (SGC) &  0.372 $\pm$ 0.042& 8.78 & 64 & 55.6 & 0.21 & 76.38 & 0.12  \\
$C_L^{v^{\mathrm{kSZ}}v^{\mathrm{cont}}}$(diagonal) (joint) & 0.339 $\pm$ 0.034& 10.04 & 128 & 129.25 & 0.13 & 156.34 & 0.04  \\\hline

\end{tabular}
\caption{The constraints and goodness-of-fit for the baseline NGC, SGC, and combined case. We use 16 redshift bins and only use the diagonals $C_L^{v^\alpha v^\beta}$ for $\alpha=\beta$. For the $b^\alpha_v$ model, where the velocity biases are constrained independently in each redshift bin, we find very good and consistent fits in SGC and NGC and so we combine them, and also find a good joint fit. The single-parameter model is a slightly worse fit in all cases.  }\label{tab:compare_sgcngc}
\end{table*}

\begin{figure}
    \centering
    \includegraphics[width=0.49\textwidth]{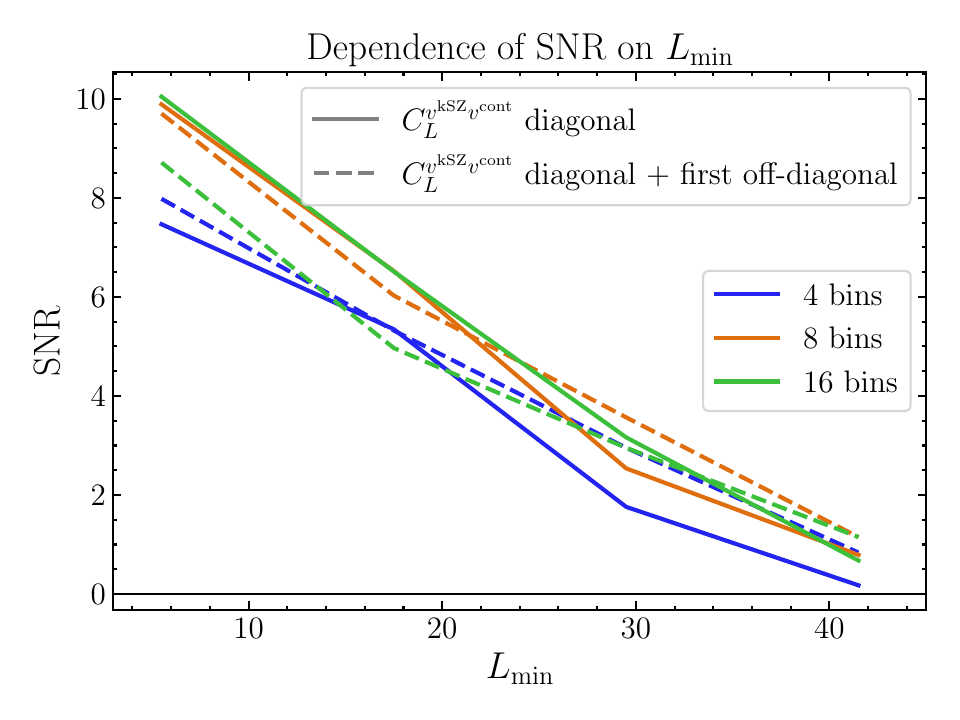}

    \caption{The dependence of the total SNR on the minimum $L$ used in the analysis, for various data combinations. We have used $L_{\mathrm{max}}=53$; using a larger $L_{\mathrm{max}}$ does not change the results.}
    \label{fig:snrminell}
\end{figure}

\begin{figure}
    \centering
    \includegraphics[width=0.49\textwidth]{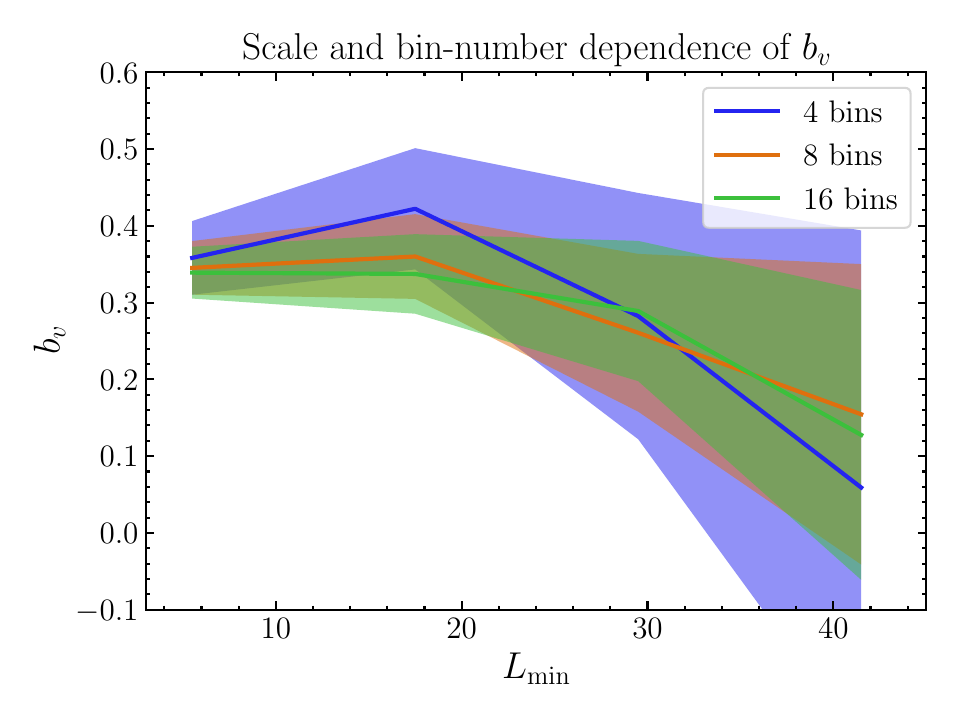}

    \caption{Dependence of the constraint on the minimum $L$ included and number of bins. In general, we see a consistent inference on $b_v$ across scales and number of bins, with most of the SNR coming from large scales. The shaded regions indicate the $1\sigma$ uncertainty.}
    \label{fig:aplotminell}
\end{figure}

\subsubsection{$L$-dependence of the signal and signal-to-noise}

We  show in Figure~\ref{fig:snrminell} the dependence of the overall signal-to-noise on the choice of minimum $L$ in the analysis, for different combinations of the datasets. The statistical significance drops to below $2\sigma$ when $L_{\mathrm{min}}\sim40$. We also show in Figure~\ref{fig:aplotminell} how the constraint changes when we go to larger $L_{\mathrm{min}}$, and to smaller number of redshift bins $N$; in general we find a consistent measurement.

\subsection{Constraints on scale-dependent bias and non-Gaussianity}\label{sec:scale_dep}

\begin{figure*}
\includegraphics[width=0.4\textwidth]
{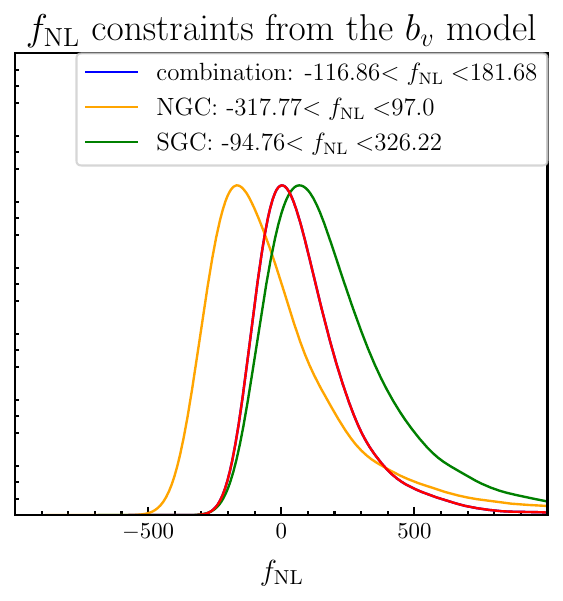}
\includegraphics[width=0.4\textwidth]{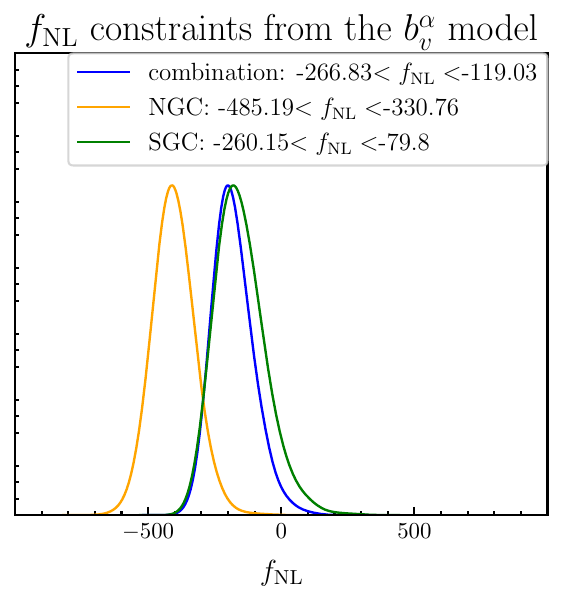}
\caption{Constraints on $f_{\mathrm{NL}}$, from the $b_v^\alpha$ model (left) and the $b_v$ model (right). The constraints quoted in the legend are 67\% confidence intervals. See the text for a discussion of the low ``NGC'' posterior on the RHS.}
\label{fig:fnlposts}
\end{figure*}

For our baseline dataset, we free up the $f_{\mathrm{NL}}$ parameter using the model described in Section~\ref{sec:fnltheory}. {Note that this part of the analysis assumes that the transfer function for the $f_{\mathrm{NL}}$ response is the same as for the standard velocity. In future work, we will test and refine this assumption using simulations including non-Gaussianity.} We show our ${f_{\mathrm{NL}}}$ posteriors in Fig.~\ref{fig:fnlposts}.

We note that in this case the $b_v^\alpha$ and $b_v$ models---where the velocity bias is varied independently in each redshift bin and where it is fixed to be equal---are conceptually different models, as the $b_v$ model allows for all of the signal to simultaneously approach 0, where the constraining power on $f_{\mathrm{NL}}$ is 0. In contrast, the $b_v^{\alpha}$ model breaks this coupling of the parameters, as the amplitudes are no longer inter-correlated. We illustrate this by showing both models in the corner plot in Fig.~\ref{fig:fnlbalpha}, where the degeneracy-breaking can be clearly seen. The full corner plot is shown in Appendix~\ref{sec:fnlcorner}.  The marginalized $f_{\mathrm{NL}}$ constraints are isolated in Fig.~\ref{fig:fnlposts}.

This behaviour is due to the fact  that the non-linearity of the model results in the sensitivity of a given redshift bin to $f_{\mathrm{NL}}$ depending on the inferred $b _v^\alpha$. We discuss this further in Appendix~\ref{sec:fnlcorner}.  This is not a fundamental problem of the analysis: in a higher signal-to-noise-ratio, $b_v$ (or $b_v^\alpha$) will be measured by the smaller-scale measurement and this degeneracy will be broken.

\begin{figure}
\includegraphics[width=\columnwidth]{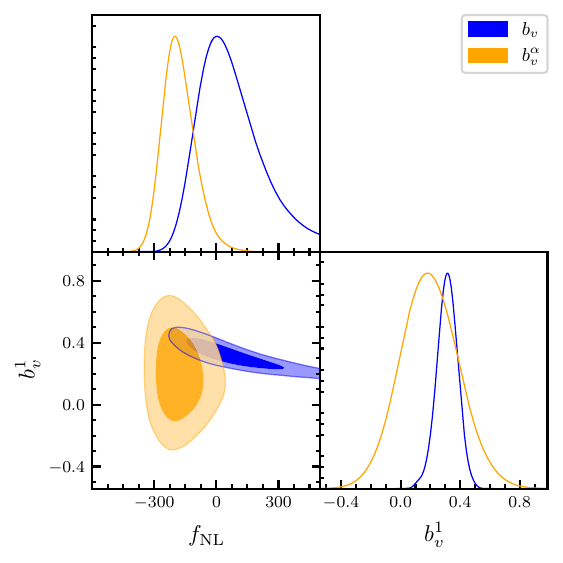}
\caption{Constraints on $f_{\mathrm{NL}}$ and the $b_v^\alpha$ models, for the case where the velocity bias is varied independently in each redshift bin (orange contours) as well as when it is fixed in each redshift bin (blue contours). We show only a subset of the full corner plot, as it is all that is necessary for illustration purposes. The degeneracy with $f_{\mathrm{NL}}$ is clearly broken in the $b_v^\alpha$ case.  We show the full corner plot in Appendix.~\ref{sec:fnlcorner}.}\label{fig:fnlbalpha}
\end{figure}

\begin{figure}
\includegraphics[width=\columnwidth]{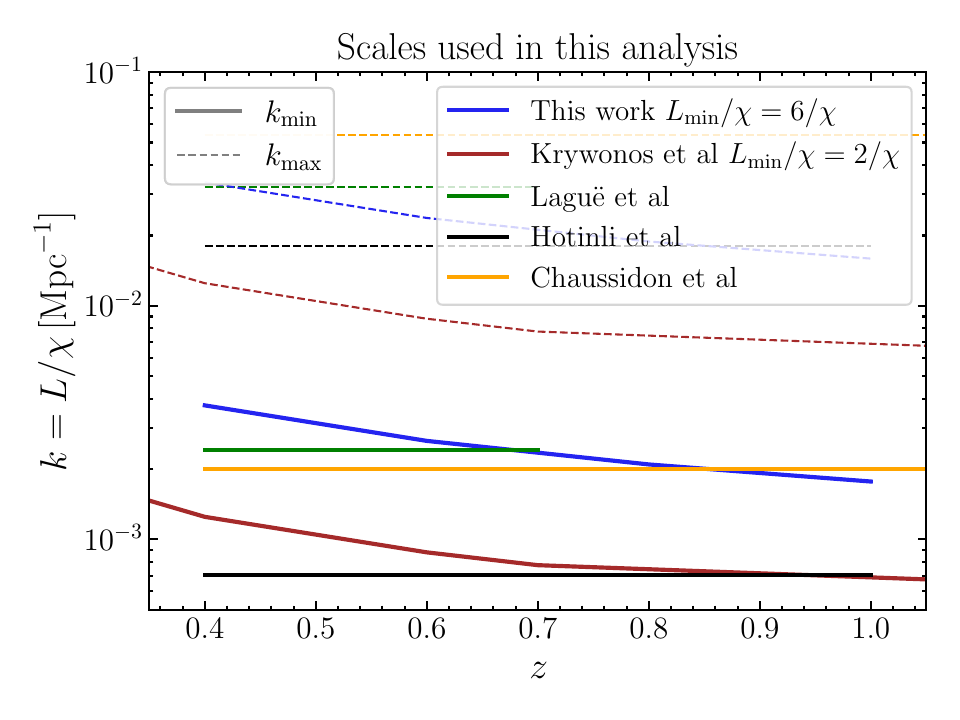}
\caption{Scales used in this analysis. We show an approximate $k$-scale calculated by $L/\chi$, and we indicate the scale cuts used by other relevant recent works for comparison---in particular, of three works which have constrained $f_{\mathrm{NL}}$ with the kSZ velocity cross-correlation~\citep{2024arXiv240805264K,Lague:2024czc,2025arXiv250621657H}, and of the three-dimensional  $f_{\mathrm{NL}}$ of the same galaxy sample\protect{~\citep{2025JCAP...06..029C}}. 
}\label{fig:kmin}
\end{figure}

\begin{figure*}
\includegraphics[width=0.4\textwidth]{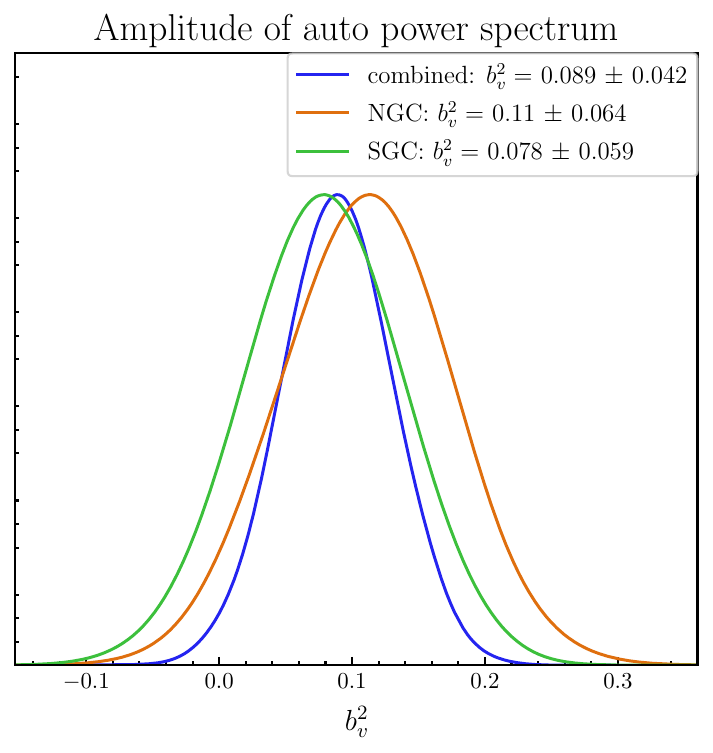}
\includegraphics[width=0.4\textwidth]{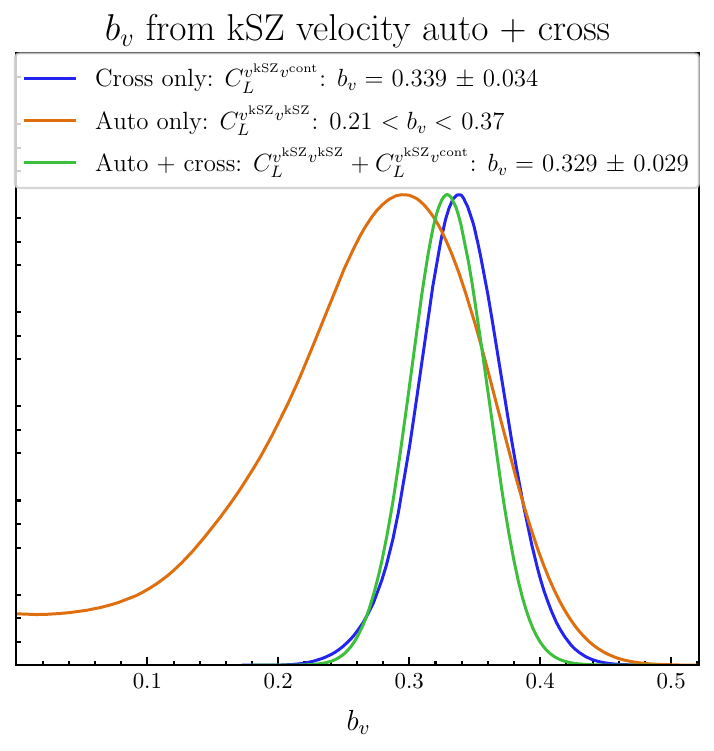}
\caption{\textit{Left}: The constraints on the amplitude of the kSZ velocity auto power spectrum, $b_v^2$. We measure this at $2.1\sigma$ to be $b_v^2=0.089\pm0.042$. \textit{Right:} The constraints on $b_v$ from the kSZ velocity auto power spectrum, its cross with the continuity-equation velocity, and their combination. Note that on the left hand side the prior is linear in the amplitude of the signal $b_v^2$ (for the purposes of assessing detection significance), while on the right hand side we impose a physical prior $b_v>0$ which is linear in $b_v$. We measure the auto power spectrum at $2\sigma$ significance. It is consistent with the cross power-spectrum, and the combined constraint is significant at $11\sigma$.}\label{fig:combwithauto}
\end{figure*}

We show the constraints on $f_{\mathrm{NL}}$ in Fig.~\ref{fig:fnlposts}. For these purposes, we have used priors on $b_v$ and $b_v^\alpha$ that allow them to be negative, but we have checked that imposing positivity priors does not significanctly affect the $f_{\mathrm{NL}}$ posteriors.  We use the constraint obtained from the $b_v^\alpha$ model as our baseline result: we find $f_{\mathrm{NL}}=-180^{+61}_{-86}$ at 67\% confidence, with $f_{\mathrm{NL}}$ consistent with zero at $2\sigma$. 

{The introduction of many poorly measured parameters has the risk of introducing  spurious prior volume effects, and so we have performed some parameter-recovery tests of our pipeline. In particular, on the replacement of our datavector with a theory vector with $f_{\mathrm{NL}}=0$ and $b_v=b_v^\alpha=0.3$, we found that for the $b_v$ model our posteriors for NGC, SGC, and the joint case are all consistent with the truth; however, for the $b_v^\alpha$  this is only the case for the joint NGC+SGC case, with the NGC-alone case showing a significant negative bias (as we see in the data). However, in we use the joint case which as our baseline result, we believe that these prior volume effects are mitigated, as in this case $b_v^\alpha$ is well-enough constrained in this case.}

The scales used in this analysis are indicated in Fig.~\ref{fig:kmin}. Note that we have used the simple Limber conversion to convert from $L,\chi$ to $k$; on the large scales we use we go beyond Limber and so this is not accurate, but gives a guideline. We defer an in-depth study of the scales from which we get information to future work. However, in this simple comparison, it is possible that our $L_{\mathrm{min}}$ prevents us from reaching the extremely large scales used in~\cite{2025arXiv250621657H}, and is a possible explanation for why we get less tight constraints than that work on $f_{\mathrm{NL}}$ with a similar signal to noise.

\subsection{Inclusion of the auto power spectrum}\label{sec:auto}

The dominant contribution to the auto power spectra is the $N^{(0)}_L$ bias. We have an analytic expectation for this, which is broadly in line with what we see at high-$L$ in the data, but a careful analysis would require a full realisation-dependent $N^{(0)}_L$ subtraction~\citep{2013MNRAS.431..609N}, as is often done in CMB lensing power spectrum analyses~\citep{2020A&A...641A...8P,2024ApJ...962..112Q}. We defer this to future work, but for now we note that if we model $N^{(0)}_L$ as scale-independent we can include the auto-points in our likelihood with the following model:
\begin{align}
C_L^{v^{\mathrm{kSZ}}_\alpha v^{\mathrm{kSZ}}_\alpha} = (b_v^\alpha{})^2 C_L^{v_\alpha v_\alpha} + a^\alpha N_L^{(0),\alpha},
\end{align}
where $a^\alpha$ is a scale-independent correction to the noise that we fit and $C_L^{v_\alpha v_\alpha}$ is the $\Lambda$CDM theory prediction. The model is fit separately in each redshift bin, and in NGC and SGC (with $a^\alpha$ different in NGC and SGC).

For the purposes of assessing detection significance, we first constrain an amplitude parameter, $(b_v^\alpha)^2$, using a prior that is linear in  $(b_v^\alpha)^2$ and allows it to be negative. When we do this, we see a hint of a signal (at 2.1$\sigma$) in the auto alone, with $b_v^2=0.089\pm0.042$, which we show on the left hand side of Figure~\ref{fig:combwithauto}. 

We also use this to place a constraint on $b_v$, with a linear prior on $b_v$ and imposing $b_v>0$; we find $0.25<b_v<0.37$ at 67\% confidence. {We get a good fit in the auto-alone case, with a best-fit $\chi^2$ of 127 for 192 datapoints and 33 parameters (one $b_v$ parameter and $16\times2$ $a^\alpha$ parameters), corresponding to a PTE of 97\%.} Previous work~\citep{2024arXiv240500809B} constrained the model above using Planck and unWISE data; note that the numerical bounds on $b_v$ from that work cannot be directly compared to the results here due to different choices of $C_\ell^{g\tau}$ model (see discussion in Section~\ref{sec:optbias}).  

\begin{figure}
\includegraphics[width=\columnwidth]{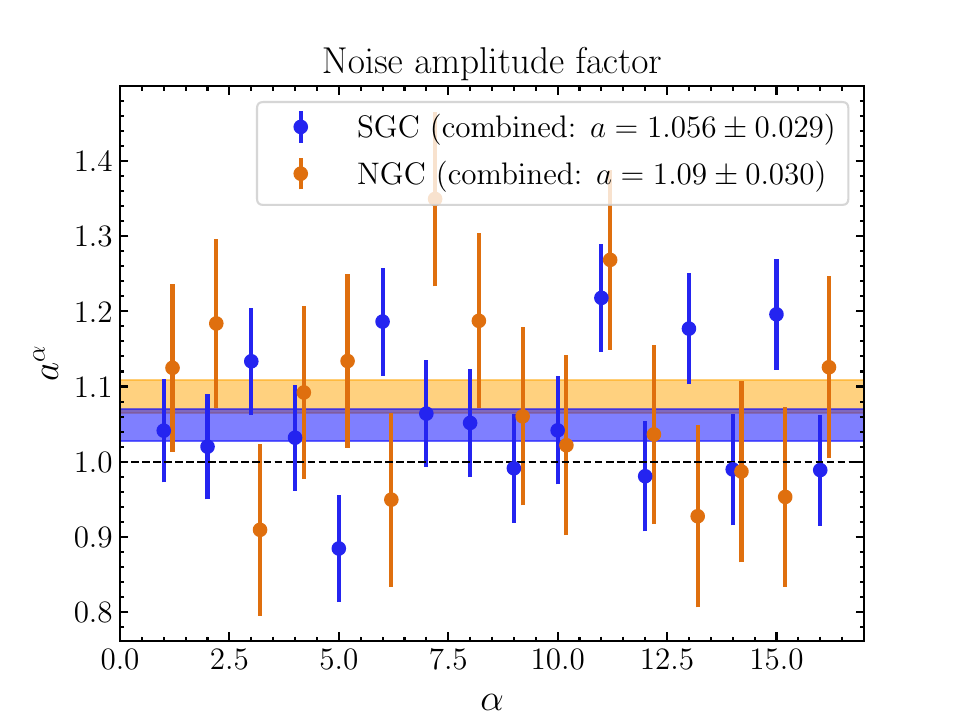}
\caption{The recovered noise amplitudes when the auto power spectrum is fit to a constant noise model. These are multiplicative amplitudes with respect to the analytical noise prediction.  
}\label{fig:noiseamp}
\end{figure}

 The auto-alone case is consistent with the cross-only constraint and so we can combine them; this gives $b_v=0.33\pm0.03$ (where we again we use a linear prior in $b_v$ as the cross spectra are sufficiently constraining), increasing our signal-to-noise ratio to $11$. 
 These are summarized in Fig.~\ref{fig:combwithauto}. The PTE for the combination is 0.04 (with a best-fit of $\chi^2$ 245.1 for 256 datapoints and 48 parameters). We show the recovered noise amplitude factors in Fig.~\ref{fig:noiseamp}.
 
 Note that in our covariance matrix we assume that $a^\alpha=1$ everywhere. As the combined $a$s are closer to $a^\alpha=1.06$ on SGC and $a^\alpha=1.1$ on NGC. As our SNR is dominated by SGC, and the errorbars on the cross-spectrum are proportional to the square root of this number, we expect the effect on our reported SNR to be minor.

\section{Tests for foreground contamination}\label{sec:fg_contamination}

\begin{figure}
\centering
\includegraphics[width=0.8\columnwidth]{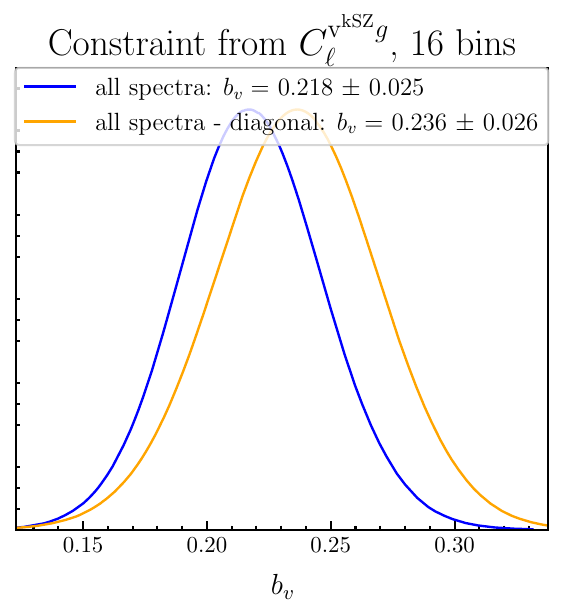}

    \caption{The shift in the $b_v$ measurement from $C_L^{v^{\mathrm{kSZ}}g}$ when the diagonals are dropped. This is an inconsistent shift (at the $2.2\sigma$ level) for a nested dataset, as quantified by the test derived in~\protect\cite{2020MNRAS.499.3410G}. }\label{fig:dropincludeautos}
\end{figure}

On small scales, the millimeter sky (on which we measure the CMB) is dominated by non-kSZ foregrounds, such as the  tSZ effect and the CIB. These are correlated with large scale structure. In this section we explore whether our datapoints are contaminated by a foreground bias. We do this by exploring the dependence of the datapoints on the choice of CMB+kSZ map: we compare the standard minimum-variance (MV) NILC map with a map with the tSZ explicitly removed by $y$-deprojection. Note that this map will have no tSZ foregrounds but a different amount of any other foregrounds; thus we don't use it as a ``foreground-cleaned'' map but instead as a ``differently-cleaned'' map. A foreground-immune signal will be consistent when measured with each map. We also use a foreground-only map (a blackbody-deprojected $y$ map) as a ``foreground-only'' map on which we expect a null result. 

In addition to performing contamination tests of our signal, we perform similar tests on an alternative signal - $C_L^{v^{\mathrm{kSZ}g}}$, the cross-correlation measured between the kSZ-estimated velocity and the galaxies themselves, as was measured in~\cite{2025arXiv250621684L}.

The velocity-velocity correlation $C_L^{v_\alpha v_\beta}$ concentrates most of the signal-to-noise in the equal-redshift diagonal ($\alpha=\beta$). In contrast, due to the non-local nature of the $v-g$ correlation, the velocity-galaxy correlation $C_L^{v_\alpha g_\beta}$ concentrates most of the signal-to-noise in the off-diagonals ($\alpha\ne\beta$) (see Section~\ref{sec:redshiftstructure} for a discussion), and very little on the diagonal; however, in practice we measure a large signal  on the diagonal.

\begin{figure}
\centering
\includegraphics[width=\columnwidth]{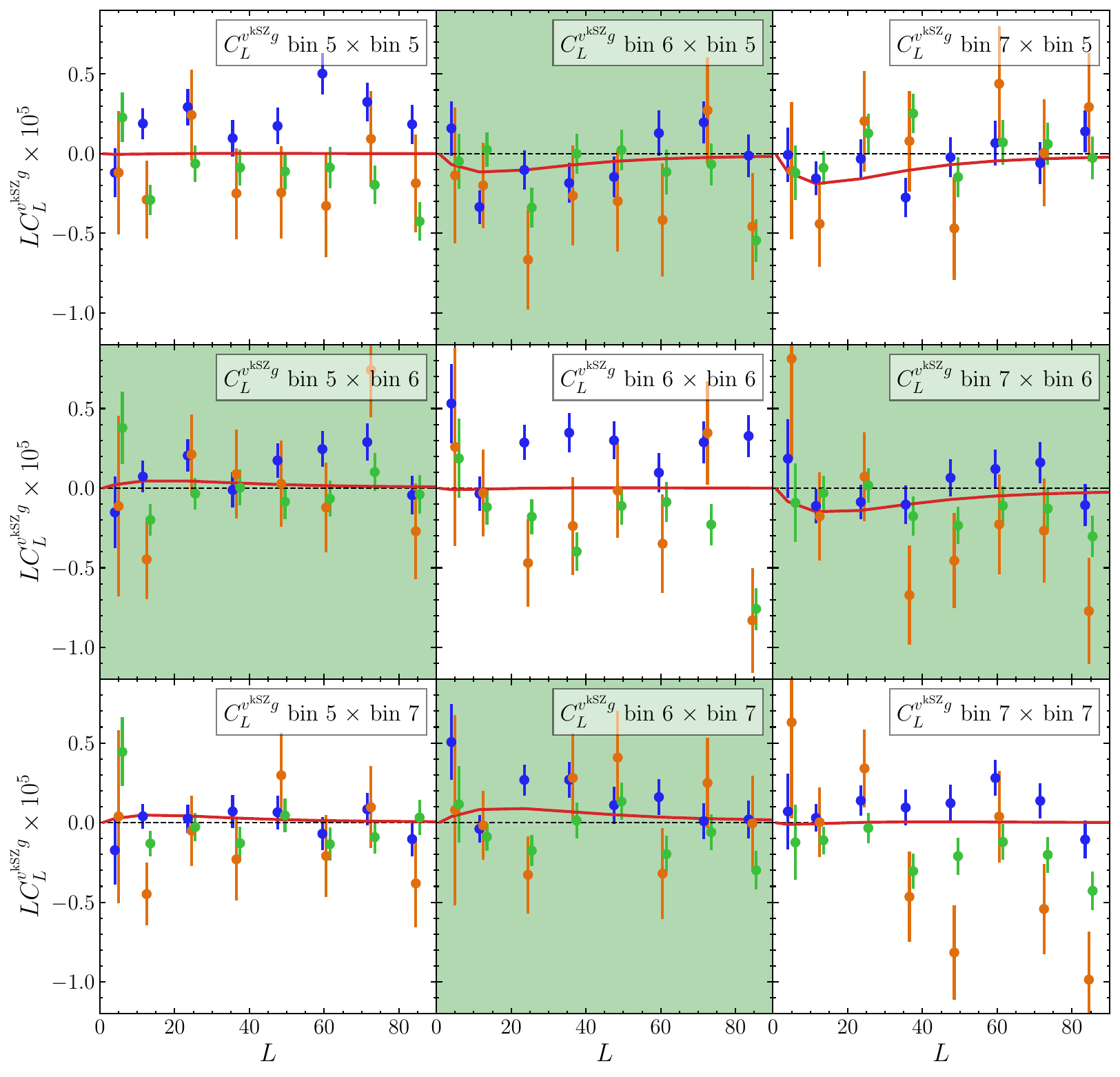}\\
\includegraphics[width=0.3\textwidth]{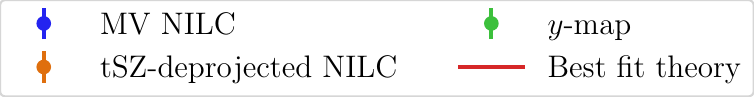}
\caption{ The  data points measured from reconstruction performed on an MV NILC (blue) versus an SZ-deprojected NILC (orange), for a subset of $C_L^{v^{\mathrm{kSZ}}g}$. We have co-added the SGC and NGC datapoints. A general observation is that these sets of points appear consistent except for a possible bias on the diagonal of $C_L^{v^{\mathrm{kSZ}}g}$.   We also show the points measured with a \textit{foreground-only} map (``$y$-map'', in green). In general these are consistent with null  on the off-diagonals, but not on the diagonal.  Note that we have highlighted the first off-diagonal (where most of the signal-to-noise is concentrated) in green to guide the eye. As argued in the text, this is evidence for foreground bias on the diagonals of the $C_L^{v^{\mathrm{kSZ}}g}$ measurement.}\label{fig:compare_ydeproj_4bins_vg}
\end{figure}

We show a subset of the $C_L^{v^{\mathrm{kSZ}}_\alpha g_\beta}$ datapoints in Fig.~\ref{fig:compare_ydeproj_4bins_vg}. Even by inspection by eye, we notice a large excess on the $\alpha=\beta$ diagonals of $C_L^{v^{\mathrm{kSZ}}g}$ (where the signal is expected to be close to zero), on almost all scales. This is \textit{not} present in $C_L^{v^{\mathrm{kSZ}}v^{\mathrm{cont}}}$. After performing tests with differently cleaned maps, we attribute this to foreground contamination of the form $\left< \mathrm{FG}^S \delta^S \delta^L\right>$, where $\mathrm{FG}^S$ indicates a small-scale foreground in the CMB+kSZ map. We note that contamination with this behaviour – affecting  $C_L^{v^{\mathrm{kSZ}}g}$ but not  $C_L^{v^{\mathrm{kSZ}}v^{\mathrm{cont}}}$ at equal redshifts – would arise if foreground contamination to the kSZ velocity reconstruction depended directly on the large-scale density field (this could result from a small-scale non-Gaussian CMB foreground coupling with the small-scale galaxies within the quadratic estimator). {Contamination of this form would not correlate with the continuity-equation velocity on the equal-redshift bins due to the suppressed density-velocity signal at equal redshift.}

\begin{figure}
\centering
\includegraphics[width=\columnwidth]{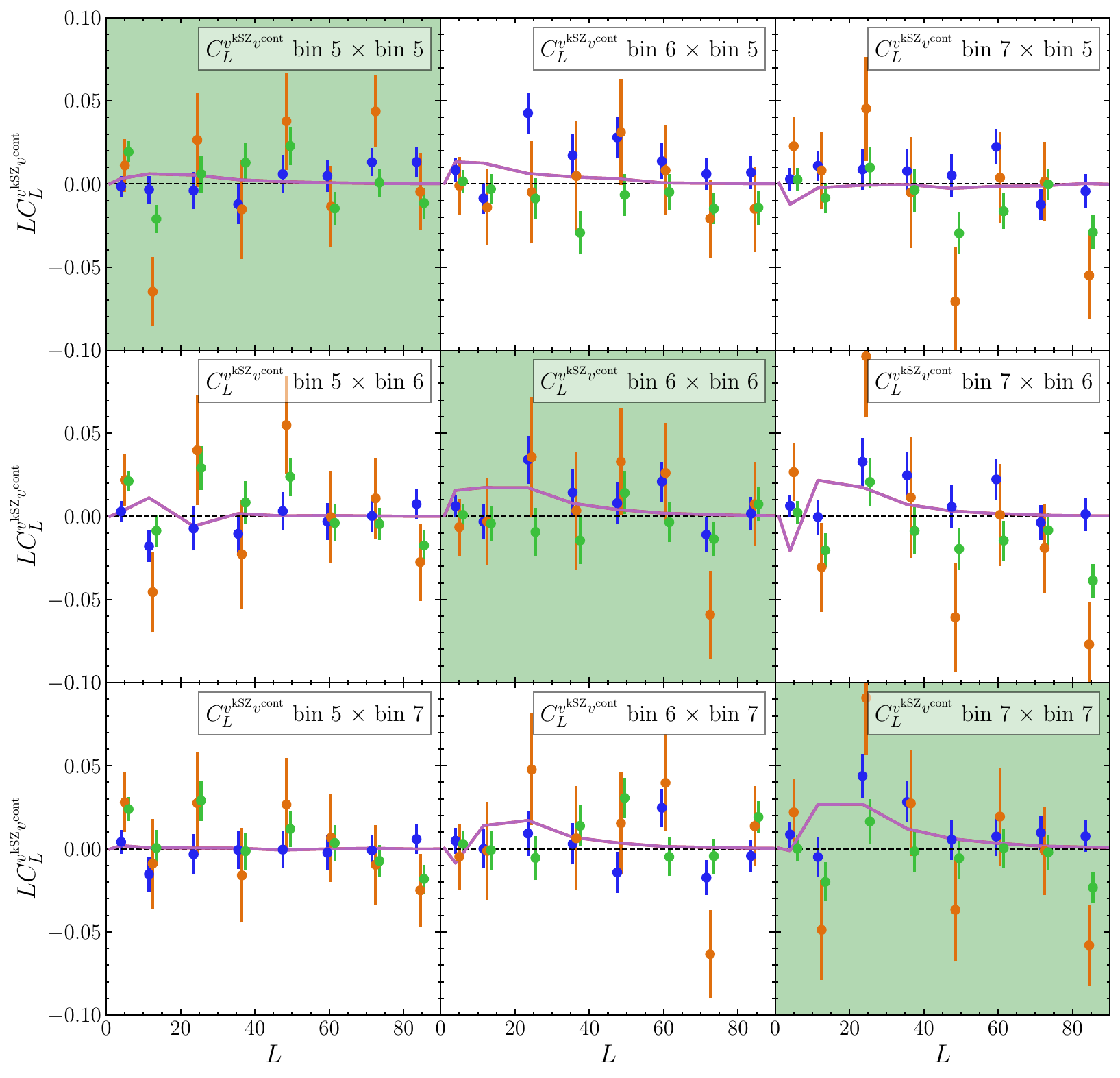}
\includegraphics[width=0.3\textwidth]{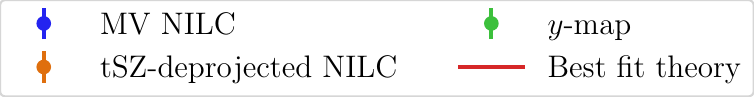}
\caption{The NGC data points measured from reconstruction performed on an MV NILC (blue) versus an SZ-deprojected NILC (orange), for a subset of $C_L^{v^{\mathrm{kSZ}}v^{\mathrm{cont}}}$. A general observation is that these sets of points appear consistent with each other. We also show the points measured with a \textit{foreground-only} map (``$y$-map'', in green).  In general these are consistent with null.  Note that we have highlighted the diagonal (where most of the signal-to-noise is concentrated) in green to guide the eye.  This is consistent with no contribution from foreground bias on the diagonals of the $C_L^{v^{\mathrm{kSZ}}v^{\mathrm{cont}}}$ measurement, which we use as our baseline dataset. }\label{fig:compare_ydeproj_4bins_vv}
\end{figure}

We perform some tests of the signal in order to check this hypothesis. In particular,  we show a subset of $C_L^{v^{\mathrm{kSZ}}g}$ in Fig.~\ref{fig:compare_ydeproj_4bins_vg} and a subset of $C_L^{v^{\mathrm{kSZ}}v^{\mathrm{cont}}}$ in Fig.~\ref{fig:compare_ydeproj_4bins_vv}, including the some off-diagonals, when we use the MV NILC and the SZ-deprojected NILC; we also show some foreground-only (null-signal) points, where the reconstruction was performed with the $y$-map replacing the temperature map. For the $y$-map case, we use the same filters in the velocity reconstruction as for the MV map, but for the SZ-deprojected map we replace the $C_L^{TT}$ filter with the measured power of the SZ-deprojected map. Thus, the measured power spectrum of the SZ-deprojected map is well aproximated by the analytical $N^{(0)}_L$ calculated with the new filters, but the $y$-map signal is not, due to a mismatch in estimated and true power. We rescale the outputs by a constant factor such that the observed power matches the analytical $N^{(0)}_L$.\footnote{We have also performed this test for the case where we recalculate the filters to match the observed power spectra of the $y$-map and use the analytical $N^{(0)}_L$  in the covariance calculation. Conclusions are qualitatively similar.}

For the MV NILC map and the SZ-deprojected map, we calculate the covariance as we do for the main analysis.  We use the mask-decoupled theoretical covariance matrix assuming the appropriate analytical estimate of $N_L^{(0)}$ for the auto power spectra of the kSZ reconstructions, and the measured power for the auto power spectra of the large-scale galaxies or continuity-equation velocity. For the $y$-map case, we use the $N_L^{(0)}$ of the MV-reconstruction case, remembering that we have rescaled the outputs such that their power matches this.

 In general, it is hard to make strict conclusions by eye as the SZ-deprojected points are noisier. However, there is clear evidence for a systematic inconsistency between the MV NILC and SZ-deprojected NILC points on the diagonal of $C_L^{{v^{\mathrm{kSZ}}g}}$ and not elsewhere. Additionally, the ``$y$-map'' points are not null here but are null elsewhere.

To quantify this and further test for foreground contamination, we explore constraints on an overall velocity bias parameter in different regimes in Appendix~\ref{sec:biasconstraints}. Here, we simply show the biggest shifts in the constraints: the final constraints from $C_\ell^{v^{\mathrm{kSZ}}g}$ shift by $\sim1\sigma$ when the diagonal is dropped, as is shown in Figure~\ref{fig:dropincludeautos}. ~\cite{2020MNRAS.499.3410G} devised a useful test to check whether the shift in a parameter when analyzed with a subset of a larger dataset as compared to the full dataset is within the expected fluctuation range: in particular, we expect fluctuations of order $\sigma_\Delta^2\sim\sigma_a^2-\sigma_b^2$. By this metric, there is a significant shift in the inferred $b_v$ when going from the full dataset to one with no diagonal; the shift is a $2.2\sigma$ anomaly. Given this information and our knowledge of the shifts or consistency in datapoints and $b_v$ inference for differently-cleaned maps and our tests with foreground-only maps, we conclude that at the current level of sensitivity the diagonal of $C_L^{vg}$ is foreground-dominated. 

{We propose reducing the foreground contribution to $C_L^{vg}$ by forming the estimator $C_L^{v_\alpha g_\beta}-C_L^{v_\beta g_\alpha}$, which could be  more immune to foreground bias. We will explore this in future work.}

\section{Conclusion}\label{sec:conclusion}

We have measured the velocity field with the small-scale kSZ signal in ACT through its anisotropic cross-correlation with the DESI LRG galaxy sample. We have measured a $10\sigma$ cross-correlation signal with an alternative velocity field estimate from the continuity equation, and found a good fit to the data for our model. We find a low velocity bias with respect to our theory, which was created to match the Illustris-TNG simulations and a HOD prescription, indicating at high significance that this theory is an incomplete model of the true astrophysics. This is consistent with recent work which used a similar data combination to directly constrain this model~\citep{2024arXiv240707152H}, and other recent works which constrained a slightly different model~\citep{2025arXiv250621684L,2025arXiv250621657H}.

We perform several consistency tests of our measurement, including making a measurement with randomly shuffled galaxies and confirming consistency with zero; and various frequency-dependent tests. We see no evidence for frequency-dependent foregrounds in our $C_L^{v^{\mathrm{kSZ}}v^{\mathrm{cont}}}$ statistic. This stands in contrast to what we see in a $C_L^{v^{\mathrm{kSZ}}g}$ measurement that we perform for comparison, where we find evidence for some contamination { on the equal-redshift spectra}. However, we note that these spectra contribute only a very small amount to the signal-to-noise that only contribute a small amount to the total signal-to-noise.

The signal that we measure is sensitive to scale-dependent bias, and so we constrain the local primordial non-Gaussianity parameter $f_{\mathrm{NL}}$ to be $f_{\mathrm{NL}}=-180^{+61}_{-86}$ at 67\% confidence, and consistent with zero at $2\sigma$. Constraints on primordial non-Gaussianity are a main goal of the kSZ velocity reconstruction programme, and it would be of interest to compare the information in the 2-dimensional reconstruction performed here to that in a recent complementary 3-dimensional reconstruction~\cite{2025arXiv250621657H}, which found similar signal-to-noise on a smaller sky area and yet significantly tighter constraints on $f_{\mathrm{NL}}$. We have suggested that this may be due to our probing of different maximum scales in our analysis. An analytic study of the scales most important for $f_{\mathrm{NL}}$ constraints would be of interest but beyond the scope of this paper.

There are several ways in which our analysis was suboptimal, and it would be of interest to improve it. First, by performing a quadratic-maximum-likelihood analysis following~\cite{2025arXiv250621684L,2025arXiv251005215K}; second, by more optimal weighting of the galaxies by their redshift error following~\cite{2025arXiv250621657H}. 

We additionally found a {2.1$\sigma$} hint of the auto power spectrum signal in $C_L^{v^{\mathrm{kSZ}}v^{\mathrm{kSZ}}}$, and it would be of interest to follow up on this with a more robust analysis. We note that foreground biases to this statistic may be a problem, which we have shown can be somewhat mitigated by removal of tSZ clusters but it would be well-motivated to investigate this further.

A key aspect of this paper is our demonstration of our pipeline on $N$-body simulations. The continuity-equation velocity reconstruction process required this for our theoretical model; in this respect, the $C_L^{v^{\mathrm{kSZ}}g }$ signal, which can be modelled cleanly through analytic calculations, is much more attractive than $C_L^{v^{\mathrm{kSZ}}v^\mathrm{cont} }$.  
The kSZ era of CMB measurements continues to develop, and it will be imperative to understand better and control systematics such as the foreground biases investigated here in future analyses.  Detailed studies for experiment-specific setups will be of high interest for current and upcoming surveys.

\section*{Acknowledgements}

We thank Edmond Chaussidon, Simone Ferraro, Selim Hotinli, and Kendrick Smith for useful discussions. Support for ACT was through the U.S.~National Science Foundation through awards AST-0408698, AST-0965625, and AST-1440226 for the ACT project, as well as awards PHY-0355328, PHY-0855887 and PHY-1214379. Funding was also provided by Princeton University, the University of Pennsylvania, and a Canada Foundation for Innovation (CFI) award to UBC. ACT operated in the Parque Astron\'omico Atacama in northern Chile under the auspices of the Agencia Nacional de Investigaci\'on y Desarrollo (ANID). The development of multichroic detectors and lenses was supported by NASA grants NNX13AE56G and NNX14AB58G. Detector research at NIST was supported by the NIST Innovations in Measurement Science program. Computing for ACT was performed using the Princeton Research Computing resources at Princeton University, the National Energy Research Scientific Computing Center (NERSC), and the Niagara supercomputer at the SciNet HPC Consortium. SciNet is funded by the CFI under the auspices of Compute Canada, the Government of Ontario, the Ontario Research Fund–Research Excellence, and the University of Toronto. We thank the Republic of Chile for hosting ACT in the northern Atacama, and the local indigenous Licanantay communities whom we follow in observing and learning from the night sky.

FMcC acknowledges support from the European Research Council (ERC) under the European Union's Horizon 2020 research and innovation programme (Grant agreement No.~851274). 
The Flatiron Institute is a division of the Simons Foundation.

This work made use of python, \texttt{numpy}\footnote{\url{https://numpy.org}}~\citep{harris2020array}, \texttt{matplotlib}\footnote{\url{https://matplotlib.org}}~\citep{Hunter:2007}, \texttt{scipy}\footnote{\url{https://scipy.org}}~\citep{2020SciPy-NMeth}, \texttt{astropy}\footnote{\url{https://www.astropy.org/index.html}}~\citep{2013A&A...558A..33A,2018AJ....156..123A,2022ApJ...935..167A}, HEALPix~\citep{2005ApJ...622..759G} and its python implementation \texttt{healpy}\footnote{\url{https://healpy.readthedocs.io/en/latest/}}~\citep{Zonca2019},  \texttt{pixell}\footnote{\url{https://pixell.readthedocs.io/en/latest/}}, \texttt{pymaster}\footnote{\url{https://namaster.readthedocs.io/en/latest/}}~\citep{2019MNRAS.484.4127A}, CAMB\footnote{\url{https://camb.info}}~\citep{2000ApJ...538..473L}, \texttt{class}\footnote{\url{http://class-code.net}}~\citep{2011arXiv1104.2932L,2011JCAP...07..034B}, \texttt{Cobaya}\footnote{\url{https://cobaya.readthedocs.io}}~\citep{2019ascl.soft10019T,2021JCAP...05..057T}, and \texttt{getDist}\footnote{\url{https://getdist.readthedocs.io/en/latest/}}~\citep{Lewis:2019xzd}.
%%%%%%%%%%%%%%%%%%%%%%%%%%%%%%%%%%%%%%%%%%%%%%%%%%


\bibliographystyle{act_titles}
\bibliography{references}

\begin{thebibliography}{110}
\expandafter\ifx\csname natexlab\endcsname\relax\def\natexlab#1{#1}\fi

\bibitem[{{Abitbol} et~al.(2025)}]{2025JCAP...08..034A}
{Abitbol}, M., et~al. 2025,
  \href{https://arxiv.org/abs/2503.00636}{arXiv:2503.00636},
  \href{https://dx.doi.org/10.1088/1475-7516/2025/08/034}{\jcap, 2025, 034,
  }{The Simons Observatory: science goals and forecasts for the enhanced Large
  Aperture Telescope}

\bibitem[{{ACTDESHSC Collaboration} et~al.(2025)}]{2025arXiv250721459A}
{ACTDESHSC Collaboration}, et~al. 2025,
  \href{https://arxiv.org/abs/2507.21459}{arXiv:2507.21459},
  \href{https://dx.doi.org/10.48550/arXiv.2507.21459}{arXiv e-prints,
  arXiv:2507.21459, }{The Atacama Cosmology Telescope: DR6 Sunyaev-Zel'dovich
  Selected Galaxy Clusters Catalog}

\bibitem[{{Ade} et~al.(2019)}]{2019JCAP...02..056A}
{Ade}, P., et~al. 2019,
  \href{https://arxiv.org/abs/1808.07445}{arXiv:1808.07445},
  \href{https://dx.doi.org/10.1088/1475-7516/2019/02/056}{\jcap, 2019, 056,
  }{The Simons Observatory: science goals and forecasts}

\bibitem[{{Alonso} et~al.(2019){Alonso}, {Sanchez}, {Slosar}, \& {LSST Dark
  Energy Science Collaboration}}]{2019MNRAS.484.4127A}
{Alonso}, D., {Sanchez}, J., {Slosar}, A., \& {LSST Dark Energy Science
  Collaboration}. 2019, \href{https://dx.doi.org/10.1093/mnras/stz093}{\mnras,
  484, 4127, }{A unified pseudo-C$_{{\ensuremath{\ell}}}$ framework}

\bibitem[{{Astropy Collaboration} et~al.(2013)}]{2013A&A...558A..33A}
{Astropy Collaboration}, et~al. 2013,
  \href{https://arxiv.org/abs/1307.6212}{arXiv:1307.6212},
  \href{https://dx.doi.org/10.1051/0004-6361/201322068}{\aap, 558, A33,
  }{Astropy: A community Python package for astronomy}

\bibitem[{{Astropy Collaboration} et~al.(2018)}]{2018AJ....156..123A}
---. 2018, \href{https://arxiv.org/abs/1801.02634}{arXiv:1801.02634},
  \href{https://dx.doi.org/10.3847/1538-3881/aabc4f}{\aj, 156, 123, }{The
  Astropy Project: Building an Open-science Project and Status of the v2.0 Core
  Package}

\bibitem[{{Astropy Collaboration} et~al.(2022)}]{2022ApJ...935..167A}
---. 2022, \href{https://arxiv.org/abs/2206.14220}{arXiv:2206.14220},
  \href{https://dx.doi.org/10.3847/1538-4357/ac7c74}{\apj, 935, 167, }{The
  Astropy Project: Sustaining and Growing a Community-oriented Open-source
  Project and the Latest Major Release (v5.0) of the Core Package}

\bibitem[{{Barreira}(2022)}]{2022JCAP...11..013B}
{Barreira}, A. 2022, \href{https://arxiv.org/abs/2205.05673}{arXiv:2205.05673},
  \href{https://dx.doi.org/10.1088/1475-7516/2022/11/013}{\jcap, 2022, 013,
  }{Can we actually constrain f$_{NL}$ using the scale-dependent bias effect?
  An illustration of the impact of galaxy bias uncertainties using the BOSS
  DR12 galaxy power spectrum}

\bibitem[{{Bennett} et~al.(1992)}]{1992ApJ...396L...7B}
{Bennett}, C.~L., et~al. 1992, \href{https://dx.doi.org/10.1086/186505}{\apjl,
  396, L7, }{Preliminary Separation of Galactic and Cosmic Microwave Emission
  for the COBE Differential Microwave Radiometers}

\bibitem[{{Bermejo-Climent} et~al.(2025)}]{2025A&A...698A.177B}
{Bermejo-Climent}, J.~R., et~al. 2025,
  \href{https://arxiv.org/abs/2412.10279}{arXiv:2412.10279},
  \href{https://dx.doi.org/10.1051/0004-6361/202453446}{\aap, 698, A177,
  }{Constraints on primordial non-Gaussianity from the cross-correlation of
  DESI luminous red galaxies and Planck CMB lensing}

\bibitem[{{Blas} et~al.(2011){Blas}, {Lesgourgues}, \&
  {Tram}}]{2011JCAP...07..034B}
{Blas}, D., {Lesgourgues}, J., \& {Tram}, T. 2011,
  \href{https://arxiv.org/abs/1104.2933}{arXiv:1104.2933},
  \href{https://dx.doi.org/10.1088/1475-7516/2011/07/034}{\jcap, 2011, 034,
  }{The Cosmic Linear Anisotropy Solving System (CLASS). Part II: Approximation
  schemes}

\bibitem[{{Bloch} \& {Johnson}(2024)}]{2024arXiv240500809B}
{Bloch}, R. \& {Johnson}, M.~C. 2024,
  \href{https://arxiv.org/abs/2405.00809}{arXiv:2405.00809},
  \href{https://dx.doi.org/10.48550/arXiv.2405.00809}{arXiv e-prints,
  arXiv:2405.00809, }{Kinetic Sunyaev Zel'dovich velocity reconstruction from
  Planck and unWISE}

\bibitem[{{Calafut} et~al.(2021)}]{2021PhRvD.104d3502C}
{Calafut}, V., et~al. 2021,
  \href{https://arxiv.org/abs/2101.08374}{arXiv:2101.08374},
  \href{https://dx.doi.org/10.1103/PhysRevD.104.043502}{\prd, 104, 043502,
  }{The Atacama Cosmology Telescope: Detection of the pairwise kinematic
  Sunyaev-Zel'dovich effect with SDSS DR15 galaxies}

\bibitem[{{Carron} et~al.(2022){Carron}, {Mirmelstein}, \&
  {Lewis}}]{2022JCAP...09..039C}
{Carron}, J., {Mirmelstein}, M., \& {Lewis}, A. 2022,
  \href{https://arxiv.org/abs/2206.07773}{arXiv:2206.07773},
  \href{https://dx.doi.org/10.1088/1475-7516/2022/09/039}{\jcap, 2022, 039,
  }{CMB lensing from Planck PR4 maps}

\bibitem[{{Cartis} et~al.(2018{\natexlab{a}}){Cartis}, {Fiala}, {Marteau}, \&
  {Roberts}}]{2018arXiv180400154C}
{Cartis}, C., {Fiala}, J., {Marteau}, B., \& {Roberts}, L. 2018{\natexlab{a}},
  \href{https://arxiv.org/abs/1804.00154}{arXiv:1804.00154}, arXiv e-prints,
  arXiv:1804.00154, {Improving the Flexibility and Robustness of Model-Based
  Derivative-Free Optimization Solvers}

\bibitem[{{Cartis} et~al.(2018{\natexlab{b}}){Cartis}, {Roberts}, \&
  {Sheridan-Methven}}]{2018arXiv181211343C}
{Cartis}, C., {Roberts}, L., \& {Sheridan-Methven}, O. 2018{\natexlab{b}},
  \href{https://arxiv.org/abs/1812.11343}{arXiv:1812.11343}, arXiv e-prints,
  arXiv:1812.11343, {Escaping local minima with derivative-free methods: a
  numerical investigation}

\bibitem[{{Cayuso} et~al.(2023){Cayuso}, {Bloch}, {Hotinli}, {Johnson}, \&
  {McCarthy}}]{2023JCAP...02..051C}
{Cayuso}, J., {Bloch}, R., {Hotinli}, S.~C., {Johnson}, M.~C., \& {McCarthy},
  F. 2023, \href{https://arxiv.org/abs/2111.11526}{arXiv:2111.11526},
  \href{https://dx.doi.org/10.1088/1475-7516/2023/02/051}{\jcap, 2023, 051,
  }{Velocity reconstruction with the cosmic microwave background and galaxy
  surveys}

\bibitem[{{Chaussidon} et~al.(2025)}]{2025JCAP...06..029C}
{Chaussidon}, E., et~al. 2025,
  \href{https://arxiv.org/abs/2411.17623}{arXiv:2411.17623},
  \href{https://dx.doi.org/10.1088/1475-7516/2025/06/029}{\jcap, 2025, 029,
  }{Constraining primordial non-Gaussianity with DESI 2024 LRG and QSO samples}

\bibitem[{{Chen} \& {Wright}(2009)}]{2009ApJ...694..222C}
{Chen}, X. \& {Wright}, E.~L. 2009,
  \href{https://arxiv.org/abs/0809.4025}{arXiv:0809.4025},
  \href{https://dx.doi.org/10.1088/0004-637X/694/1/222}{\apj, 694, 222,
  }{Extragalactic Point Source Search in Five-Year WMAP 41, 61, and 94 Ghz
  Maps}

\bibitem[{{Contreras} et~al.(2019){Contreras}, {Johnson}, \&
  {Mertens}}]{2019JCAP...10..024C}
{Contreras}, D., {Johnson}, M.~C., \& {Mertens}, J.~B. 2019,
  \href{https://arxiv.org/abs/1904.10033}{arXiv:1904.10033},
  \href{https://dx.doi.org/10.1088/1475-7516/2019/10/024}{\jcap, 2019, 024,
  }{Towards detection of relativistic effects in galaxy number counts using kSZ
  tomography}

\bibitem[{{Cooray} \& {Sheth}(2002)}]{2002PhR...372....1C}
{Cooray}, A. \& {Sheth}, R. 2002,
  \href{https://arxiv.org/abs/astro-ph/0206508}{astro-ph/0206508},
  \href{https://dx.doi.org/10.1016/S0370-1573(02)00276-4}{\physrep, 372, 1,
  }{Halo models of large scale structure}

\bibitem[{{Coulton} et~al.(2024{\natexlab{a}})}]{2024PhRvD.109f3530C}
{Coulton}, W., et~al. 2024{\natexlab{a}},
  \href{https://arxiv.org/abs/2307.01258}{arXiv:2307.01258},
  \href{https://dx.doi.org/10.1103/PhysRevD.109.063530}{\prd, 109, 063530,
  }{Atacama Cosmology Telescope: High-resolution component-separated maps
  across one third of the sky}

\bibitem[{{Coulton} et~al.(2024{\natexlab{b}})}]{2024arXiv240113033C}
{Coulton}, W.~R., et~al. 2024{\natexlab{b}},
  \href{https://arxiv.org/abs/2401.13033}{arXiv:2401.13033},
  \href{https://dx.doi.org/10.48550/arXiv.2401.13033}{arXiv e-prints,
  arXiv:2401.13033, }{The Atacama Cosmology Telescope: A search for late-time
  anisotropic screening of the Cosmic Microwave Background}

\bibitem[{{Dalal} et~al.(2008){Dalal}, {Dor{\'e}}, {Huterer}, \&
  {Shirokov}}]{2008PhRvD..77l3514D}
{Dalal}, N., {Dor{\'e}}, O., {Huterer}, D., \& {Shirokov}, A. 2008,
  \href{https://arxiv.org/abs/0710.4560}{arXiv:0710.4560},
  \href{https://dx.doi.org/10.1103/PhysRevD.77.123514}{\prd, 77, 123514,
  }{Imprints of primordial non-Gaussianities on large-scale structure:
  Scale-dependent bias and abundance of virialized objects}

\bibitem[{{de Putter} et~al.(2017){de Putter}, {Gleyzes}, \&
  {Dor{\'e}}}]{2017PhRvD..95l3507D}
{de Putter}, R., {Gleyzes}, J., \& {Dor{\'e}}, O. 2017,
  \href{https://arxiv.org/abs/1612.05248}{arXiv:1612.05248},
  \href{https://dx.doi.org/10.1103/PhysRevD.95.123507}{\prd, 95, 123507, }{Next
  non-Gaussianity frontier: What can a measurement with {\ensuremath{\sigma}}
  (f$_{NL}$){\ensuremath{\lesssim}}1 tell us about multifield inflation?}

\bibitem[{{Delabrouille} et~al.(2009){Delabrouille}, {Cardoso}, {Le Jeune},
  {Betoule}, {Fay}, \& {Guilloux}}]{2009A&A...493..835D}
{Delabrouille}, J., {Cardoso}, J.~F., {Le Jeune}, M., {Betoule}, M., {Fay}, G.,
  \& {Guilloux}, F. 2009,
  \href{https://arxiv.org/abs/0807.0773}{arXiv:0807.0773},
  \href{https://dx.doi.org/10.1051/0004-6361:200810514}{\aap, 493, 835, }{A
  full sky, low foreground, high resolution CMB map from WMAP}

\bibitem[{{Deutsch} et~al.(2018){Deutsch}, {Dimastrogiovanni}, {Johnson},
  {M{\"u}nchmeyer}, \& {Terrana}}]{2018PhRvD..98l3501D}
{Deutsch}, A.-S., {Dimastrogiovanni}, E., {Johnson}, M.~C., {M{\"u}nchmeyer},
  M., \& {Terrana}, A. 2018,
  \href{https://arxiv.org/abs/1707.08129}{arXiv:1707.08129},
  \href{https://dx.doi.org/10.1103/PhysRevD.98.123501}{\prd, 98, 123501,
  }{Reconstruction of the remote dipole and quadrupole fields from the kinetic
  Sunyaev Zel'dovich and polarized Sunyaev Zel'dovich effects}

\bibitem[{{Dey} et~al.(2019)}]{2019AJ....157..168D}
{Dey}, A., et~al. 2019,
  \href{https://arxiv.org/abs/1804.08657}{arXiv:1804.08657},
  \href{https://dx.doi.org/10.3847/1538-3881/ab089d}{\aj, 157, 168, }{Overview
  of the DESI Legacy Imaging Surveys}

\bibitem[{{Eisenstein} et~al.(2007){Eisenstein}, {Seo}, {Sirko}, \&
  {Spergel}}]{ESSS07}
{Eisenstein}, D.~J., {Seo}, H.-J., {Sirko}, E., \& {Spergel}, D.~N. 2007,
  \href{https://arxiv.org/abs/astro-ph/0604362}{astro-ph/0604362},
  \href{https://dx.doi.org/10.1086/518712}{\apj, 664, 675, }{Improving
  Cosmological Distance Measurements by Reconstruction of the Baryon Acoustic
  Peak}

\bibitem[{{Fabbian} et~al.(2025){Fabbian}, {Alonso}, {Storey-Fisher}, \&
  {Cornish}}]{2025arXiv250420992F}
{Fabbian}, G., {Alonso}, D., {Storey-Fisher}, K., \& {Cornish}, T. 2025,
  \href{https://arxiv.org/abs/2504.20992}{arXiv:2504.20992},
  \href{https://dx.doi.org/10.48550/arXiv.2504.20992}{arXiv e-prints,
  arXiv:2504.20992, }{Constraints on primordial non-Gaussianity from Quaia}

\bibitem[{{Fang} et~al.(2020){Fang}, {Krause}, {Eifler}, \&
  {MacCrann}}]{2020JCAP...05..010F}
{Fang}, X., {Krause}, E., {Eifler}, T., \& {MacCrann}, N. 2020,
  \href{https://arxiv.org/abs/1911.11947}{arXiv:1911.11947},
  \href{https://dx.doi.org/10.1088/1475-7516/2020/05/010}{\jcap, 2020, 010,
  }{Beyond Limber: efficient computation of angular power spectra for galaxy
  clustering and weak lensing}

\bibitem[{{Garrison} et~al.(2021){Garrison}, {Eisenstein}, {Ferrer},
  {Maksimova}, \& {Pinto}}]{2021MNRAS.508..575G}
{Garrison}, L.~H., {Eisenstein}, D.~J., {Ferrer}, D., {Maksimova}, N.~A., \&
  {Pinto}, P.~A. 2021,
  \href{https://arxiv.org/abs/2110.11392}{arXiv:2110.11392},
  \href{https://dx.doi.org/10.1093/mnras/stab2482}{\mnras, 508, 575, }{The
  ABACUS cosmological N-body code}

\bibitem[{{Giannantonio} \& {Percival}(2014)}]{2014MNRAS.441L..16G}
{Giannantonio}, T. \& {Percival}, W.~J. 2014,
  \href{https://arxiv.org/abs/1312.5154}{arXiv:1312.5154},
  \href{https://dx.doi.org/10.1093/mnrasl/slu036}{\mnras, 441, L16, }{Using
  correlations between cosmic microwave background lensing and large-scale
  structure to measure primordial non-Gaussianity.}

\bibitem[{{Giri} \& {Smith}(2022)}]{2022JCAP...09..028G}
{Giri}, U. \& {Smith}, K.~M. 2022,
  \href{https://arxiv.org/abs/2010.07193}{arXiv:2010.07193},
  \href{https://dx.doi.org/10.1088/1475-7516/2022/09/028}{\jcap, 2022, 028,
  }{Exploring KSZ velocity reconstruction with N-body simulations and the halo
  model}

\bibitem[{{G{\'o}rski} et~al.(2005){G{\'o}rski}, {Hivon}, {Banday}, {Wandelt},
  {Hansen}, {Reinecke}, \& {Bartelmann}}]{2005ApJ...622..759G}
{G{\'o}rski}, K.~M., {Hivon}, E., {Banday}, A.~J., {Wandelt}, B.~D., {Hansen},
  F.~K., {Reinecke}, M., \& {Bartelmann}, M. 2005,
  \href{https://arxiv.org/abs/astro-ph/0409513}{astro-ph/0409513},
  \href{https://dx.doi.org/10.1086/427976}{\apj, 622, 759, }{HEALPix: A
  Framework for High-Resolution Discretization and Fast Analysis of Data
  Distributed on the Sphere}

\bibitem[{{Gratton} \& {Challinor}(2020)}]{2020MNRAS.499.3410G}
{Gratton}, S. \& {Challinor}, A. 2020,
  \href{https://arxiv.org/abs/1911.07754}{arXiv:1911.07754},
  \href{https://dx.doi.org/10.1093/mnras/staa2996}{\mnras, 499, 3410,
  }{Understanding parameter differences between analyses employing nested data
  subsets}

\bibitem[{{Guachalla} et~al.(2025)}]{2025PhRvD.112j3512G}
{Guachalla}, B.~R., et~al. 2025,
  \href{https://arxiv.org/abs/2503.19870}{arXiv:2503.19870},
  \href{https://dx.doi.org/10.1103/lqbj-wcqj}{\prd, 112, 103512, }{Backlighting
  extended gas halos around luminous red galaxies: Kinematic Sunyaev-Zel'dovich
  effect from DESI Y1 and ACT data}

\bibitem[{{Hadzhiyska} et~al.(2024){Hadzhiyska}, {Ferraro}, {Ried Guachalla},
  \& {Schaan}}]{2024PhRvD.109j3534H}
{Hadzhiyska}, B., {Ferraro}, S., {Ried Guachalla}, B., \& {Schaan}, E. 2024,
  \href{https://arxiv.org/abs/2312.12434}{arXiv:2312.12434},
  \href{https://dx.doi.org/10.1103/PhysRevD.109.103534}{\prd, 109, 103534,
  }{Velocity reconstruction in the era of DESI and Rubin/LSST. II. Realistic
  samples on the light cone}

\bibitem[{{Hadzhiyska} et~al.(2022){Hadzhiyska}, {Garrison}, {Eisenstein}, \&
  {Bose}}]{2022MNRAS.509.2194H}
{Hadzhiyska}, B., {Garrison}, L.~H., {Eisenstein}, D., \& {Bose}, S. 2022,
  \href{https://arxiv.org/abs/2110.11413}{arXiv:2110.11413},
  \href{https://dx.doi.org/10.1093/mnras/stab3066}{\mnras, 509, 2194, }{The
  halo light-cone catalogues of ABACUSSUMMIT}

\bibitem[{{Hadzhiyska} et~al.(2025{\natexlab{a}}){Hadzhiyska}, {Sailer}, \&
  {Ferraro}}]{2025arXiv250617379H}
{Hadzhiyska}, B., {Sailer}, N., \& {Ferraro}, S. 2025{\natexlab{a}},
  \href{https://arxiv.org/abs/2506.17379}{arXiv:2506.17379},
  \href{https://dx.doi.org/10.48550/arXiv.2506.17379}{arXiv e-prints,
  arXiv:2506.17379, }{Mapping the gas density with the kinematic
  Sunyaev-Zel'dovich and patchy screening effects: a self-consistent
  comparison}

\bibitem[{{Hadzhiyska} et~al.(2025{\natexlab{b}})}]{2024arXiv240707152H}
{Hadzhiyska}, B., et~al. 2025{\natexlab{b}},
  \href{https://arxiv.org/abs/2407.07152}{arXiv:2407.07152},
  \href{https://dx.doi.org/10.1103/kclp-x5j1}{\prd, 112, 083509, }{Evidence for
  large baryonic feedback at low and intermediate redshifts from kinematic
  Sunyaev-Zel'dovich observations with ACT and DESI photometric galaxies}

\bibitem[{{Hand} et~al.(2012)}]{2012PhRvL.109d1101H}
{Hand}, N., et~al. 2012,
  \href{https://arxiv.org/abs/1203.4219}{arXiv:1203.4219},
  \href{https://dx.doi.org/10.1103/PhysRevLett.109.041101}{\prl, 109, 041101,
  }{Evidence of Galaxy Cluster Motions with the Kinematic Sunyaev-Zel'dovich
  Effect}

\bibitem[{Harris et~al.(2020)}]{harris2020array}
Harris, C.~R., et~al. 2020,
  \href{https://dx.doi.org/10.1038/s41586-020-2649-2}{Nature, 585, 357, }Array
  programming with {NumPy}, \url{https://doi.org/10.1038/s41586-020-2649-2}

\bibitem[{{Hill} et~al.(2016){Hill}, {Ferraro}, {Battaglia}, {Liu}, \&
  {Spergel}}]{2016PhRvL.117e1301H}
{Hill}, J.~C., {Ferraro}, S., {Battaglia}, N., {Liu}, J., \& {Spergel}, D.~N.
  2016, \href{https://arxiv.org/abs/1603.01608}{arXiv:1603.01608},
  \href{https://dx.doi.org/10.1103/PhysRevLett.117.051301}{\prl, 117, 051301,
  }{Kinematic Sunyaev-Zel'dovich Effect with Projected Fields: A Novel Probe of
  the Baryon Distribution with Planck, WMAP, and WISE Data}

\bibitem[{{Hilton} et~al.(2021)}]{2021ApJS..253....3H}
{Hilton}, M., et~al. 2021,
  \href{https://arxiv.org/abs/2009.11043}{arXiv:2009.11043},
  \href{https://dx.doi.org/10.3847/1538-4365/abd023}{\apjs, 253, 3, }{The
  Atacama Cosmology Telescope: A Catalog of >4000
  Sunyaev-Zel{\textquoteright}dovich Galaxy Clusters}

\bibitem[{{Hivon} et~al.(2002){Hivon}, {G{\'o}rski}, {Netterfield}, {Crill},
  {Prunet}, \& {Hansen}}]{2002ApJ...567....2H}
{Hivon}, E., {G{\'o}rski}, K.~M., {Netterfield}, C.~B., {Crill}, B.~P.,
  {Prunet}, S., \& {Hansen}, F. 2002,
  \href{https://dx.doi.org/10.1086/338126}{\apj, 567, 2, }{MASTER of the Cosmic
  Microwave Background Anisotropy Power Spectrum: A Fast Method for Statistical
  Analysis of Large and Complex Cosmic Microwave Background Data Sets}

\bibitem[{{Hotinli} et~al.(2025){Hotinli}, {Smith}, \&
  {Ferraro}}]{2025arXiv250621657H}
{Hotinli}, S.~C., {Smith}, K.~M., \& {Ferraro}, S. 2025,
  \href{https://arxiv.org/abs/2506.21657}{arXiv:2506.21657},
  \href{https://dx.doi.org/10.48550/arXiv.2506.21657}{arXiv e-prints,
  arXiv:2506.21657, }{Velocity Reconstruction from KSZ: Measuring $f_{NL}$ with
  ACT and DESILS}

\bibitem[{Hunter(2007)}]{Hunter:2007}
Hunter, J.~D. 2007, \href{https://dx.doi.org/10.1109/MCSE.2007.55}{Computing in
  Science \& Engineering, 9, 90, }Matplotlib: A 2D graphics environment

\bibitem[{{Kim} et~al.(2024)}]{2024JCAP...12..022K}
{Kim}, J., et~al. 2024,
  \href{https://arxiv.org/abs/2407.04606}{arXiv:2407.04606},
  \href{https://dx.doi.org/10.1088/1475-7516/2024/12/022}{\jcap, 2024, 022,
  }{The Atacama Cosmology Telescope DR6 and DESI: structure formation over
  cosmic time with a measurement of the cross-correlation of CMB lensing and
  luminous red galaxies}

\bibitem[{{Krolewski} et~al.(2024)}]{2024JCAP...03..021K}
{Krolewski}, A., et~al. 2024,
  \href{https://arxiv.org/abs/2305.07650}{arXiv:2305.07650},
  \href{https://dx.doi.org/10.1088/1475-7516/2024/03/021}{\jcap, 2024, 021,
  }{Constraining primordial non-Gaussianity from DESI quasar targets and Planck
  CMB lensing}

\bibitem[{{Krywonos} et~al.(2024){Krywonos}, {Hotinli}, \&
  {Johnson}}]{2024arXiv240805264K}
{Krywonos}, J., {Hotinli}, S.~C., \& {Johnson}, M.~C. 2024,
  \href{https://arxiv.org/abs/2408.05264}{arXiv:2408.05264},
  \href{https://dx.doi.org/10.48550/arXiv.2408.05264}{arXiv e-prints,
  arXiv:2408.05264, }{Constraints on cosmology beyond $\Lambda$CDM with kinetic
  Sunyaev Zel'dovich velocity reconstruction}

\bibitem[{{Kusiak} et~al.(2021){Kusiak}, {Bolliet}, {Ferraro}, {Hill}, \&
  {Krolewski}}]{2021PhRvD.104d3518K}
{Kusiak}, A., {Bolliet}, B., {Ferraro}, S., {Hill}, J.~C., \& {Krolewski}, A.
  2021, \href{https://arxiv.org/abs/2102.01068}{arXiv:2102.01068},
  \href{https://dx.doi.org/10.1103/PhysRevD.104.043518}{\prd, 104, 043518,
  }{Constraining the baryon abundance with the kinematic Sunyaev-Zel'dovich
  effect: Projected-field detection using P l a n c k , W M A P , and u n W I S
  E}

\bibitem[{{Kvasiuk} et~al.(2025){Kvasiuk}, {Lai}, {M{\"u}nchmeyer}, \&
  {Smith}}]{2025arXiv251005215K}
{Kvasiuk}, Y., {Lai}, A., {M{\"u}nchmeyer}, M., \& {Smith}, K.~M. 2025,
  \href{https://arxiv.org/abs/2510.05215}{arXiv:2510.05215}, arXiv e-prints,
  arXiv:2510.05215, {QML-FAST - A Fast Code for low-$\ell$ Tomographic Maximum
  Likelihood Power Spectrum Estimation}

\bibitem[{Lagu\"e et~al.(2024)Lagu\"e, Madhavacheril, Smith, Ferraro, \&
  Schaan}]{Lague:2024czc}
Lagu\"e, A., Madhavacheril, M.~S., Smith, K.~M., Ferraro, S., \& Schaan, E.
  2024, \href{https://arxiv.org/abs/2411.08240}{arXiv:2411.08240}, {Constraints
  on local primordial non-Gaussianity with 3d Velocity Reconstruction from the
  Kinetic Sunyaev-Zeldovich Effect}

\bibitem[{{Lagu{\"e}} et~al.(2025){Lagu{\"e}}, {Madhavacheril}, {Smith},
  {Ferraro}, \& {Schaan}}]{2025PhRvL.134o1003L}
{Lagu{\"e}}, A., {Madhavacheril}, M.~S., {Smith}, K.~M., {Ferraro}, S., \&
  {Schaan}, E. 2025, \href{https://arxiv.org/abs/2411.08240}{arXiv:2411.08240},
  \href{https://dx.doi.org/10.1103/PhysRevLett.134.151003}{\prl, 134, 151003,
  }{Constraints on Local Primordial Non-Gaussianity with 3D Velocity
  Reconstruction from the Kinetic Sunyaev-Zeldovich Effect}

\bibitem[{{Lai} et~al.(2025){Lai}, {Kvasiuk}, \&
  {M{\"u}nchmeyer}}]{2025arXiv250621684L}
{Lai}, A. C.~M., {Kvasiuk}, Y., \& {M{\"u}nchmeyer}, M. 2025,
  \href{https://arxiv.org/abs/2506.21684}{arXiv:2506.21684},
  \href{https://dx.doi.org/10.48550/arXiv.2506.21684}{arXiv e-prints,
  arXiv:2506.21684, }{KSZ Velocity Reconstruction with ACT and DESI-LS using a
  Tomographic QML Power Spectrum Estimator}

\bibitem[{{Lesgourgues}(2011)}]{2011arXiv1104.2932L}
{Lesgourgues}, J. 2011,
  \href{https://arxiv.org/abs/1104.2932}{arXiv:1104.2932},
  \href{https://dx.doi.org/10.48550/arXiv.1104.2932}{arXiv e-prints,
  arXiv:1104.2932, }{The Cosmic Linear Anisotropy Solving System (CLASS) I:
  Overview}

\bibitem[{Lewis(2019)}]{Lewis:2019xzd}
Lewis, A. 2019, \href{https://arxiv.org/abs/1910.13970}{arXiv:1910.13970},
  {GetDist: a Python package for analysing Monte Carlo samples},
  \url{https://getdist.readthedocs.io}

\bibitem[{{Lewis} et~al.(2000){Lewis}, {Challinor}, \&
  {Lasenby}}]{2000ApJ...538..473L}
{Lewis}, A., {Challinor}, A., \& {Lasenby}, A. 2000,
  \href{https://arxiv.org/abs/astro-ph/9911177}{astro-ph/9911177},
  \href{https://dx.doi.org/10.1086/309179}{\apj, 538, 473, }{Efficient
  Computation of Cosmic Microwave Background Anisotropies in Closed
  Friedmann-Robertson-Walker Models}

\bibitem[{{Limber}(1953)}]{1953ApJ...117..134L}
{Limber}, D.~N. 1953, \href{https://dx.doi.org/10.1086/145672}{\apj, 117, 134,
  }{The Analysis of Counts of the Extragalactic Nebulae in Terms of a
  Fluctuating Density Field.}

\bibitem[{{Liu} et~al.(2025){Liu}, {Hadzhiyska}, {Ferraro}, {Bose}, \&
  {Hern{\'a}ndez-Aguayo}}]{Liu:2025}
{Liu}, R.~H., {Hadzhiyska}, B., {Ferraro}, S., {Bose}, S., \&
  {Hern{\'a}ndez-Aguayo}, C. 2025,
  \href{https://arxiv.org/abs/2504.11794}{arXiv:2504.11794},
  \href{https://dx.doi.org/10.48550/arXiv.2504.11794}{arXiv e-prints,
  arXiv:2504.11794, }{Fast Baryonic Field Painting for Sunyaev-Zel'dovich
  Analyses: Transfer Function vs. Hybrid Effective Field Theory}

\bibitem[{{Louis} et~al.(2025)}]{2025arXiv250314452L}
{Louis}, T., et~al. 2025,
  \href{https://arxiv.org/abs/2503.14452}{arXiv:2503.14452},
  \href{https://dx.doi.org/10.48550/arXiv.2503.14452}{arXiv e-prints,
  arXiv:2503.14452, }{The Atacama Cosmology Telescope: DR6 Power Spectra,
  Likelihoods and $\Lambda$CDM Parameters}

\bibitem[{{MacCrann} et~al.(2024)}]{2024ApJ...966..138M}
{MacCrann}, N., et~al. 2024,
  \href{https://arxiv.org/abs/2304.05196}{arXiv:2304.05196},
  \href{https://dx.doi.org/10.3847/1538-4357/ad2610}{\apj, 966, 138, }{The
  Atacama Cosmology Telescope: Mitigating the Impact of Extragalactic
  Foregrounds for the DR6 Cosmic Microwave Background Lensing Analysis}

\bibitem[{{Madhavacheril} et~al.(2019){Madhavacheril}, {Battaglia}, {Smith}, \&
  {Sievers}}]{2019PhRvD.100j3532M}
{Madhavacheril}, M.~S., {Battaglia}, N., {Smith}, K.~M., \& {Sievers}, J.~L.
  2019, \href{https://arxiv.org/abs/1901.02418}{arXiv:1901.02418},
  \href{https://dx.doi.org/10.1103/PhysRevD.100.103532}{\prd, 100, 103532,
  }{Cosmology with the kinematic Sunyaev-Zeldovich effect: Breaking the optical
  depth degeneracy with fast radio bursts}

\bibitem[{{Madhavacheril} et~al.(2024)}]{2024ApJ...962..113M}
{Madhavacheril}, M.~S., et~al. 2024,
  \href{https://arxiv.org/abs/2304.05203}{arXiv:2304.05203},
  \href{https://dx.doi.org/10.3847/1538-4357/acff5f}{\apj, 962, 113, }{The
  Atacama Cosmology Telescope: DR6 Gravitational Lensing Map and Cosmological
  Parameters}

\bibitem[{{Maksimova} et~al.(2021){Maksimova}, {Garrison}, {Eisenstein},
  {Hadzhiyska}, {Bose}, \& {Satterthwaite}}]{2021MNRAS.508.4017M}
{Maksimova}, N.~A., {Garrison}, L.~H., {Eisenstein}, D.~J., {Hadzhiyska}, B.,
  {Bose}, S., \& {Satterthwaite}, T.~P. 2021,
  \href{https://arxiv.org/abs/2110.11398}{arXiv:2110.11398},
  \href{https://dx.doi.org/10.1093/mnras/stab2484}{\mnras, 508, 4017,
  }{ABACUSSUMMIT: a massive set of high-accuracy, high-resolution N-body
  simulations}

\bibitem[{{McCarthy} \& {Johnson}(2020)}]{2020PhRvD.102d3520M}
{McCarthy}, F. \& {Johnson}, M.~C. 2020,
  \href{https://arxiv.org/abs/1907.06678}{arXiv:1907.06678},
  \href{https://dx.doi.org/10.1103/PhysRevD.102.043520}{\prd, 102, 043520,
  }{Remote dipole field reconstruction with dusty galaxies}

\bibitem[{{McCarthy} et~al.(2023){McCarthy}, {Madhavacheril}, \&
  {Maniyar}}]{2023PhRvD.108h3522M}
{McCarthy}, F., {Madhavacheril}, M.~S., \& {Maniyar}, A.~S. 2023,
  \href{https://arxiv.org/abs/2210.01049}{arXiv:2210.01049},
  \href{https://dx.doi.org/10.1103/PhysRevD.108.083522}{\prd, 108, 083522,
  }{Constraints on primordial non-Gaussianity from halo bias measured through
  CMB lensing cross-correlations}

\bibitem[{{McCarthy} et~al.(2025)}]{2025JCAP...05..057M}
{McCarthy}, F., et~al. 2025,
  \href{https://arxiv.org/abs/2410.06229}{arXiv:2410.06229},
  \href{https://dx.doi.org/10.1088/1475-7516/2025/05/057}{\jcap, 2025, 057,
  }{The Atacama Cosmology Telescope: Large-scale velocity reconstruction with
  the kinematic Sunyaev-Zel'dovich effect and DESI LRGs}

\bibitem[{{M.J.D.}(2009)}]{bobyqa}
{M.J.D.}, P. 2009, Technical Report 2009/NA06, DAMTP, University of Cambridge),
  The BOBYQA algorithm for bound constrained optimization without derivatives

\bibitem[{{M{\"u}nchmeyer} et~al.(2019){M{\"u}nchmeyer}, {Madhavacheril},
  {Ferraro}, {Johnson}, \& {Smith}}]{2019PhRvD.100h3508M}
{M{\"u}nchmeyer}, M., {Madhavacheril}, M.~S., {Ferraro}, S., {Johnson}, M.~C.,
  \& {Smith}, K.~M. 2019,
  \href{https://arxiv.org/abs/1810.13424}{arXiv:1810.13424},
  \href{https://dx.doi.org/10.1103/PhysRevD.100.083508}{\prd, 100, 083508,
  }{Constraining local non-Gaussianities with kinetic Sunyaev-Zel'dovich
  tomography}

\bibitem[{{Naess} et~al.(2025)}]{2025arXiv250314451N}
{Naess}, S., et~al. 2025,
  \href{https://arxiv.org/abs/2503.14451}{arXiv:2503.14451},
  \href{https://dx.doi.org/10.48550/arXiv.2503.14451}{arXiv e-prints,
  arXiv:2503.14451, }{The Atacama Cosmology Telescope: DR6 Maps}

\bibitem[{{Namikawa} et~al.(2013){Namikawa}, {Hanson}, \&
  {Takahashi}}]{2013MNRAS.431..609N}
{Namikawa}, T., {Hanson}, D., \& {Takahashi}, R. 2013,
  \href{https://arxiv.org/abs/1209.0091}{arXiv:1209.0091},
  \href{https://dx.doi.org/10.1093/mnras/stt195}{\mnras, 431, 609,
  }{Bias-hardened CMB lensing}

\bibitem[{{Nelson} et~al.(2019)}]{2019ComAC...6....2N}
{Nelson}, D., et~al. 2019,
  \href{https://arxiv.org/abs/1812.05609}{arXiv:1812.05609},
  \href{https://dx.doi.org/10.1186/s40668-019-0028-x}{Computational
  Astrophysics and Cosmology, 6, 2, }{The IllustrisTNG simulations: public data
  release}

\bibitem[{{Okamoto} \& {Hu}(2003)}]{2003PhRvD..67h3002O}
{Okamoto}, T. \& {Hu}, W. 2003,
  \href{https://arxiv.org/abs/astro-ph/0301031}{astro-ph/0301031},
  \href{https://dx.doi.org/10.1103/PhysRevD.67.083002}{\prd, 67, 083002,
  }{Cosmic microwave background lensing reconstruction on the full sky}

\bibitem[{{Planck Collaboration}(2013)}]{2013A&A...557A..52P}
{Planck Collaboration}. 2013,
  \href{https://arxiv.org/abs/1212.4131}{arXiv:1212.4131},
  \href{https://dx.doi.org/10.1051/0004-6361/201220941}{\aap, 557, A52,
  }{Planck intermediate results. XI. The gas content of dark matter halos: the
  Sunyaev-Zeldovich-stellar mass relation for locally brightest galaxies}

\bibitem[{{Planck Collaboration}(2020{\natexlab{a}})}]{2020A&A...641A...6P}
---. 2020{\natexlab{a}},
  \href{https://arxiv.org/abs/1807.06209}{arXiv:1807.06209},
  \href{https://dx.doi.org/10.1051/0004-6361/201833910}{\aap, 641, A6, }{Planck
  2018 results. VI. Cosmological parameters}

\bibitem[{{Planck Collaboration}(2020{\natexlab{b}})}]{2020A&A...641A...8P}
---. 2020{\natexlab{b}},
  \href{https://arxiv.org/abs/1807.06210}{arXiv:1807.06210},
  \href{https://dx.doi.org/10.1051/0004-6361/201833886}{\aap, 641, A8, }{Planck
  2018 results. VIII. Gravitational lensing}

\bibitem[{{Planck Collaboration}(2020{\natexlab{c}})}]{2020A&A...643A..42P}
---. 2020{\natexlab{c}},
  \href{https://arxiv.org/abs/2007.04997}{arXiv:2007.04997},
  \href{https://dx.doi.org/10.1051/0004-6361/202038073}{\aap, 643, A42,
  }{Planck intermediate results. LVII. Joint Planck LFI and HFI data
  processing}

\bibitem[{{Planck Collaboration}
  et~al.(2020{\natexlab{a}})}]{2020A&A...641A...1P}
{Planck Collaboration}, et~al. 2020{\natexlab{a}},
  \href{https://arxiv.org/abs/1807.06205}{arXiv:1807.06205},
  \href{https://dx.doi.org/10.1051/0004-6361/201833880}{\aap, 641, A1, }{Planck
  2018 results. I. Overview and the cosmological legacy of Planck}

\bibitem[{{Planck Collaboration}
  et~al.(2020{\natexlab{b}})}]{2020A&A...641A...9P}
---. 2020{\natexlab{b}},
  \href{https://arxiv.org/abs/1905.05697}{arXiv:1905.05697},
  \href{https://dx.doi.org/10.1051/0004-6361/201935891}{\aap, 641, A9, }{Planck
  2018 results. IX. Constraints on primordial non-Gaussianity}

\bibitem[{{Qu} et~al.(2024)}]{2024ApJ...962..112Q}
{Qu}, F.~J., et~al. 2024,
  \href{https://arxiv.org/abs/2304.05202}{arXiv:2304.05202},
  \href{https://dx.doi.org/10.3847/1538-4357/acfe06}{\apj, 962, 112, }{The
  Atacama Cosmology Telescope: A Measurement of the DR6 CMB Lensing Power
  Spectrum and Its Implications for Structure Growth}

\bibitem[{{Remazeilles} et~al.(2011){Remazeilles}, {Delabrouille}, \&
  {Cardoso}}]{2011MNRAS.410.2481R}
{Remazeilles}, M., {Delabrouille}, J., \& {Cardoso}, J.-F. 2011,
  \href{https://arxiv.org/abs/1006.5599}{arXiv:1006.5599},
  \href{https://dx.doi.org/10.1111/j.1365-2966.2010.17624.x}{\mnras, 410, 2481,
  }{CMB and SZ effect separation with constrained Internal Linear Combinations}

\bibitem[{{Ried Guachalla} et~al.(2024){Ried Guachalla}, {Schaan},
  {Hadzhiyska}, \& {Ferraro}}]{2024PhRvD.109j3533R}
{Ried Guachalla}, B., {Schaan}, E., {Hadzhiyska}, B., \& {Ferraro}, S. 2024,
  \href{https://arxiv.org/abs/2312.12435}{arXiv:2312.12435},
  \href{https://dx.doi.org/10.1103/PhysRevD.109.103533}{\prd, 109, 103533,
  }{Velocity reconstruction in the era of DESI and Rubin/LSST. I. Exploring
  spectroscopic, photometric, and hybrid samples}

\bibitem[{{Roper} et~al.(2025){Roper}, {Cai}, \&
  {Peacock}}]{2025arXiv251012553R}
{Roper}, F.~A., {Cai}, Y.-C., \& {Peacock}, J.~A. 2025,
  \href{https://arxiv.org/abs/2510.12553}{arXiv:2510.12553},
  \href{https://dx.doi.org/10.48550/arXiv.2510.12553}{arXiv e-prints,
  arXiv:2510.12553, }{Mass dependence of halo baryon fractions from the kinetic
  Sunyaev-Zeldovich effect}

\bibitem[{{Sailer} et~al.(2025)}]{2025JCAP...06..008S}
{Sailer}, N., et~al. 2025,
  \href{https://arxiv.org/abs/2407.04607}{arXiv:2407.04607},
  \href{https://dx.doi.org/10.1088/1475-7516/2025/06/008}{\jcap, 2025, 008,
  }{Cosmological constraints from the cross-correlation of DESI Luminous Red
  Galaxies with CMB lensing from Planck PR4 and ACT DR6}

\bibitem[{{Schaan} et~al.(2016)}]{2016PhRvD..93h2002S}
{Schaan}, E., et~al. 2016,
  \href{https://arxiv.org/abs/1510.06442}{arXiv:1510.06442},
  \href{https://dx.doi.org/10.1103/PhysRevD.93.082002}{\prd, 93, 082002,
  }{Evidence for the kinematic Sunyaev-Zel'dovich effect with the Atacama
  Cosmology Telescope and velocity reconstruction from the Baryon Oscillation
  Spectroscopic Survey}

\bibitem[{{Schaan} et~al.(2021)}]{2021PhRvD.103f3513S}
---. 2021, \href{https://arxiv.org/abs/2009.05557}{arXiv:2009.05557},
  \href{https://dx.doi.org/10.1103/PhysRevD.103.063513}{\prd, 103, 063513,
  }{Atacama Cosmology Telescope: Combined kinematic and thermal
  Sunyaev-Zel'dovich measurements from BOSS CMASS and LOWZ halos}

\bibitem[{{Schutt} et~al.(2024){Schutt}, {Maniyar}, {Schaan}, {Coulton}, \&
  {Mishra}}]{2024PhRvD.109j3539S}
{Schutt}, T., {Maniyar}, A.~S., {Schaan}, E., {Coulton}, W.~R., \& {Mishra}, N.
  2024, \href{https://arxiv.org/abs/2401.13040}{arXiv:2401.13040},
  \href{https://dx.doi.org/10.1103/PhysRevD.109.103539}{\prd, 109, 103539,
  }{New temperature inversion estimator to detect CMB patchy screening by
  large-scale structure}

\bibitem[{{Seljak}(2009)}]{2009PhRvL.102b1302S}
{Seljak}, U. 2009, \href{https://arxiv.org/abs/0807.1770}{arXiv:0807.1770},
  \href{https://dx.doi.org/10.1103/PhysRevLett.102.021302}{\prl, 102, 021302,
  }{Extracting Primordial Non-Gaussianity without Cosmic Variance}

\bibitem[{{Slosar} et~al.(2008){Slosar}, {Hirata}, {Seljak}, {Ho}, \&
  {Padmanabhan}}]{2008JCAP...08..031S}
{Slosar}, A., {Hirata}, C., {Seljak}, U., {Ho}, S., \& {Padmanabhan}, N. 2008,
  \href{https://arxiv.org/abs/0805.3580}{arXiv:0805.3580},
  \href{https://dx.doi.org/10.1088/1475-7516/2008/08/031}{\jcap, 2008, 031,
  }{Constraints on local primordial non-Gaussianity from large scale structure}

\bibitem[{{Smith} et~al.(2018){Smith}, {Madhavacheril}, {M{\"u}nchmeyer},
  {Ferraro}, {Giri}, \& {Johnson}}]{2018arXiv181013423S}
{Smith}, K.~M., {Madhavacheril}, M.~S., {M{\"u}nchmeyer}, M., {Ferraro}, S.,
  {Giri}, U., \& {Johnson}, M.~C. 2018,
  \href{https://arxiv.org/abs/1810.13423}{arXiv:1810.13423},
  \href{https://dx.doi.org/10.48550/arXiv.1810.13423}{arXiv e-prints,
  arXiv:1810.13423, }{KSZ tomography and the bispectrum}

\bibitem[{{Soergel} et~al.(2016)}]{2016MNRAS.461.3172S}
{Soergel}, B., et~al. 2016,
  \href{https://arxiv.org/abs/1603.03904}{arXiv:1603.03904},
  \href{https://dx.doi.org/10.1093/mnras/stw1455}{\mnras, 461, 3172,
  }{Detection of the kinematic Sunyaev-Zel'dovich effect with DES Year 1 and
  SPT}

\bibitem[{{Stein} et~al.(2020){Stein}, {Alvarez}, {Bond}, {van Engelen}, \&
  {Battaglia}}]{2020JCAP...10..012S}
{Stein}, G., {Alvarez}, M.~A., {Bond}, J.~R., {van Engelen}, A., \&
  {Battaglia}, N. 2020,
  \href{https://arxiv.org/abs/2001.08787}{arXiv:2001.08787},
  \href{https://dx.doi.org/10.1088/1475-7516/2020/10/012}{\jcap, 2020, 012,
  }{The Websky extragalactic CMB simulations}

\bibitem[{{Sunyaev} \& {Zeldovich}(1970)}]{1970Ap&SS...7....3S}
{Sunyaev}, R.~A. \& {Zeldovich}, Y.~B. 1970,
  \href{https://dx.doi.org/10.1007/BF00653471}{\apss, 7, 3, }{Small-Scale
  Fluctuations of Relic Radiation}

\bibitem[{{Sunyaev} \& {Zeldovich}(1980)}]{1980MNRAS.190..413S}
---. 1980, \href{https://dx.doi.org/10.1093/mnras/190.3.413}{\mnras, 190, 413,
  }{The velocity of clusters of galaxies relative to the microwave background -
  The possibility of its measurement.}

\bibitem[{{Tanimura} et~al.(2021){Tanimura}, {Zaroubi}, \&
  {Aghanim}}]{2021A&A...645A.112T}
{Tanimura}, H., {Zaroubi}, S., \& {Aghanim}, N. 2021,
  \href{https://arxiv.org/abs/2007.02952}{arXiv:2007.02952},
  \href{https://dx.doi.org/10.1051/0004-6361/202038846}{\aap, 645, A112,
  }{Direct detection of the kinetic Sunyaev-Zel'dovich effect in galaxy
  clusters}

\bibitem[{{Tegmark} \& {de Oliveira-Costa}(2001)}]{2001PhRvD..64f3001T}
{Tegmark}, M. \& {de Oliveira-Costa}, A. 2001,
  \href{https://arxiv.org/abs/astro-ph/0012120}{astro-ph/0012120},
  \href{https://dx.doi.org/10.1103/PhysRevD.64.063001}{\prd, 64, 063001, }{How
  to measure CMB polarization power spectra without losing information}

\bibitem[{{Terrana} et~al.(2017){Terrana}, {Harris}, \&
  {Johnson}}]{2017JCAP...02..040T}
{Terrana}, A., {Harris}, M.-J., \& {Johnson}, M.~C. 2017,
  \href{https://arxiv.org/abs/1610.06919}{arXiv:1610.06919},
  \href{https://dx.doi.org/10.1088/1475-7516/2017/02/040}{\jcap, 2017, 040,
  }{Analyzing the cosmic variance limit of remote dipole measurements of the
  cosmic microwave background using the large-scale kinetic Sunyaev Zel'dovich
  effect}

\bibitem[{{Tishue} et~al.(2025){Tishue}, {Shiveshwarkar}, \&
  {Holder}}]{2025arXiv251025821T}
{Tishue}, A.~J., {Shiveshwarkar}, C., \& {Holder}, G. 2025,
  \href{https://arxiv.org/abs/2510.25821}{arXiv:2510.25821},
  \href{https://dx.doi.org/10.48550/arXiv.2510.25821}{arXiv e-prints,
  arXiv:2510.25821, }{The kSZ optical depth degeneracy and future constraints
  on local primordial non-Gaussianity}

\bibitem[{{Torrado} \& {Lewis}(2019)}]{2019ascl.soft10019T}
{Torrado}, J. \& {Lewis}, A. 2019, {Cobaya: Bayesian analysis in cosmology},
  Astrophysics Source Code Library, record ascl:1910.019

\bibitem[{{Torrado} \& {Lewis}(2021)}]{2021JCAP...05..057T}
---. 2021, \href{https://arxiv.org/abs/2005.05290}{arXiv:2005.05290},
  \href{https://dx.doi.org/10.1088/1475-7516/2021/05/057}{\jcap, 2021, 057,
  }{Cobaya: code for Bayesian analysis of hierarchical physical models}

\bibitem[{Virtanen et~al.(2020)}]{2020SciPy-NMeth}
Virtanen, P., et~al. 2020,
  \href{https://dx.doi.org/10.1038/s41592-019-0686-2}{Nature Methods, 17, 261,
  }{{SciPy} 1.0: Fundamental Algorithms for Scientific Computing in Python}

\bibitem[{{White}(2015)}]{2015MNRAS.450.3822W}
{White}, M. 2015, \href{https://arxiv.org/abs/1504.03677}{arXiv:1504.03677},
  \href{https://dx.doi.org/10.1093/mnras/stv842}{\mnras, 450, 3822,
  }{Reconstruction within the Zeldovich approximation}

\bibitem[{{Yuan} et~al.(2024)}]{2024MNRAS.530..947Y}
{Yuan}, S., et~al. 2024,
  \href{https://arxiv.org/abs/2306.06314}{arXiv:2306.06314},
  \href{https://dx.doi.org/10.1093/mnras/stae359}{\mnras, 530, 947, }{The DESI
  one-per cent survey: exploring the halo occupation distribution of luminous
  red galaxies and quasi-stellar objects with ABACUSSUMMIT}

\bibitem[{{Zeldovich} \& {Sunyaev}(1969)}]{1969Ap&SS...4..301Z}
{Zeldovich}, Y.~B. \& {Sunyaev}, R.~A. 1969,
  \href{https://dx.doi.org/10.1007/BF00661821}{\apss, 4, 301, }{The Interaction
  of Matter and Radiation in a Hot-Model Universe}

\bibitem[{Zhang(2010)}]{Zhang_2010}
Zhang, P. 2010,
  \href{https://dx.doi.org/10.1111/j.1745-3933.2010.00899.x}{Monthly Notices of
  the Royal Astronomical Society: Letters, 407, L36, }The dark flow induced
  small-scale kinetic Sunyaev--Zel'dovich effect,
  \url{http://dx.doi.org/10.1111/j.1745-3933.2010.00899.x}

\bibitem[{{Zhou} et~al.(2023{\natexlab{a}})}]{2023JCAP...11..097Z}
{Zhou}, R., et~al. 2023{\natexlab{a}},
  \href{https://arxiv.org/abs/2309.06443}{arXiv:2309.06443},
  \href{https://dx.doi.org/10.1088/1475-7516/2023/11/097}{\jcap, 2023, 097,
  }{DESI luminous red galaxy samples for cross-correlations}

\bibitem[{{Zhou} et~al.(2023{\natexlab{b}})}]{2023AJ....165...58Z}
---. 2023{\natexlab{b}},
  \href{https://arxiv.org/abs/2208.08515}{arXiv:2208.08515},
  \href{https://dx.doi.org/10.3847/1538-3881/aca5fb}{\aj, 165, 58, }{Target
  Selection and Validation of DESI Luminous Red Galaxies}

\bibitem[{Zonca et~al.(2019)Zonca, Singer, Lenz, Reinecke, Rosset, Hivon, \&
  Gorski}]{Zonca2019}
Zonca, A., Singer, L., Lenz, D., Reinecke, M., Rosset, C., Hivon, E., \&
  Gorski, K. 2019, \href{https://dx.doi.org/10.21105/joss.01298}{Journal of
  Open Source Software, 4, 1298, }{healpy: equal area pixelization and
  spherical harmonics transforms for data on the sphere in Python}

\end{thebibliography}

%%%%%%%%%%%%%%%%%%%%%%%%%%%%%%%%%%%%%%%%%%%%%%%%%%

%%%%%%%%%%%%%%%%% APPENDICES %%%%%%%%%%%%%%%%%%%%%

\appendix

\section{Correlation structure of the binned galaxies}\label{sec:corrstructure_smallscale}

\begin{figure}
\includegraphics[width=\columnwidth]{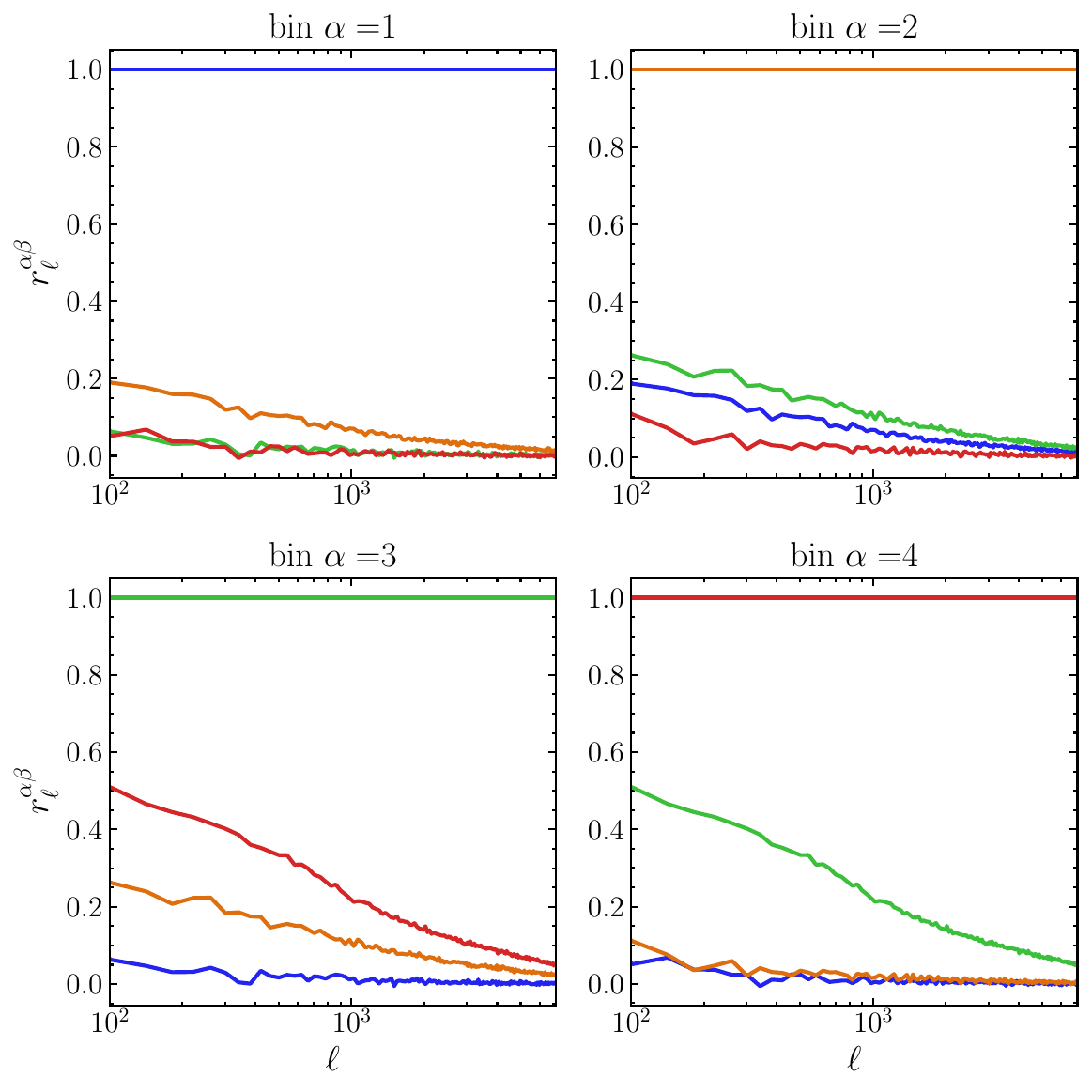}
\includegraphics[width=\columnwidth]{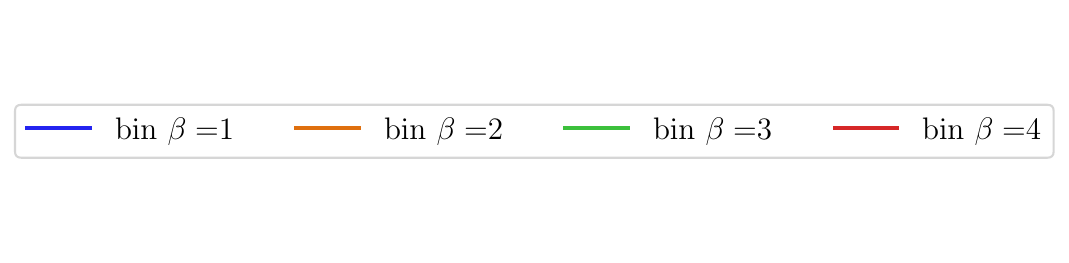}
\caption{Inter-bin correlation structure of galaxy overdensity maps for the $N=4$ bin case.}\label{fig:smallsmallcor_4bins}
\end{figure}

\begin{figure*}
\includegraphics[width=\textwidth]{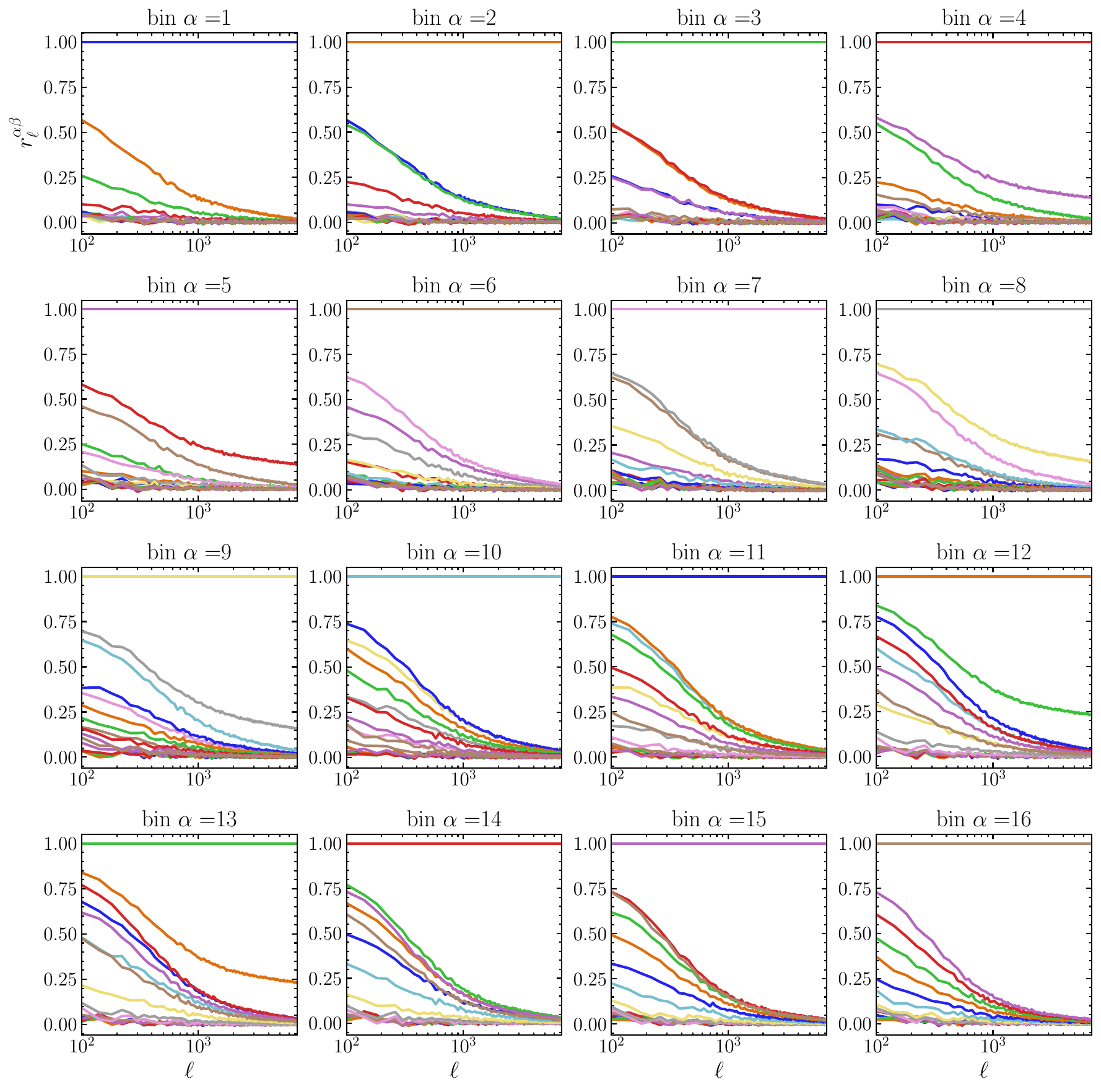}
\includegraphics[width=\textwidth]{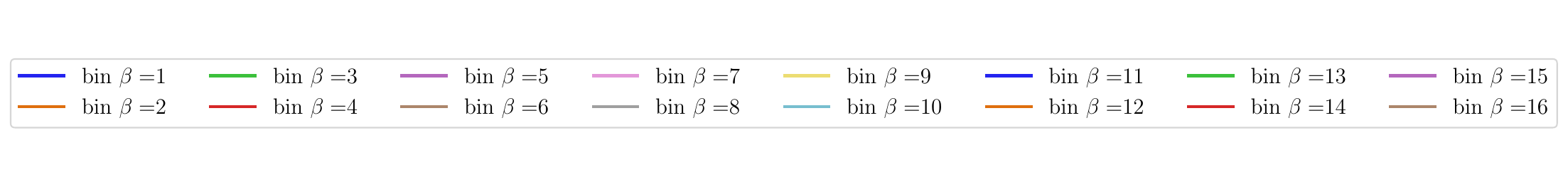}
\caption{Inter-bin correlation structure of galaxy overdensity maps for the $N=16$ bin case.}\label{fig:smallsmallcor_16bins}

\end{figure*}

When we split the galaxies into more redshift bins, these bins have non-zero correlation sourced by their overlapping redshift kernels due to the photo-z uncertainty. We show in this Appendix some plots of the off-diagonal correlation structure, specifically by plotting $r_\ell^{\alpha\beta}\equiv\frac{C_\ell^{g_\alpha g_\beta}}{\sqrt{C_\ell^{g_\alpha g_\alpha}C_\ell^{g_\beta g_\beta}}}$. We show this for 4 bins in Fig.~\ref{fig:smallsmallcor_4bins} and for 16 bins in Fig.~\ref{fig:smallsmallcor_16bins}.  The correlation is clear on large scales, and falls of on small scales due to the shot noise, although the underlying signals remain correlated.

\section{kSZ velocity reconstruction theory}\label{sec:kzsvelrectheory}

The (two-dimensional)\footnote{Throughout, we refer to fields as two-dimensional if they depend on the two-dimensional coordinate $\hat n$; three-dimensional if they depend on the three-dimensional coordinates $(\chi, \hat n)$; and 2+1-dimensional if they depend on the quasi-three-dimensional redshift-binned tomographic coordinates $\left(\alpha,\hat n\right)$. Thus both the three-dimensional vector field $\hat v(\chi, \hat n)$ and the scalar field $v_r(\chi, \hat n)$ are three-dimensional in our language.} kSZ temperature anisotropy $\Delta T^{\mathrm{kSZ}}$  in direction $\hat n$ is of the form
\begin{align}
\frac{\Delta T^{\rm {kSZ}}(\hat n)}{\bar T} = &- \sigma_T\int d\chi  a(\chi)  v_r (\chi,\hat n) n_e(\chi,\hat n) ,\label{ksz_def}\\
\equiv& - \int d\chi    v_r (\chi,\hat n)\dot \tau(\chi, \hat n) \label{diffoptdetph}
\end{align}
where $\bar T$ is the mean CMB temperature; $\sigma_T$ is the Thomson scattering cross section; $a(\chi)$ is the scale-factor at comoving distance $\chi$; $v_r(\chi, \hat n) \equiv \vec v(\chi, \hat n) \cdot \hat n$ is the three-dimensional radial velocity field at $(\chi, \hat n )$; $n_e(\chi, \hat n)$ is the electron density at $(\chi, \hat n)$; and $\dot \tau(\chi, \hat n)\equiv\frac{d\tau }{d\chi}(\chi, \hat n)$ is the differential optical depth.

The (2+1)-dimensional galaxy number density field $\delta^{g}_\alpha(\hat n)$ is 
\begin{align}
{\delta ^{g}_\alpha}(\hat n) = \frac{\int {d\chi }\frac{dN^\alpha}{d\chi}\delta^g(\chi, \hat n)}{\int d\chi \frac{d N^\alpha}{d\chi}}\label{dndz_gal_def}
\end{align}
where $\frac{dN^\alpha}{d\chi}$ is the redshift distribution of the galaxies in bin $\alpha$.  {When the large-scale velocity is held fixed (i.e., to the realization of the velocity field in our Universe}), the galaxies and the kSZ are correlated  at the two-point level according to \footnote{In this equation we restore a factor of $(-1)^m$ that was erroneously omitted in Equation (6) of~\cite{2025JCAP...05..057M}.}
\begin{align}
 \left<\frac{\Delta T^{\mathrm{kSZ}}_{\ell m} }{\bar T}\delta^{g^\alpha}_{\ell^\prime m^\prime}\right>_{\mathrm{fixed\,} v} =&  \sum_{\ell_1 m_1 \ell_2 m_2}(-1)^{m_1}W^{\ell_1 \ell_2 \ell}_{m_1 m_2 -m}\times\nonumber
 \\&\int d\chi \left<v_r{}_{\ell_1 m_1}(\chi)\dot \tau  _{\ell_2 m_2} (\chi)
 (\delta ^{g}_\alpha)_{\ell^\prime m^\prime}\right>,\label{generalbispectrum}
\end{align}
where $W^{\ell_1 \ell_2 \ell_3}_{m_1 m_2 m_3}$ is the following mode-coupling matrix:
\begin{align}
W^{\ell_1 \ell_2 \ell_3}_{m_1 m_2 m_3} = \sqrt{\frac{(2 \ell_1+1)(2 \ell_2+1)(2 \ell_3+1)}{4\pi}}&\wignerJ{\ell_1}{\ell_2}{\ell_3}{0}{0}{0}\times\nonumber\\
&\wignerJ{\ell_1}{\ell_2}{\ell_3}{m_1}{m_2}{m_3}.
\end{align}
Here, $\wignerJ{\ell_1}{\ell_2}{\ell_3}{m_1}{m_2}{m_3}$ are the Wigner 3-J symbols. In the squeezed $L\equiv \ell_1\ll\ell_2,\ell_3$ limit (which dominates), this can be simplified and approximated as
\begin{align}
\left<\frac{\Delta T^{\mathrm{kSZ}}_{\ell m} }{\bar T}(\delta ^{g}_\alpha)_{\ell^\prime m^\prime}\right> \simeq     \int d\chi \sum_{LM} (-1)^{M}\wignerJ{\ell}{\ell^\prime}{L}{m}{m^\prime}{-M}     f_{\ell  \ell^\prime L}\times \nonumber\\
 C_{\ell^\prime}^{\dot\tau {g_\alpha}} (\chi) v_r{} _{LM}(\chi)\label{twopointnondiag}
\end{align}
where 
\begin{equation}
f_{\ell  \ell^\prime L} =  \sqrt{\frac{(2\ell+1)(2\ell^\prime+1)(2L+1)}{4\pi}}\wignerJ{\ell}{\ell^\prime}{L}{0}{0}{0}
\end{equation}
and 
\begin{align}
\left<\dot \tau_{\ell m}(\chi) (\delta ^{g}_\alpha)_{\ell^\prime m^\prime}\right>\equiv C_\ell ^{\dot \tau g_\alpha}(\chi) \delta_{\ell \ell^\prime}\delta_{m m^\prime},
\end{align}
resulting from the statistical isotropy of the small-scale statistics of the galaxies and the electrons. Equation~\eqref{twopointnondiag} is the basis for kSZ velocity reconstruction: the only ``unknown'' in this expression is $v_{r}{}_{LM}(\chi)$, as the left hand side is directly measurable from data and we can model $C_\ell ^{\dot \tau g_\alpha}(\chi) $. Given $N$ independent galaxy overdensity maps labelled by $\alpha$, we can create $N$ velocity estimates
\begin{equation}
(v_\alpha)_{LM}\sim  W^\prime{}^{L \ell_1 \ell_2 }_{M m_1 m_2 }\left<T_{\ell_1 m_1}^{\mathrm{kSZ}} (\delta ^{g}_\alpha)_{\ell_2 m_2}\right>
\end{equation}
where $W^\prime$  are  weights that depend on $C_\ell^{\dot \tau g_\alpha}(\chi)$. The optimal weights have been derived several times in the literature (see, \textit{e.g.}~\citealt{2018PhRvD..98l3501D,2020PhRvD.102d3520M,2023JCAP...02..051C,2024arXiv240500809B}). The real-space estimator we use (see Equation~\eqref{vkSZ_recon}) is a real-space implementation of this optimal QE. Note that the final quantity estimated is a \textit{projected} quantity (\textit{i.e.}, with the redshift-dependence integrated out)
\begin{align}
(v_\alpha) _{LM}=\int W^{v^{\mathrm{kSZ}}_{\alpha}}_{LM}(\chi) v_{r,_{LM}}(\chi) d\chi
\end{align}
where $ W^{v^{\mathrm{kSZ}}_{\alpha}}_{LM}(\chi)$ is a projection kernel. In practice, this is scale-independent on large scales such that $ W^{v^{\mathrm{kSZ}}_{\alpha}}_{LM}(\chi)= W^{v^{\mathrm{kSZ}}_{\alpha}}(\chi)$. The projection kernel depends on $C_\ell^{\dot\tau g}$:
\begin{align}
W^{v^{\mathrm{kSZ}}_{\alpha}}(\chi) \approx \frac{C_{\ell=\bar \ell}^{\dot \tau g_\alpha}(\chi)}{C_{\ell=\bar \ell}^{\tau g_\alpha}},\label{correct_w}
\end{align}
 where $\bar \ell$ is some characteristic high-$\ell$ multipole which dominates the signal-to-noise~\citep{2024arXiv240500809B}, and
\begin{equation}
    C_\ell^{\tau g_\alpha}\equiv\int d\chi C_\ell^{\dot \tau g_\alpha}(\chi)
\end{equation}
is the isotropic optical depth-galaxy cross power (with $\tau$ the total optical depth as opposed to the differential optical depth $\dot \tau$ such that $\tau \equiv \int \dot \tau d\chi$). If we make the assumption that the electrons evolve slowly in each redshift bin compared to the galaxy density, this can be further approximated as
\begin{align}
W_{LM}^{v_r^{\mathrm{kSZ,\alpha}}}(\chi) \approx \frac{\frac{dN^\alpha}{d\chi}}{\int d\chi \frac{dN^\alpha}{d\chi}},\label{ksz_proj}
\end{align}
and so the velocity we estimate is the mean velocity projected along the redshift distribution of the galaxies used in the reconstruction.

\section{Description of simulations}\label{sec:abacus_description}

We use $N$-body simulations populated with galaxies using a halo occupation distribution (HOD) to validate our pipeline and complete our theory model as described in the text (Section~\ref{sec:simdem}). In this Section, we describe how we create the simulations.

\subsubsection*{Lightcone simulations}

The  \texttt{AbacusSummit} simulations are a set of $N$-body simulations run with \texttt{Abacus}~\citep{2021MNRAS.508..575G}, with LRG-like galaxies added according to the HOD prescription of~\citep{2024MNRAS.530..947Y}. There are 25 ``\texttt{base}'' \texttt{AbacusSummit} lightcone simulations and 2 ``\texttt{huge}'' simulations. The ``\texttt{huge}'' simulations cover enough volume to be full-sky; the ``\texttt{base}'' simulations cover only 1/8 of the sky, and so cannot be used to fully cover the survey geometry we are interested in. Thus we use one of the  ``\texttt{huge}'' simulations to test our pipeline, and create six quasi-independent simulations from it by rotating the sky.

\subsubsection*{$\tau$ and kSZ simulations}

We create a $\tau$ map  using the transfer function method of \citet{Liu:2025}, in which the dark matter density field from an $N$-body simulation is mapped to the corresponding gas field using a transfer function calibrated on a hydrodynamical simulation; we use the gas–dark matter transfer function measured from the TNG300-1 simulation~\citep{2019ComAC...6....2N} at $z=0.5$. 

Concretely, we use dark matter density shells from the \texttt{AbacusSummit} lightcone (with shell spacing $\sim10 \ h^{-1}{\rm Mpc}$ between $0.3<z<1.1$), and apply the gas–dark matter transfer function measured from the TNG300-1 simulation~\citep{2019ComAC...6....2N} at $z=0.5$ to obtain $\delta_{\rm gas}$. The resulting $\tau$ shells are then summed (after converting to the appropriate units) to yield the redshift-evolving $\tau(\hat{n},z)$ map at \texttt{nside}=8192.

We create a kSZ map by additionally multiplying each  $\tau$ shell by the line-of-sight halo velocity field before summing, with the velocity field estimated from the particle velocities within each pixel.

\subsubsection*{Small-scale DESI DR9 LRG Galaxy simulations}

We add galaxies according to a HOD  described in~\cite{2024PhRvD.109j3534H}.We downsample the galaxies to follow the $\frac{dN}{dz}$ measured from the data.   We add photo-$z$ noise of width $0.027\times(1+z)$to the galaxies corresponding to the extended LRG sample.

For the small-scale two-dimensional galaxy overdensity maps, we create tomographically binned galaxy overdensity maps according to the same prescription we do on the data, by applying the photo-$z$ boundaries in Table~\ref{tab:redshiftboundaries} and projecting to 2+1 dimensions following the description in Section~\eqref{sec:projection_gals}, although on the full sky. We cross-correlate this maps with the $\tau$ map to  estimate the $C_\ell^{g\tau}$ used in our kSZ reconstruction pipeline.

\subsubsection*{Large-scale DESI DR10 LRG Galaxy simulations}

To reproduce the photometric redshift errors and $\frac{dN}{dz}$ distribution of the DR10 Extended LRG sample, we add Gaussian noise of width $\sigma_z=0.02\times(1+z)$ to the observed (redshift-space) redshift of each object and downsample the galaxy catalog to match the observed $\frac{dN}{dz}$. We then apply a mask so that the mock footprint closely matches that of the data. Finally, we divide the mock catalog into four redshift bins, matched to the data both in range and $\frac{dN}{dz}$.

\subsubsection*{2-dimensional velocity field simulations}
We create an estimate of the ``true'' velocity field that we wish to estimate $v^{\mathrm{true}}$ by projecting the true halo velocities of the objects according to  
\begin{align}
\hat v^{\mathrm{true},\alpha}(\hat n) = \frac{\sum _i v_i \frac{dN^\alpha}{dz}(z^i)}{\int \frac{dN}{dz}},
\end{align}
where the sum is over all objects $i$ in the pixel at $(\hat n)$ and $\frac{dN}{dz^\alpha}(z^i)$ is the redshift kernel of the velocity field we wish to estimate evaluated at the redshift of the object $i$. Again, we normalize by the $\frac{dN}{dz}$ of the entire simulation. We estimate $\hat v^{\mathrm{true}}$ \textit{before} downsampling as it is a cosmological quantity that does not depend on the details of the survey used to estimate it.

The quantity that $\hat v^{\mathrm{true}}$ is estimating can be modelled as the following projection of the underlying velocity field:
\begin{align}
v^{\mathrm{true},\alpha}(\hat n) = \int d\chi  \frac{dN^\alpha}{d\chi}v (\chi, \hat n).
\end{align}

\section{Auto power spectra of the continuity-equation velocity}\label{sec:boryanauto}

\begin{figure*}
\includegraphics[width=0.8\textwidth]
{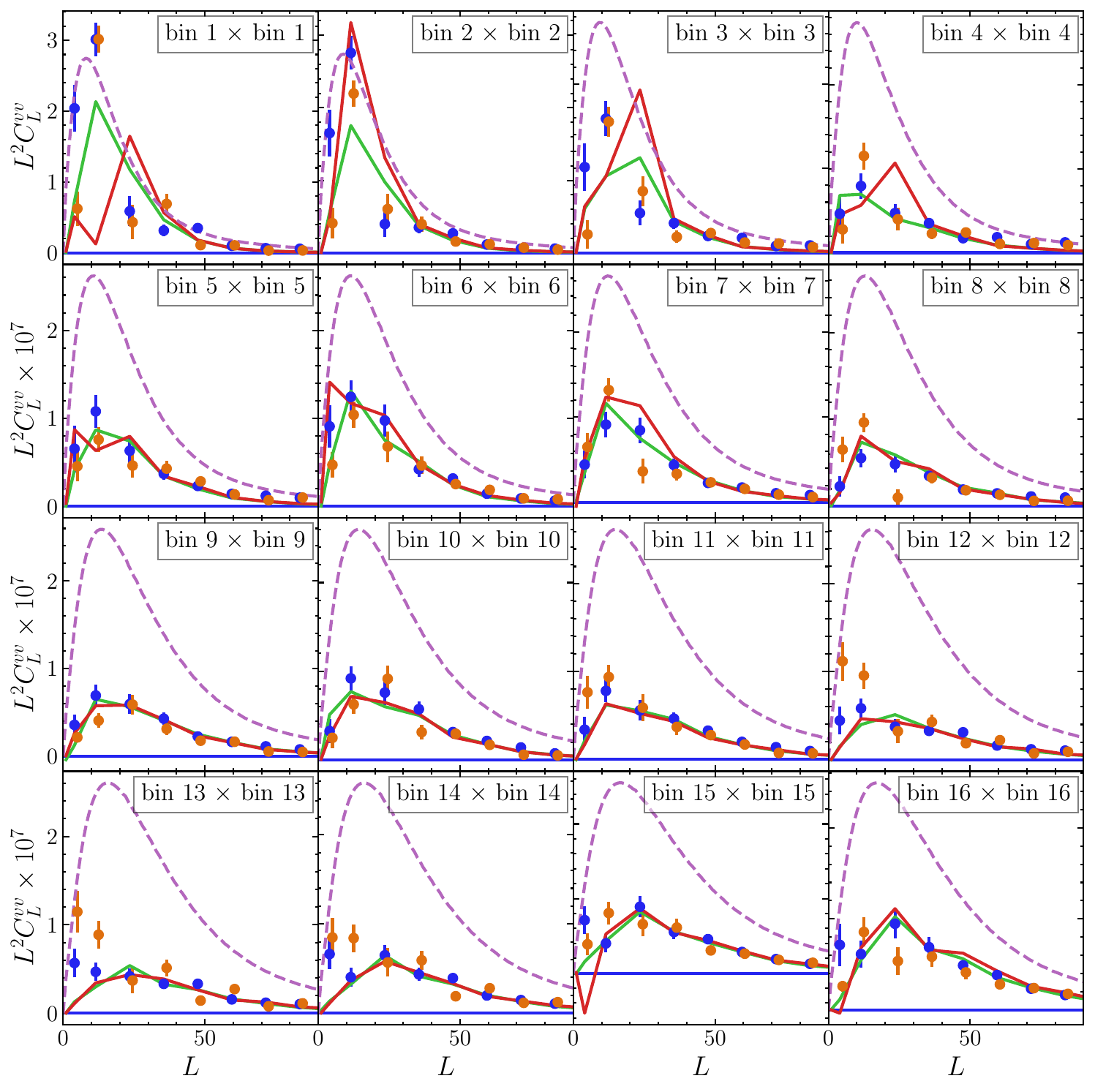}
\includegraphics[width=0.6\textwidth]{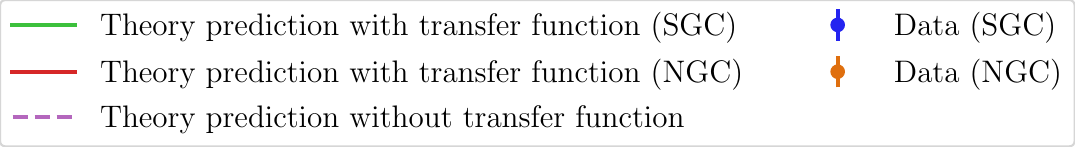}
\caption{The measured auto power spectra of the continuity-equation velocities. We show the $\Lambda$CDM prediction in dashed purple; the continuity equation reconstruction suppresses the power. We calibrate this suppression with simulations and we show the mean from the simulations in red and green. The simulations capture the power in the data well.}\label{fig:boryanaauto16bins}
\end{figure*}

In this Appendix we show the measured auto power spectra of the continuity-equation auto power spectra, and show that our model is sensible. In Fig.~\ref{fig:boryanaauto16bins} we show the measurement as well as the prediction from theory and simulations. The $\Lambda$CDM theory is indicated in dashed purple; the mean from the simulations is shown in red and green for is shown in for NGC and SGC respectively. The power suppression with respect to theory seen in the simulations seems to match what is seen in the data.

\section{Stability of constraints for various data combinations}\label{sec:biasconstraints}

We found evidence for a foreground bias in the diagonal of $C_L^{v^{\mathrm{kSZ}}g}$. In this Appendix we explore how such a bias may propagate into the amplitude constraints. 

\begin{figure*}

\includegraphics[width=0.32\textwidth]
{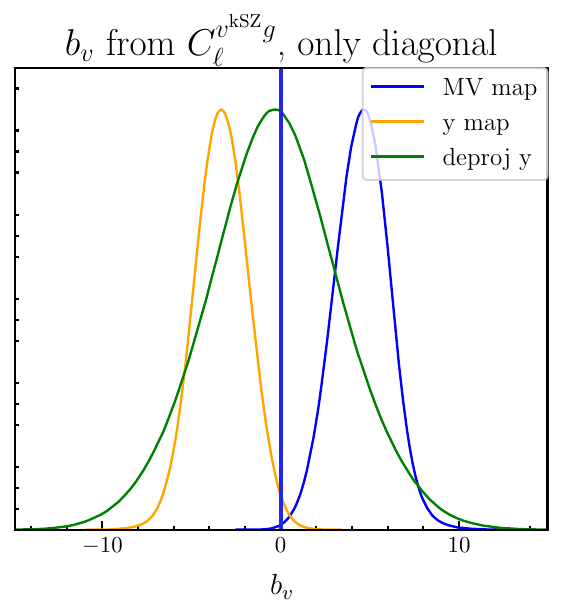}
\includegraphics[width=0.32\textwidth]{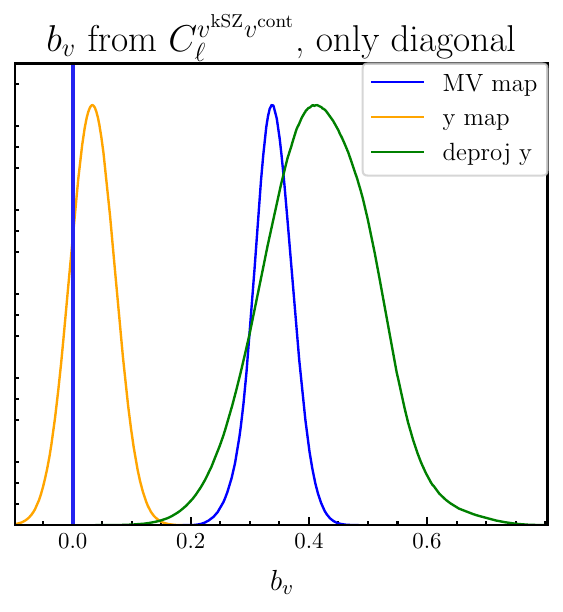}
\includegraphics[width=0.32\textwidth]{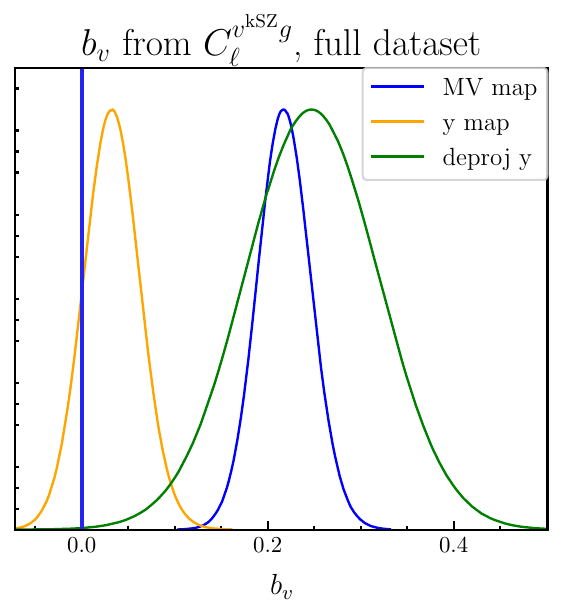}\\
\includegraphics[width=0.32\textwidth]{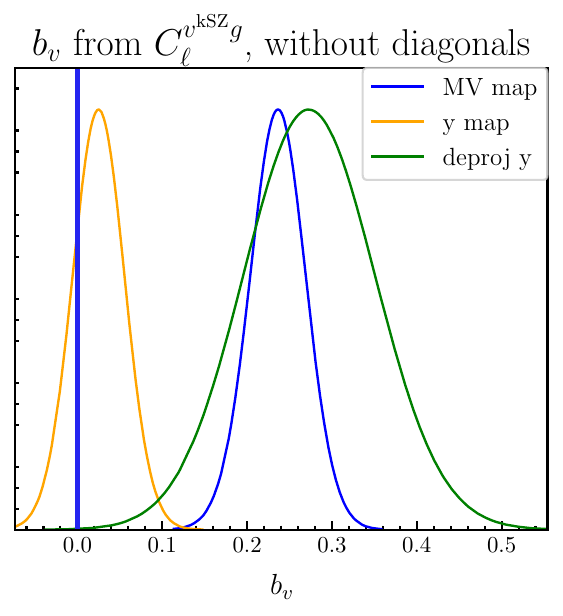}
\includegraphics[width=0.32\textwidth]
{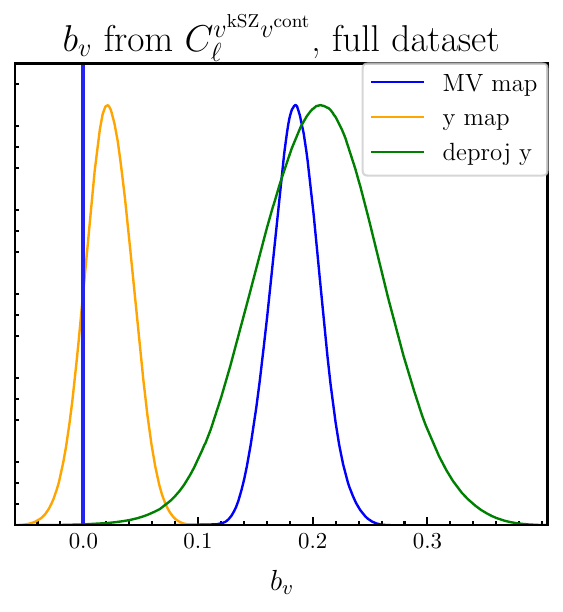}
\includegraphics[width=0.32\textwidth]{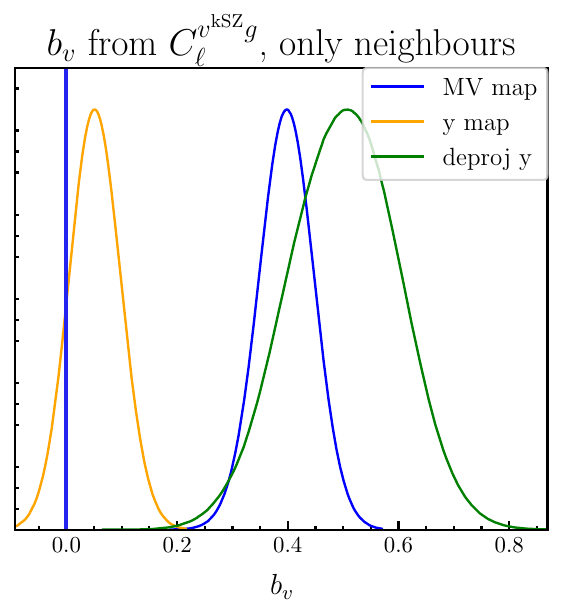}

\caption{The constraints on the overall amplitude $b_v$ for several data combinations. In particular, we are testing consistency for different signal maps and also immunity to foregrounds. The yellow curves,labelled ``$y$ map'', should be null, as the temperature leg used for velocity reconstruction is expected to have no kSZ signal and only foregrounds. At the same time, the blue and green curves --labelled ``MV NILC'' and ``deproj $y$'' respectively--should be consistent, as we expect these are both unbiased measurements of the kSZ signal but with different amounts of foregrounds and noise (in particular, the $y$-deproj map has no tSZ, at the expense of possibly higher CIB, and it also has significantly higher noise). For all of the $C_L^{v^{\mathrm{kSZ}v^{\mathrm{cont}}}}$ cases that we look at --both the diagonal-only and the diagonal+neighbours case--these tests pass. However, for the $C_L^{v^{\mathrm{kSZ}g}}$ diagonals (the first plot), these tests fail. We see a strong negative signal with the $y$-map. Because this subset of the data contributes only a small amount of the overall signal-to-noise, it is not significant enough to leak into the full measurement (the third plot), but it is clear that these points should not be included and combined with the rest of the points as they are not consistent. We also inspect the constraint from an ``only neighbours'' formulation of $C_L^{v^{\mathrm{kSZ}}g}$, as there is a lot of redshift overlap between neighbouring bins.}\label{fig:bv_diagetc}
\end{figure*}

We show in Figure~\ref{fig:bv_diagetc} the constraints on one overall velocity bias parameter for the $C_L^{v^{\mathrm{kSZ}}v^{\mathrm{cont}}}$ case and for the  $C_L^{v^{\mathrm{kSZ}}g}$, where we consider the diagonals separately to the remaining points. We see here that we see signals consistent with the presence of foregrounds in the $C_L^{v^{\mathrm{kSZ}}g}$ measurement, as we see a  non-zero signal when we use the $y$ map for the velocity reconstruction, and an inconsistent signal when we use the MV NILC and the $y$-deprojected NILC. Interestingly, the bias is not enough to be noticeable in the overall dataset.

\section{Dependence of auto power spectra on cluster masking for the 16-bin case}\label{sec:clustermasking_auto}
\begin{figure*}
\includegraphics[width=\textwidth]{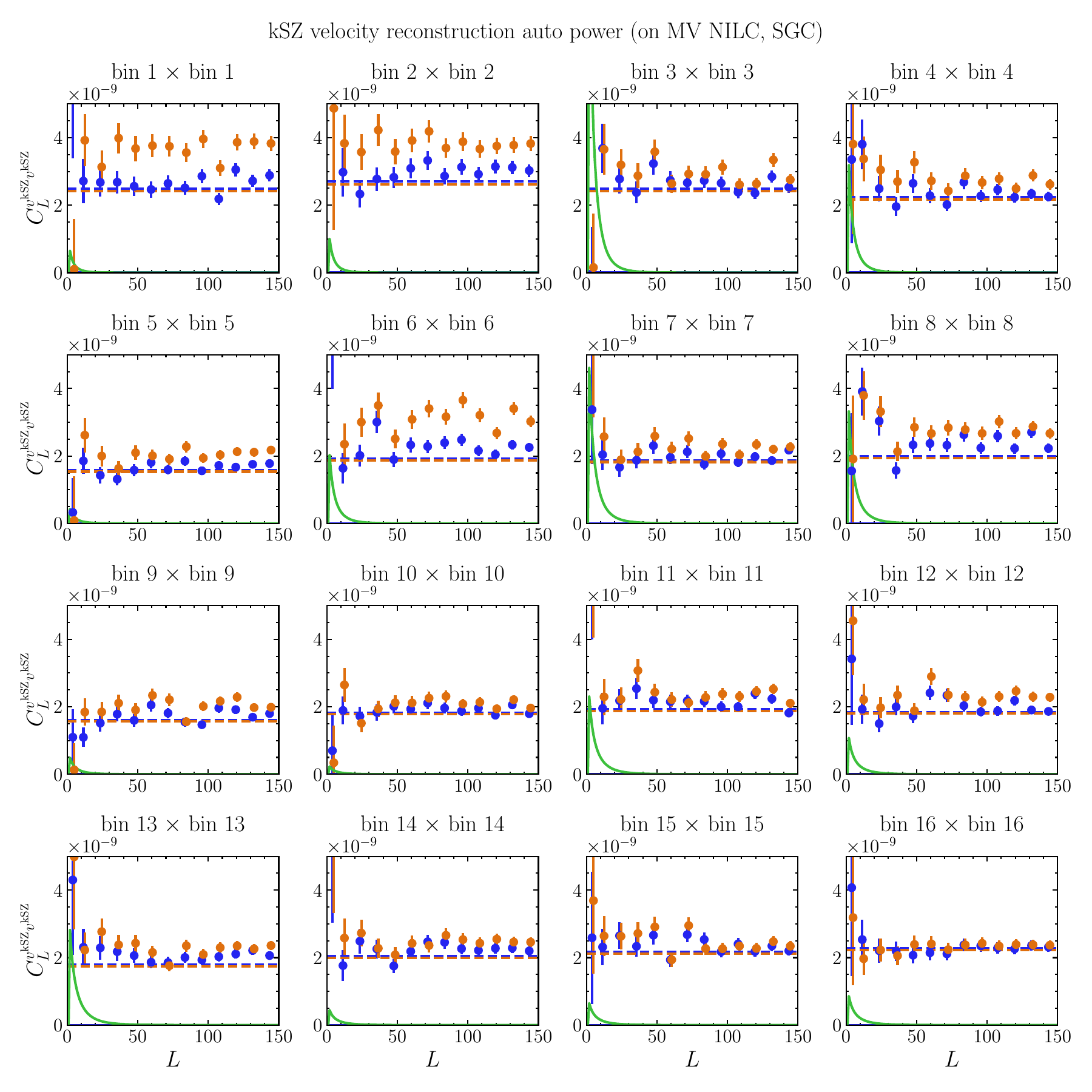}
\includegraphics[width=0.7\textwidth]{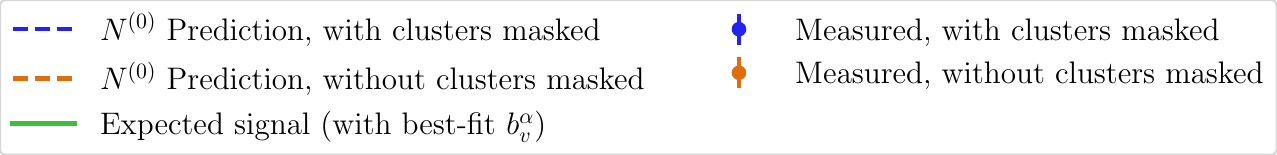}
\caption{Auto power spectra of the kSZ reconstructed velocity for the $N=16$, SGC bin case, with the unconstrained NILC. The reconstruction with clusters masked has power that agrees with the analytic prediction; as in the 4-bin case, the one with clusters retained does Note that in the higher redshift bins, even the unmasked reconstruction agrees with theory; this is not unexpected given that the clusters are mostly at lower $z$.  Note that we have slightly offset the orange ``measured, without clusters masked'' points for slightly better comparison with the other points.}\label{fig:auto_tsz_16bins_sgc}
\end{figure*}

\begin{figure*}
\includegraphics[width=\textwidth]{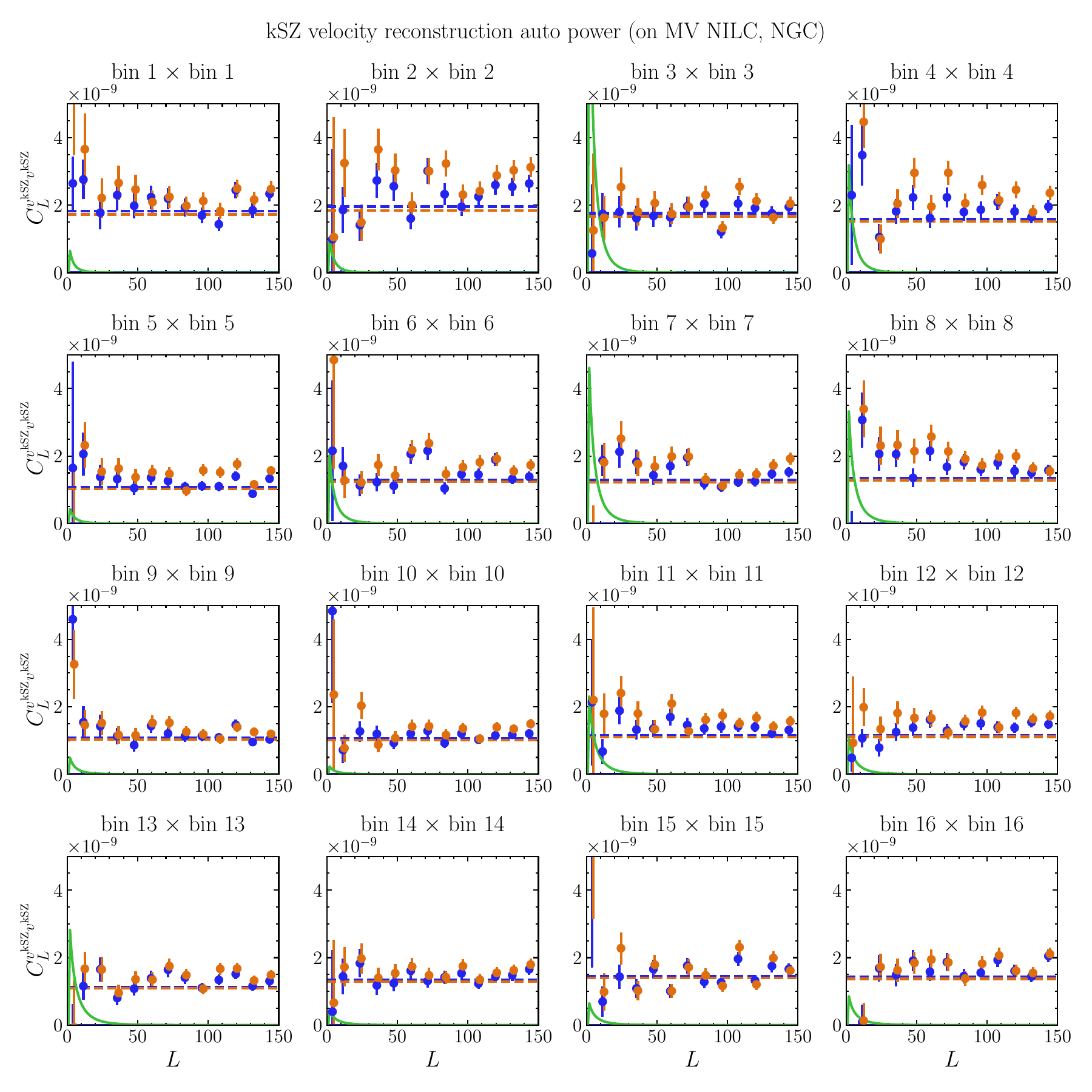}
\includegraphics[width=0.7\textwidth]{Images/ksz_auto_allbins_autos_compareclustersmask_margyerr16_bins_sepbins_deltaell12_ngc_legend.pdf}
\caption{Auto power spectra of the kSZ reconstructed velocity for the $N=16$, NGC bin case, with the unconstrained NILC. The reconstruction with clusters masked has power that agrees with the analytic prediction; as in the 4-bin case, the one with clusters retained does Note that in the higher redshift bins, even the unmasked reconstruction agrees with theory; this is not unexpected given that the clusters are mostly at lower $z$.  Note that we have slightly offset the orange ``measured, without clusters masked'' points for slightly better comparison with the other points.}\label{fig:auto_tsz_16bins_ngc}
\end{figure*}

As discussed in Section~\ref{sec:autospectra}, we find that the observed offset between the measured auto power spectrum of the kSZ-reconstructed velocity and the analytic prediction is mitigated when tSZ clusters are masked before performing the kSZ velocity reconstruction. In this Appendix we show explicitly the effect of this procedure for the baseline $N=16$ bin case. This is shown for SGC in Fig.~\ref{fig:auto_tsz_16bins_sgc} and for NGC in Fig.~\ref{fig:auto_tsz_16bins_ngc}.

\section{Degeneracies in the $f_{\mathrm{NL}}$ analysis and full $f_{\mathrm{NL}}$ corner plot}\label{sec:fnlcorner}

\begin{figure*}
\includegraphics[width=\textwidth]{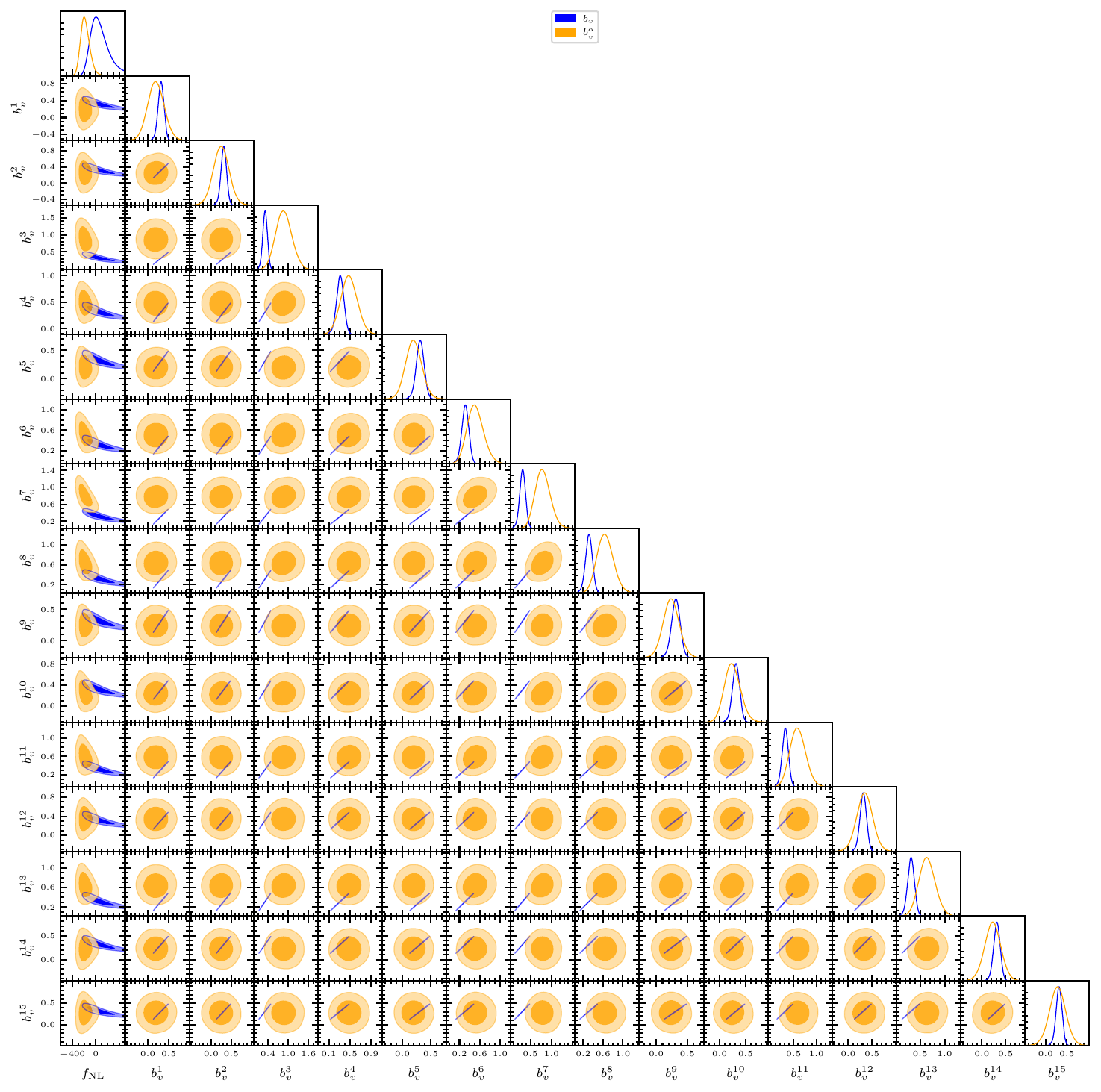}
\caption{Full corner plot for the $f_{\mathrm{NL}}$ analysis}\label{fig:fullcorner}
\end{figure*}

\begin{figure*}
\includegraphics[width=\columnwidth]{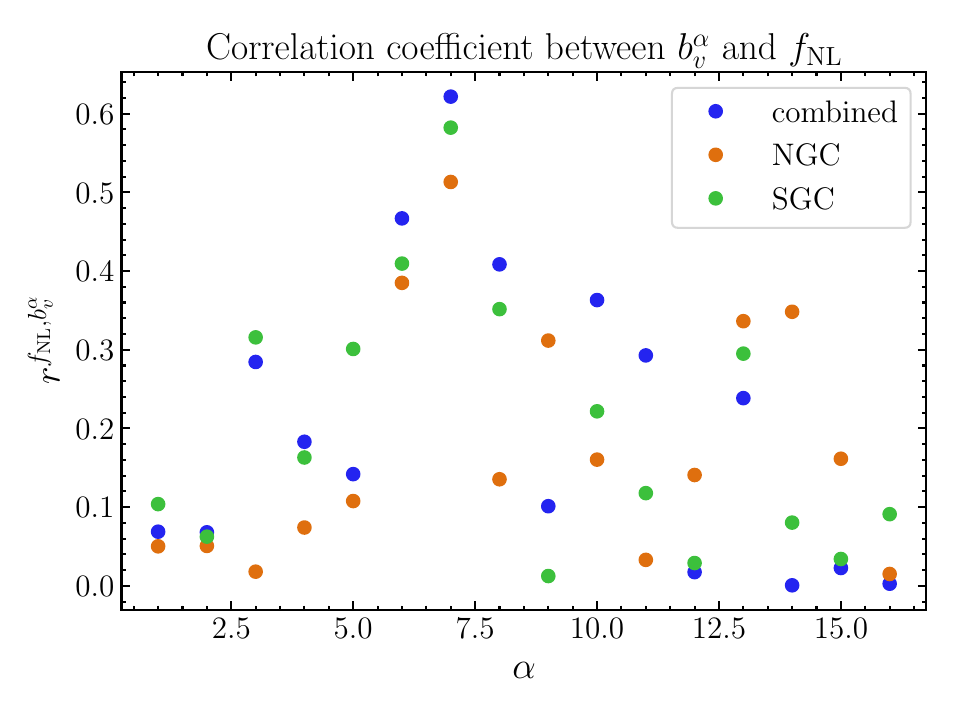}
\includegraphics[width=\columnwidth]{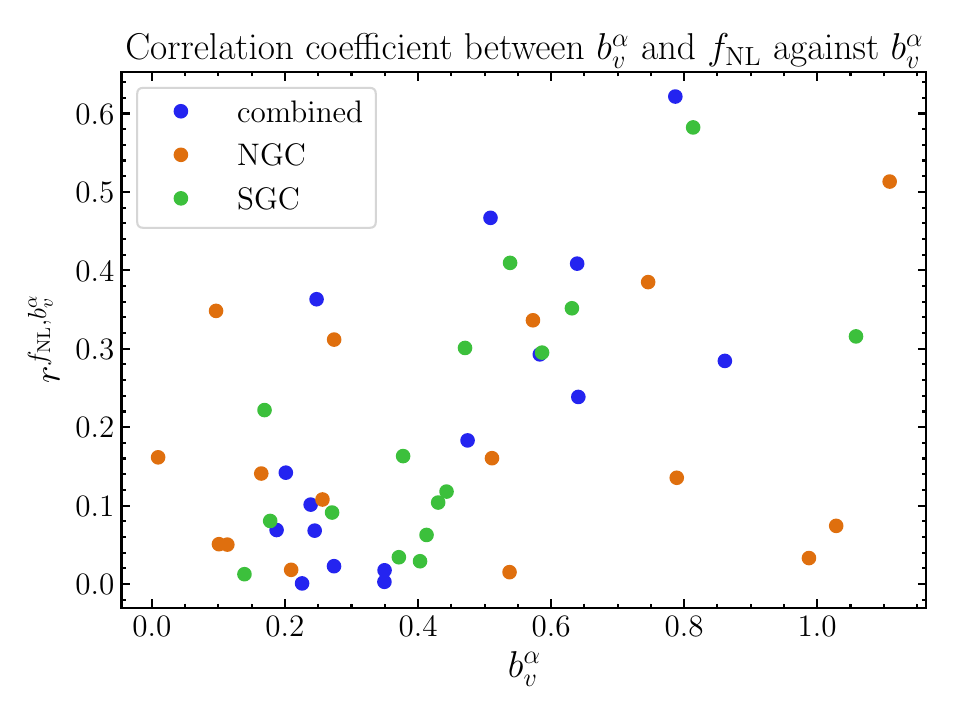}

\caption{The correlation coefficient between the velocity bias and $f_{\mathrm{NL}}$,plotted against bin number on the left to indicate which bins are dominating the constraint, and against value of the velocity bias on the right. We see that bin 7 is the most correlated with ${f_\mathrm{NL}}$, and we see a positive correlation between $r^{b_v^\alpha f_{\mathrm{NL}}}$ and $b_v^\alpha$. }\label{fig:corrfnlalpha}
\end{figure*}

The full corner plot for the $f_{\mathrm{NL}}$ analysis is shown in Fig.~\ref{fig:fullcorner}. The use of a multiple-$b_v^\alpha$ model reduces the strong correlation between $b_v$ and $f_{\mathrm{NL}}$ that is clearly evident int he single-$b_v$ model; however, the $b_v^\alpha$ remain correlated with $f_{\mathrm{NL}}$, even if the correlation is not as strong as in the single-$b_v$ case. Indeed, the bins with higher $b_v^\alpha$ are most correlated with $f_{\mathrm{NL}}$, as they are precisely the subsets of the dataset where we have most constraining power; this is indicated in Fig.~\ref{fig:corrfnlalpha} where we extract the correlation coefficient between each $b_v^\alpha$ parameter and $\mathrm{f_{\mathrm{NL}}}$ and plot this against the posterior mean of $b_v^\alpha$. A positive correlation is evident in this plot.

% Don't change these lines
\bsp	% typesetting comment
\label{lastpage}
\end{document}